\documentclass[a4paper]{article}
\usepackage{graphicx} 

\usepackage{amsmath}
\usepackage{amssymb}
\usepackage{mathtools}
\usepackage{braket}
\usepackage{simpler-wick}
\usepackage{dsfont}
\usepackage[bitstream-charter]{mathdesign}

\DeclareSymbolFont{usualmathcal}{OMS}{cmsy}{m}{n}
\DeclareSymbolFontAlphabet{\mathcal}{usualmathcal}

\newcommand{\R}	{\mathds{R}}
\newcommand{\C}	{\mathbb{C}}
\newcommand{\Z}	{\mathbb{Z}}
\newcommand{\N}	{\mathbb{N}}

\DeclareMathOperator{\tr}{\tr}

\renewcommand{\tr}	{\mathrm{tr}}

\newcommand\id		{\mathbf{1}}

\newtheorem{theorem}{Theorem}[section]
\newtheorem{definition}{Definition}[section]

\newif\ifsol
\soltrue 

\renewcommand{\epsilon}{\varepsilon}

\renewcommand{\tilde}{\widetilde}

\newcommand\be{\begin{equation}}
\newcommand\ee{\end{equation}}
\newcommand\bet{\begin{tabular}}
\newcommand\eet{\end{tabular}}

\newcommand\ve{\varepsilon}

\newcommand\m{\mu}

\DeclarePairedDelimiter{\abs}{\lvert}{\rvert}

\definecolor{SkyBlue}{rgb}{0.53,0.81,0.92}

\usepackage[most]{tcolorbox}
\usepackage{adjustbox}
\tcbset{colback=SkyBlue!35!white, colframe=black!50!black, 
        highlight math style= {enhanced, 
            colframe=black,colback=black!20!white,boxsep=0pt,breakable}
        }
        
\tikzset{
    partial ellipse/.style args={#1:#2:#3}{
        insert path={+ (#1:#3) arc (#1:#2:#3)}
    }
}

\usepackage{CJKutf8}

\newcommand\ie{\begin{equation}\begin{aligned}}
\newcommand\fe{\end{aligned}\end{equation}}

\title{Introduction to\\ Abelian Anyons
in Planar Systems}
\author{Pieralberto Marchetti$^{a,b}$  \\
        \small $^{a}$ Dipartimento di Fisica e Astronomia, Università di Padova, Italy \\
        \small $^{b}$  INFN, Sezione di Padova, Italy\\
        \small pieralberto.marchetti@unipd.it\\
        \small 
        pieralberto.marchetti@pd.infn.it
}
\date{}
\begin{document}
\maketitle

\begin{abstract}

This paper is a review of the theory of abelian anyons in planar systems at an introductory level and with focus on the formalism of quantum field theory, but with the aim of clarify the connections between the mathematical structure and the theoretical and
experimental physical aspects of these particle excitations.\footnote{This review is an extended version of the lectures notes presented at the Young Researchers School 2024 Maynooth "Topological aspects of low-dimensional quantum physics" }

\end{abstract} \hspace{10pt}

\section{Introduction}

What are anyons? 
They are particle excitations with statistics other than bosonic and fermionic and (by default) spin neither integer nor half-integer that exist in the spatial dimension $d=2$, where the name anyon was originally introduced; see e.g. \cite{Wilczek} and \cite{khare2005fractional}, where one can also find many references to topics not discussed in this review, such as  statistical mechanics, and \cite{comtet2000aspects}, particularly the contribution of Myrheim, "Anyons", with a rich bibliography. More recently the name has also been used for particle excitations with similar statistics in the spatial dimension $d=1$,  but they will not be discussed here. 

More precisely, if in quantum systems the complex-valued wave-function of identical particle excitations for oriented exchanges acquires different and inverse phase factors for the two orientations these particles are called abelian anyons. The same
phase factors appear in the framework of quantum field theory if we perform an equal-time oriented exchange of the fields creating the corresponding
particle excitations.
Since the oriented exchanges generate the braid groups, the corresponding statistics is called braid statistics. Anyons can only exist in $d<3$ space dimensions, where the exchanges can be oriented.

 In this review we consider only the case of two-dimensional space, more precisely $\R^2$ or its contractible subsets and the focus of the review is on the quantum field theory of abelian anyons. 

The existence of this type of particles in $d=2$ was first conjectured in the late 1970s by Leeinaas and Myrheim \cite{Leinaas:1977fm} (and later independently in \cite{Goldin:1981sm}), but a concrete physical example occurred only in the 1980s in a two-dimensional electron system with the fractional quantum Hall effect (see e.g. \cite{prange1989quantum}, \cite{stormer1999nobel}), the first effect which was explained only through anyons, for which Tsui, Störmer (experimentalists) and Laughlin (theorist) received  the Nobel Prize in Physics in 1998.
The experimental verification of their statistics is even more recent: in 2020 \cite{nakamura2020direct}!

A generalization of abelian anyons is obtained by substituting the phases in oriented exchanges with unitary matrices of higher dimension, the corresponding particle excitations are called non-abelian anyons (and sometimes plektons) and we just mention them in the last section although they could play an important role
 in quantum computing (see e.g. \cite{nayak2008non} and for a very recent experimental attempt in this direction \cite{aghaee2024interferometric}); their treatment would require a lot of additional work.
 
 Although anyons in physical systems arise naturally only in condensed matter, we also discuss them in the relativistic approach, following a time-honored tradition, which actually provided deep insights even for condensed matter as exemplified in \cite{Wilczek}. The relativistic framework of quantum field theory also has a clearer mathematical foundation that allows for a rigorous and model-independent discussion of their properties, as outlined in the last section.

 The plan of the paper is as follows:
In section 2 we show the emergence of braid statistics in $d=2$ from general considerations. In section 3 we discuss the spin and in section 4 we introduce a quantum-mechanical model of abelian anyons. 
In section 5 we discuss their treatment in quantum field theory using the abelian Chern-Simons gauge fields and in section 6 we give a brief description of the system in which such particle excitations are experimentally proven to appear in thehttps://arxiv.org/ physical world: the fractional
quantum Hall effect.
Finally, in section 7 we outline general properties of anyons using the algebraic quantum field theory approach. We discuss in particular: the most general structure of statistics in $d=2$ compatible with the physical principles of local quantum field theory leading to a non-abelian generalization of abelian anyons corresponding to higher dimensional representations of braid groups, the localizability of the quantum fields of anyons, the relation with parity-breaking and the spin-statistics connection.

A mathematical definition of  the currents used in the text and a discussion on vortex/ particle duality are deferred to appendices. 

 (Except in special cases where it is useful to have explicit dimensional universal constants, we set the velocity of light $c=1$ and the Planck constant $\hbar = 1 $.

 In many formulas, when it is obvious, the dependence on space-time coordinates is not explicitly written.
 
 The drawings for this review were made by  Farbod Rassouli.)

\section{Anyons and braids: statistics in d=2}

A key question concerning anyons is: 
What is different in $d=$ 2 or 1 spatial dimensions compared to $d = 3$, that allows them to exist?

In this section we show how the braid statistics appears and we characterize the braid groups, then we discuss the wave functions for abelian anyons. 

Quantum statistics results from the fact that identical quantum particles are indistinguishable, so they can be exchanged and the state of the system, containing the physical information, remains unchanged.

Let us first consider Quantum Mechanics for two identical elementary particles and use the position-spin representation $ \{\vec X, S_z \}$. Since the space of positions is a continuum we can make the exchange "visible" by drawing two paths without intersections that perform the exchange [see Fig.1 (a)].

Where is the difference between $d=2$ and $d=3$?

In $d=3$ the exchanges clockwise and counterclockwise (said negatively and positively oriented) can be deformed into each other continuously without ever intersecting the paths (we will come back to this last condition in section 4). For example, if the paths lie on a plane [as in Fig.1 (a)], it is sufficient to rotate the plane by an angle $\pi$  around an axis passing through the positions of the two particles [see Fig.1 (b)]. In $d=2$ this is impossible and therefore we can assign a well-defined orientation to the exchanges.
\begin{figure}[h]
\centering
\includegraphics[width=0.9\textwidth]{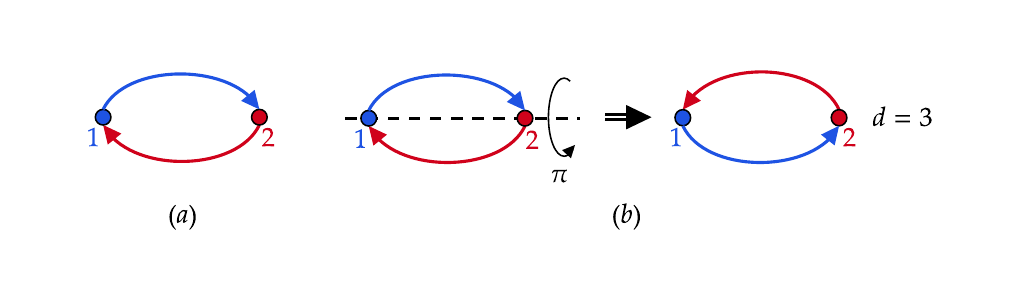}
\caption{(a) An oriented exchange (b) Its deformation into an exchange with opposite orientation  in $d=3$.}
\end{figure}

As well known, exchanges between $N$ identical particles  generate the group of permutations, $S_N$, exchanges with orientation generate a different group: the group of braids, $B_N$ \cite{artin1947theory}.
We can get an idea of this group and motivate its name by simply drawing the paths of exchanges in a three-dimensional space in which, for example, the vertical dimension represents the increasing parameter of the parametrized paths and requiring that the paths do not intersect \cite{wu1984general}\cite{wu1984multiparticle}.
Let’s formalize this construction.

\subsection{Geometric definition of the braid group}

We choose $N$ distinct points $x_1, \dots, x_N$ in $\R^2$ and consider $N$
maps $\gamma_1, \dots, \gamma_N$ from $[0,1]$ to the slab $\R^2 \times [0,1]$ in $\R^3$ with the properties [see Fig.2 (a) for $N = 4$]:

a)
$\gamma_i(0) = x_i, \quad \gamma_i(1) = x_{\sigma(i)}$
where $\sigma$ is an arbitrary permutation of ${1,..., N}$.

b)
$\frac{d \gamma_i^3 (t)}{d t}>0 \quad \forall t \in [0,1]$.

c) For
$i \neq j, \gamma_i(t) \neq \gamma_j(s), \quad \forall t \in[0,1],\quad  s \in [0,1]$.

Now we identify all the maps with properties a),b),c) related to each other by an
ambient isotopy, that is two of these braids are considered equivalent if there is a homotopy relation (with parameter $u \in [0,1]$) on the paths $ {\gamma_i(u,t)}$ such that $ \forall u$ the graph of ${\gamma_i(u,t)}$ is a braid. The resulting object is $B_N$, the braid group on $N$ strands.

The subgroup of the braid group corresponding to the trivial permutation (i.e. $\sigma = \id$) of points $x_1, \dots, x_N$ is called pure braid group and often denoted with $P_N$.

\textit{Remark} One defines the \textit{projection of a braid} $B$ as the image of $B$ under a projection on a plane, so that:
(1) The preimage of each point in the projection contains at most two points of the braid.
(2) There are only finitely many points in the projection with two preimages.
These points are called "crossings" and more precisely "overcrossing" if in the crossing of the strands $j$ and $j+1$ the strand $j$ is projected above the $j+1$ and "undercrossing" the opposite situation [see Fig.2 (b)].

\begin{figure}[h]
\centering
\includegraphics[width=0.9\textwidth]{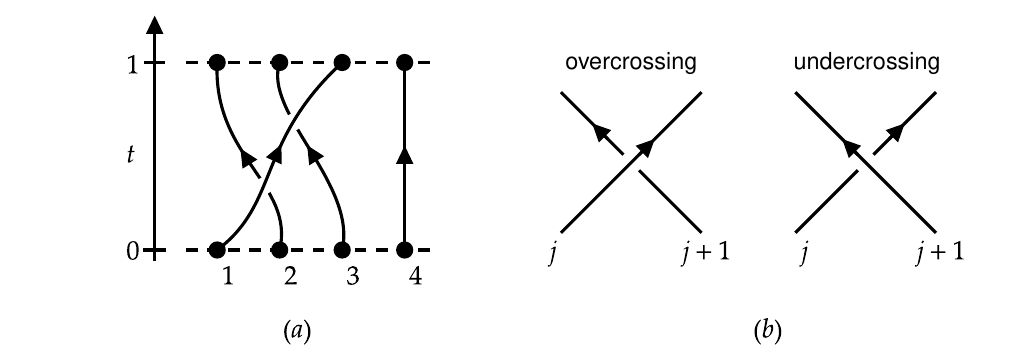}
\caption{(a) A braid with $N=4$ (b) Overcrossing and undercrossing.}
\end{figure}

\subsection{Homotopic definition of the braid group}

Braid groups can also be defined as the fundamental group of certain configuration spaces.
Let $M$ be a connected manifold, $M^{\times N}$ its product space $N$ times and $(M)^\circ_N = M^{\times N} \setminus D_N$ where $D_N = \{x_1, \dots, x_N | x_i \in M, x_i = x_j, i \neq j, i,j = 1,  \dots, N \}$. Quotienting for the permutations of $N$ points, i.e. considering $M_N= (M)^\circ_N/S_N$ if $d=2$ we have for the fundamental homotopy group $\pi_1(M_N)  \simeq B_N(M)$, the group of braids on $N$ strands in $M$. In particular $B_N(\R^2) \simeq B_N$, the group of geometric braids defined above. 

Let’s show this: Fix a base point in $(\R^2)^\circ_N$, $(x^0_1,  \dots, x^0_N)$, and let $[x^0_1,  \dots, x^0_N]$ be the corresponding $N$-ple not ordered in $\R^2_N$, such that
\begin{eqnarray}
       p : (\R^2)^\circ_N \rightarrow  \R^2_N \nonumber\\
      (x^0_1, \dots, x^0_N) \mapsto [x^0_1, \dots, x^0_N]
\end{eqnarray}

An element of $\pi_1(\R^2_N)$ is fully identified by a path $(x(t)_1, \dots, x(t)_N), t \in [0,1]$ with $[x(0)_1, \dots, x(0)_N] =[x(1)_1, \dots, x(1)_N]$. Because $N$-ples of un-ordered points can differ from each other only by a permutation, for some $\sigma \in S_N$,
$$ (x(1)_1,  \dots, x(1)_N) =(x(0)_{ \sigma(1)}, \dots, x(0)_ { \sigma(N)}).$$
It is easy to see that the graph of this path in $(\R^2)^\circ_N$ is a geometric braid and the equivalence relation between the geometric braids is exactly the homotopy relation in $\pi_1(\R^2_N)$.

Note that if the same analysis is carried out in $d \geq 3$, we obtain $\pi_1(\R^d_N) \simeq S_N$, consistent with the fact that the statistics correspond to the representations of the permutation group. In $d=1$ we get  $\pi_1( \R_N) \simeq \id$ so that the notion of statistics of particles is void in $d=1$. Even taking into account coincident points   using an orbifold formalism leads to the same conclusion, see \cite{harshman2022topological}.

\subsection{Algebraic definition of the Braid group}

There is also an algebraic definition:
Each braid with $N$ strands in $B_N$ can be generated by composition and inversion from $ (N-1)$ oriented exchanges $\sigma_i$ between strings $i$ and $i+1$ and $B_N$ admits the presentation
\begin{eqnarray}
\label{YBE}
 & B_N = <\sigma_i, i =1, \dots, N-1| \sigma_i \sigma_j = \sigma_j \sigma_i \ \text{for} \ |i-j| \geq 2, \nonumber \\
 &\sigma_i \sigma_{i+1} \sigma_i =
 \sigma_{i+1} \sigma_i \sigma_{i+1}>   
\end{eqnarray}

The last relation is also called the Yang-Baxter equation \cite{yang1967some} \cite{onsager1944crystal} \cite{baxter1972partition} (without spectral parameter) and plays an important role in many areas of Physics, for example integrability. Using the geometric representation of braids it can be drawn as in Fig.3.

\begin{figure}[h]
\centering
\includegraphics[width=0.9\textwidth]{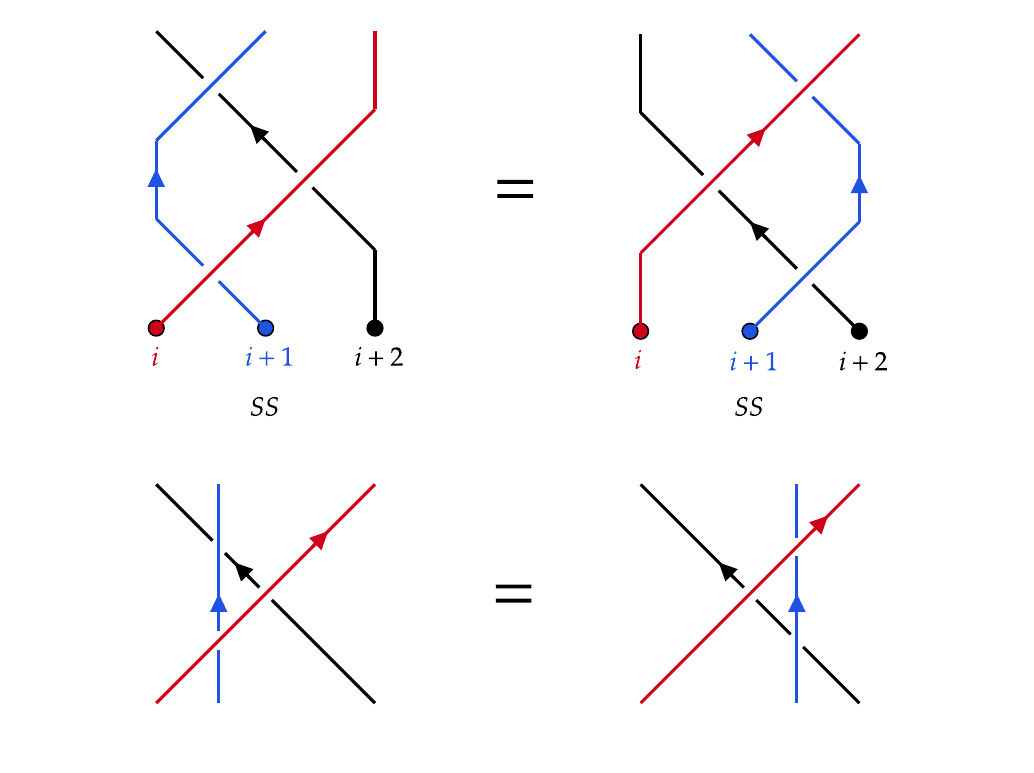}
\caption{The Yang-Baxter relation for braids represented in two equivalent ways.}
\end{figure}

Notice that if we impose in the previous presentation the additional relation $\sigma_i^2 = \id \forall i$ we get the permutation group $S_N$, consistent with the fact that now the exchanges are not oriented.

\subsection{Anyon wave functions}

After the previous digression on braid groups let's return to our problem: the statistics of identical particles in $d =2$ and consider first them in Quantum Mechanics (QM) in the representation ${\vec X}$, where the degree of freedom of spin has been frozen (for example by a strong magnetic field).

The wave function of $N$ identical particles is of the form $\psi( x_1, \dots, x_N)$ and is assumed that $x_i \neq x_j$ for $i \neq j$, so it is defined in $(\R^2)^\circ_N$. How does it change if we run the previously discussed $\gamma_i(t)$ paths, defining a geometric braid, bringing the points to a position obtained with a permutation, so that the state has not changed?  Clearly, since these path are unphysical, the homotopic paths must provide the same response and the non-homotopic paths only provide a phase factor. Therefore, on $\psi$ a unitary one-dimensional representation of $\pi_1(\R^2_N)$ acts.

Such representations are characterized by a parameter $\theta \in [0,1)$; in fact if $\chi$ denotes a representation, we  write its action on the generators of $B_N$ as
$$
\chi: \sigma_j \mapsto e^{i 2 \pi \theta_j}, j=1,\dots N-1,
$$ for a suitable $\theta_j \in [0,1)$.  Using the relation in Eq.(\ref{YBE})  we derive
$$
(e^{i 2 \pi \theta_j})^2 e^{i 2 \pi \theta_{j+1}} = ( e^{i 2 \pi \theta_{j+1}})^2 e^{i 2 \pi \theta_j},
$$ 
so that $\theta_j =\theta_{j+1} \equiv \theta$
, i.e. all one-dimensional representations of $B_N$ are of the form $\chi_\theta : \sigma_j \mapsto e^{i 2 \pi \theta}, j=1,\dots, N-1$ for some $\theta \in [0,1)$.

This argument would in principle mean that $\theta$ could depend on $N$, and should therefore be more precisely denoted by $\theta_N$. However, considerations of locality in quantum mechanics suggest that in reality 
must be independent of $N$. In fact, imagine that only short range interactions act on identical anyons (apart from their statistical interaction), and two of them are arbitrarily distant from the others $N-2$, then by locality an exchange of the two distant anyons should not affect the others, so that $\theta_N =\theta_2 \equiv \theta$.
This can be rigorously proved by using Einstein causality in relativistic theories, as discussed in sections 5 and 7. So for anyons the statistics parameter $\theta$ is not just 0 or 1/2 but ANY number in [0,1) except precisely 0 or 1/2, hence the name ANYons.

\textit{Remark} If there are more than one species of abelian anyons in the system, their statistics is described by abelian representations of the groupoid of colored braids, see sect. 7.5.
It may also happen that while the statistics of each species is not anyonic, the mutual statistics between different species is anyonic. The most famous example of this situation is the "Toric code", a lattice model introduced by Kitaev in a seminal paper for quantum computation using braid statistics. \cite{kitaev2003fault}.

As seen above, the configuration space of the wave functions is non-simply connected, it is therefore natural to define the wave functions in its universal covering space and impose the relation characterizing the statistics between different points of the covering space that have the same projection on the configuration space. We denote the quantities related to the covering space, in our case $\Tilde{\R ^2_N}$, with a $\quad \Tilde{}\quad$ . By definition two points with the same projection can be reached with a single element of $ \pi_1 (\R ^2_N)$. Let $ \chi_\theta$ be the representation of $B_N$ with $ \theta$ parameter, $ ( \tilde x_1, \dots,  \tilde x_N)$ and $( \tilde x'_1,  \dots,  \tilde x'_N)$ two points in $\Tilde{\R ^2_N}$ connected by $b \in B_N$. We impose on the wave function $ \psi$ defined in $\Tilde{\R ^2_N}$:

$$ \psi(  \tilde x'_1,  \dots,  \tilde x'_N) =  \chi_ \theta(b) \psi(  \tilde x_1,  \dots,  \tilde x_N).
$$

The Hilbert space structure on the space of pure states of $N$ anyons, ${\cal H}_N^\theta$,  is introduced by a 
positive-definite scalar product which is defined as follows. For $\psi$ and $\phi$ 
two functions on  $\Tilde{\R ^2_N}$ we define 
\begin{equation}
\label{HN}
\braket{\psi, \phi} = \int_{(R ^2)^<_N} \psi^*(  \tilde x_1,  \dots,  \tilde x_N) \phi(  \tilde x_1,  \dots,  \tilde x_N) d \tilde x_1,  \dots,  d\tilde x_N,    \end{equation}
where ${(\R^2_N)^<}$ is a fundamental domain for the action of $ \pi_1 (\R ^2_N)$ on $ \Tilde{\R ^2_N}$ (that is almost every point in $\Tilde{\R ^2_N}$ can be obtained from a point in ${(\R^2_N)^<}$ by acting with an element of $ \pi_1 (\R ^2_N$). Since $\chi_\theta$ is unitary the product appearing in the integral above is independent of the choice of the point $(\tilde x_1, \dots, \tilde x_N)$ with projection $[x_1, \dots, x_N]$.
A function on $\Tilde{\R ^2_N}$ with the desired properties under braiding can be constructed as follows: 
Represent the positions of particles $x = (x^1, x^2) \in \R^2$  
by complex numbers $z = x^1 + i x^2$ , then for $\arg z_i < \arg z_{i+1}$, set 
$$
\Theta_\theta(\id ; z_1, \dots, z_N) = \prod_{i < j} \Big( \frac{z_i-z_j}{|z_i-z_j|}\Big)^{2 \theta}.
$$ 
Let $z_i(t), t \in [0,1], i=1, \dots, N$ with $z_i(0) = z_i$ represent a braid $b \in B_N$ and set 
$$
\Theta_\theta(b; z_1, \dots, z_N)=\Theta_\theta(\id; z_1(1), \dots, z_N(1)) = e^{i 2  \theta \sum_{i<j} \arg (z_i(1)-z_j(1))}. 
$$
Then $\Theta_\theta (b; z_1, \dots, z_N), b \in B_N$ is defined in $\tilde \C_N \simeq \Tilde{\R^2_N}$ and satisfies our requirements. In fact, it acquires a phase factor $e^{\pm i 2 \pi \theta}$ under oriented exchanges, identifying the exchange of positive orientation as the anti-clockwise exchange of the $i$-th and $i+1$-th anyon.
This implies that every wave-function for anyons with $\theta$ statistics can be written as 
\begin{equation}
\label{psiN}
\psi_b (z_1, \dots, z_N) = \Theta_\theta(b; z_1, \dots, z_N) \psi_s (z_1, \dots, z_N)     
\end{equation}

with $\psi_s$ a single-valued, completely symmetric wave function. 
Note that to have a wave function defined in all $\R^{2\times N}$ the wave function $\psi$ must vanish in $D_N$, except for $\theta = 0$, so two anyons cannot occupy even potentially the same position, as for the fermions for the Pauli exclusion principle.

Similar considerations apply to the case where the space of the positions is replaced by the space of momenta $\vec P$ (see e.g. \cite{Mund:1993cf}).

The most famous wave functions for anyons are those proposed by Halperin \cite{halperin1984statistics} for vortices (or quasiparticles) introduced by Laughlin \cite{laughlin1983anomalous} for the Fractional quantum Hall effect in the plateaux at filling $\nu= \frac{1}{2 n +1}$, with positive integer $n$, see section 6.
They have the structure (\ref{psiN}) with 
\begin{equation}
\psi_s (z_1, \dots, z_N) = \prod_{i < j} |z_i-z_j|^{\nu} \prod_i \exp[-\frac{\nu|z_i|^2}{(2 \ell)^2}],    
\end{equation}
where $\ell$ is the single-particle magnetic length and $\theta = \nu/2$.
These wave functions were also a crucial step in the development of the new concept of topological order, see e.g. \cite{wen2017colloquium} or for a gentle introduction \cite{chen2019quantum}( arXiv:1508.02595), which involves long-range entanglement, in this case described by the factor $\prod_{i < j} (z_i-z_j)^{\nu}$ in the wave functions.

\textit{Remark: A bundle definition}

The wave functions on a connected  non-simply connected manifold $M$ can  be defined
as sections of a flat complex vector bundle $\cal L$ over $M$, with structure group $\pi_1(M)$ and flat connection.
 If the wave functions take values in $\C$, as considered above, then $\cal L$ is a line bundle, and
all such line bundles with connection are classified by elements of
$Hom(\pi_1(M),U(1))$,
the set of homomorphisms from $\pi_1(M)$ to $U(1)$ corresponding to the holonomy
group of the connection of the bundle.
Given the homomorphism $\chi  \in Hom(\pi_1(M),U(1))$, the bundle $\cal L$ can be constructed
as follows. Let $M_0$ be a fundamental domain of $\pi_1(M)$ in $\tilde M$ so that (almost) all other points in $\tilde M$ can be identified by pairs $( x\in M_0, \gamma \in \pi_1(M) )$.
We denote with $M_\gamma$ the image of $M_0$ under the action of $\gamma$. Then
the transition functions of $\cal L$ between $M_\gamma$ and  $M_{\gamma^\prime}$ are given by $\chi(\gamma^\prime \circ \gamma^{-1})$. A section $\psi$
of $\cal L$ corresponds to a function $\psi_\chi$ on $\tilde M$ satisfying
$$
\psi_\chi(\gamma \tilde x) = \chi(\gamma) \psi_\chi( \tilde x),
$$
for $\tilde x \in \tilde M$.
In our case $M= R^2_N$ and $\chi = \chi_\theta$.
For anyons the above bundle is trivial and as trivializing section can be chosen precisely $\Theta_\theta(b; \cdot)$ as rigorously proved in \cite{Mund:1993cf}.

\textit{Remark} For a rigorous construction of the Hamiltonian operator for free anyons, see \cite{dell1997statistics}.

For a variable number of anyons one can construct an analogous ${\cal F^\theta}$ of the Fock space for bosons and fermions by taking the direct sum of the Hilbert spaces ${\cal H}_N^\theta$ defined in (\ref{HN}) for arbitrary $N$: ${\cal F^\theta} = \oplus_{N=0}^\infty {\cal H}_N^\theta $. Note, however, that ${\cal F}^\theta$ for anyons does not have the property of bosonic or fermionic Fock spaces that its elements can be written as linear combinations of products of single-particle wave functions, due to the pre-factor $\Theta_\theta(b; z_1, \dots, z_N)$ in (\ref{psiN}).

\textit{Remark} More precisely the space ${\cal F^\theta}$ is defined as the completion of the pre-Hilbert space of finite sequences of wave functions $\Psi_N \in {\cal H}_N^\theta, N=0,...,N_{max}$, with ${\cal H}_0^\theta \simeq \C$ and the inner product defined by
\begin{equation}
\sum_{N=0}^\infty\braket{\Psi_N,\Phi_N}    
\end{equation}
where the sum is formally extended to infinity since only a finite number of terms are non-zero. The mathematically precise definition of the representation of the braid group, $B_\infty$, on  ${\cal F^\theta}$ is given e.g. in \cite{wenzl1990representations}.

\section{Anyons and spin}

In Relativistic Quantum Field Theories (RQFT)  Pauli for free fields \cite{pauli1940connection} and  Wightman in general \cite{streater2000pct} have proved the famous spin-statistics theorem, on which we will return, which proves that particles with bosonic statistics have integer spin and particles with fermionic statistics have half-integer spin. In quantum mechanics this is taken as a postulate.
The fact that anyons have no bosonic nor fermionic statistics raises the question of what spin these particles may possess.

\subsection{Spin in quantum mechanics}

Elementary particle excitations in quantum mechanics are characterized by the irreducibility of the algebra generated by the observables  position, momentum and spin. By Wigner theorem \cite{wigner2012group} the spin in $d>1$ is in bijective correspondence with the irreducible projective representations of the group of rotations  $SO(d)$ acting on the vector rays of Hilbert space, $ \cal H$, which represents in standard quantum formalism the space of pure states (with maximum information) of the system. Bargmann's theorem \cite{bargmann1947irreducible} proves that such (continous) representations are in turn in bijective correspondence with the irreducible unitary (strongly continuous) representations of the universal covering group $ \Tilde{SO(d)}$ on the vectors of $ \cal H$. In $d = 3$ the covering group is isomorphic to $SU(2)$, whose irreducible representations are labeled by $\N/2$, so that the spin can only be integer or half-integer. But in $d=2$ the covering group of $SO(2)$ is isomorphic to $\R$. Therefore, all unitary irreducible
representations of $\R$ are of the form: $ r \in \R \mapsto e^{i s r}$
where $s$ is identified with the spin of the particle and therefore is ANY real number. 

\subsection{Spin in relativistic theories}

Combining quantum mechanics with relativity, however, the above characterization fails, as the localization concept is missing. A heuristic way of looking at it is as follows: Suppose that we want to measure the position $\vec x$ of a relativistic quantum particle with
an accuracy $\Delta \vec x$. For the uncertainty principle this implies an uncertainty $\Delta \vec p$ in momentum and,  consequently, $\Delta E$ in energy. However in a relativistic quantum theory of particles of mass $m$ if $E > 2m c^2$,  particle-antiparticle pairs can be created. Which particle are you measuring then the coordinate of ?  Greater is the accuracy with which one tries to measure the position of a particle, more are the pairs that can be created. So there cannot be an observable with spectrum the position, because it is not well defined and it is therefore impossible to characterize elementary particles as in quantum mechanics.

Wigner’s idea \cite{wigner1939unitary} was to suggest that the space $\cal H$ of states of the elementary relativistic quantum particle is no
more the representation space of an irreducible representation of the algebra generated by position, momentum and spin, but the representation space of an irreducible representation of the symmetries of space-time. Subsequently
it was added that $\cal H$ could also be the representation space of internal symmetries (charge, color, ... ).

These irreducible representations therefore mathematically identify what we mean by an elementary relativistic quantum particle.

According to Wigner's definition, the Hilbert space of states of a
free relativistic quantum particle carries a projective irreducible representation of the restricted (proper orthochronous) Poincarè group ${\cal P}^\uparrow_+(1,d) = R^{1,d} \rtimes S0_+(1,d)$, where the subindex $+$ denotes the constraint that the $00$ component of the $SO(1,d)$ matrices is positive. Bargmann has shown  that all the (continuous) projective representations
of ${\cal P}^\uparrow_+$ are induced by (strongly continuous) unitary representations of its universal covering $\Tilde{ {\cal P}^\uparrow_+}$ which is isomorphic to  $R^{1,d} \rtimes \Tilde{S0_+(1,d)}$.

From a general theorem of Mackey \cite{mackey1952induced} on representations of
semi-direct products of groups follows that all unitary irreducible representations of $\Tilde{ {\cal P}^\uparrow_+(1,d)}$
are classified by
an orbit of one point $p$ in $R^{1,d}$ under the action of the group $\Tilde{S0_+(1,d)}$ and
an irreducible representation of the subgroup $H$ of $\Tilde{S0_+(1,d)}$ which leaves invariant a
point of such an orbit.
For a fixed orbit, all these subgroups are isomorphic to the same group, called
the little group.
The massive particles correspond to orbits characterized by the mass hyperboloid 
$V_m = \{ p^\mu | p^\mu p_\mu = m^2, p^0 >0 \}$
where $m \in \R^+$ is identified with the mass of the particle.  Spin characterizes the
irreducible representations of their little group, $H_m$, which is still isomorphic to the covering group of $SO(d)$, so that the spin classification of quantum mechanic is recovered and again in $D=2+1$ space-time dimensions the spin can be ANY real number.

Based on the standard spin-statistics theorem, a relation is predicted between the two-dimensional spin $s$ and the statistics parameter $\theta$, but to see heuristically how it arises in quantum mechanics it is convenient to take a look at the path-integral formulation.

\section{Path-integral  and Chern-Simons}

In this section we briefly discuss the path-integral formalism for quantum mechanics of abelian anyons (see also e.g. \cite{forte1992quantum}, \cite{lerda2008anyons}) using abelian Chern-Simons gauge fields, with a comment also on their lattice regularization and the spin-statistics connection. 

Let $M$ be the configuration space of a quantum system, with coordinates $q$ and Lagrangian ${\cal L} (q, \dot q)$. According to Feynman the amplitude for the system to evolve from the configuration $q_0$ at time $t_0$ to the 
configuration $q_1$ at time $t_1 > t_0$  is given by the kernel
$$
K(q_1, t_1; q_0, t_0) = \int_{q(t_0)=q_0, q(t_1)= q_1} {\cal D} q e^{\frac{i}{\hbar} \int_{t_0}^{t_1} dt {\cal L} (q(t), \dot q(t))}.
$$

\textit{Remark} Following a common practice among physicists, in this review we use a notation not mathematically rigorous for the path-integral. In the case of a free particle of mass $m$ in $\R^d$ with Lagrangian  ${\cal L}(\dot x(t)) = \frac{m}{2} \dot x^2$ a mathematically precise formulation involves a Wick rotation to the Euclidean space-time and the interpretation of the formal expression
$$
\int_{x(t= 0)= x_0, x(t_1)= x_1} {\cal D} x e^{-\frac{1}{\hbar} \int_{0}^{t_1} dt {\cal L} (\dot x(t))}
$$
in terms of the conditional Wiener measure $d W^{t_1 \hbar m}_{x_0,x_1}(x)$ on the Brownian paths from $x_0$ to $x_1$ related to the heat kernel by
$$
e^{\frac{t_1 \hbar}{  m} \Delta}(x_0,x_1) = \int d W^{\frac{t_1 \hbar}{  m}}_{x_0,x_1}(x),
$$
where $\Delta$ is the $d$-dimensional Laplacian. Interactions can then be introduced through the Feynman-Kac formula (see e.g. \cite{glimm2012quantum} for a mathematically rigorous discussion of path-integral formalism). Note that in $d=2$ the probability of intersection of two Brownian paths at fixed time vanishes (see e.g. \cite{bricmont1985statistical} and \cite{lawler2018introduction} at sect.8.5). Therefore in the continuum to obtain a braid statistics it is not necessary to impose the constraint that two virtual quantum paths do not intersect. In the lattice instead a hard core condition must be imposed.

Suppose now that $M$ is connected but non-simply connected. Each path from $q_0$ to $q_1$ in $M$ can be written as a contractible path, $\gamma_{01}$, from $q_0$ to $q_1$ plus closed paths, $\omega$ from $q_1$ to $q_1$, falling into homotopy  classes $[\omega] \equiv \alpha \in \pi_1(M)$ [see Fig.4].

\begin{figure}[h]
\centering
\includegraphics[width=0.9\textwidth]{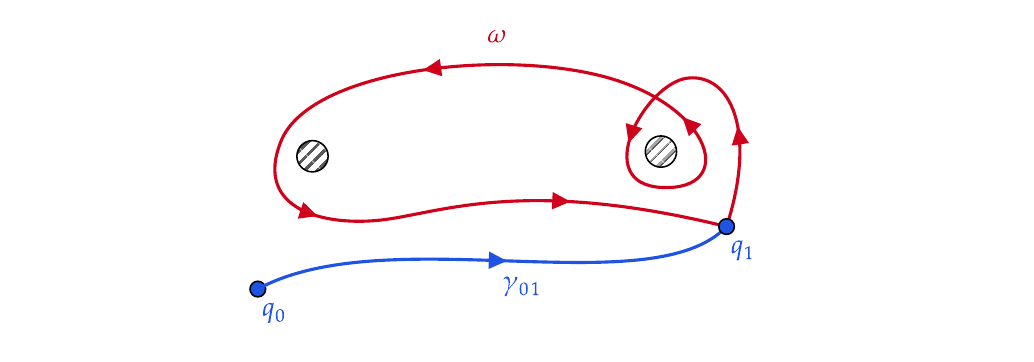}
\caption{Paths $\gamma_{01}$ and $\omega$ in a plane with darkened regions omitted.}
\end{figure}

 The corresponding paths from $q_0$ to $q_1$ are denoted by $q_\alpha(t)$. Since the sets of paths in different homotopy classes are disjoint, one can write the above amplitude as a sum on amplitudes relative to each homotopy class:
$$
K_\alpha(q_1, t_1; q_0, t_0) = \int_{q_\alpha(t_0)=q_0, q_\alpha(t_1)= q_1} {\cal D} q_\alpha e^{\frac{i}{\hbar} \int_{t_0}^{t_1} dt {\cal L} (q_\alpha(t), \dot q_\alpha(t))}.
$$
It may be noted, however, that this is not the most general combination of $K_\alpha$ allowed in $M$, but each of them can be multiplied by a complex coefficient $\chi(\alpha)$ provided that the property of composition of the amplitudes
$$
K(q_2, t_2; q_0, t_0)= \int_M d q_1 K(q_2, t_2; q_1, t_1)K(q_1, t_1; q_0, t_0).
$$
is satisfied.
Clearly a change of $\gamma_{01}$  may shift all the homotopy classes $\alpha \rightarrow \alpha + \beta$, but it cannot change the physics, so that the total amplitude can only change for a global phase $e^{i \phi(\beta)}$, i.e. 
$$\chi(\alpha + \beta) = \chi(\alpha) e^{i \phi(\beta)}.
$$
Choosing $\beta = -\alpha$ we see that $\chi(\alpha)$ can only be a phase and to satisfy the composition must obey $\chi(\alpha_1)\chi(\alpha_2)= \chi(\alpha_1+\alpha_2)$. Therefore, $\chi$ is a unitary representation of $\pi_1(M)$.

\subsection{Path-integral for anyons}

The system we are interested in consists of $N$ identical quantum particles. As discussed above, its configuration space in $\R^2$ is $\R^2_N$ and its homotopy group is $B_N$. We also know that a unitary representation of the braid group is generated by braiding particles in the multivalued function 
$$
e^{i 2 \theta \sum_{i < j} \arg (z_i(1)-z_j(1))},
$$
where we identify $q =x \in \R^2$ and $z = x^1 + i x^2$.
We then get a weight labelled by $\theta$ for the homotopy classes simply adding to the Lagrangian the  term  
\begin{equation}
\label{thetadt}
2 \theta \frac{d}{dt}\sum_{i<j} \arg (z_i(t)-z_j(t)) .    
\end{equation}
 
This is a total time derivative and the kernel applied to a wave function produces its time evolution. So the wave function of the system at $\theta =0$ is dressed at time $t$ by a phase factor 
$$
e^{i 2 \theta \sum_{i < j} \arg (z_i(t)-z_j(t))}
$$
consistent with the definition $$
\psi_b (z_1, \dots, z_N) = \Theta_b(z_1, \dots, z_N) \psi (z_1, \dots, z_N) 
$$
of the previous section. That is, the anyon wave-function can be written as a bosonic symmetric wave-function dressed by the above phase factor.

By performing the derivation in (\ref{thetadt}) and using 
$$
\partial_\mu \arg(z) = -\epsilon_{\mu \nu} \frac{x^\nu}{|z|^2} , \mu,\nu=1,2
$$
we can rewrite the additional term in $\cal L$ as
$$
\sum_i A_\mu(x_i) \frac{d x_i(t)^\mu}{dt} 
$$ 
setting 
\begin{equation}
\label{Atheta}
A_\mu(x_i) \equiv \theta \frac{\partial}{\partial x_i^\mu } \sum_{i \neq j} \arg (z_i(t)-z_j(t))= - \epsilon_{\mu \nu}  \theta \sum_{i \neq j} \frac{x_i^\nu(t)-x_j^\nu(t)}{|z_i(t)-z_j(t)|^2}.    
\end{equation}
    
The first representation  of the gauge field in (\ref{Atheta}) shows that one can see the phase factor described above, which dresses bosonic wave functions, as a singular gauge transformation.
We recognize  the new term in $\cal L$ as a minimal coupling of  particles to a gauge field $A_\mu$. This shows that the fractional $\theta$ statistics can be implemented on ordinary 
bosons by adding fictitious statistical interactions. 

\subsection{Chern-Simons}

 The above coupling to the gauge field can be made space-time covariant by introducing an independent gauge field $A_\mu(x)$ minimally coupled to the particles in a covariant way, with action the so-called abelian Chern-Simons term \cite{chern1974characteristic} action:
\begin{equation}
\label{CS1}
\frac{1}{4 \pi} \int d^3 x \epsilon_{\mu \nu\rho}A^\mu \partial^\nu A^\rho(x),
\end{equation}
 with coefficient $k$, called level, which we will then connect to $\theta$,
where $\mu,\nu,\rho=0,1,2$ and $x^0=t$. In the formalism of $p$-forms one writes the Chern-Simons action (\ref{CS1}) simply as
\begin{equation}
\label{AdA}
 \frac{1}{4 \pi} \int A \wedge dA.   
\end{equation}
The minimal coupling term then is 
\begin{equation}
\label{inter}
-\int d^3x A^\mu j_\mu
\end{equation}
with 
$$
j^\mu(x)= \sum_j \frac{d x^\mu_j(t)}{d t}\delta(\vec x -\vec x_j(t)),$$
where $\vec x_j(t)$ denotes the position in space of the $j$-th particle at time $t$.

To match the definition in terms of Chern-Simons with the previous definition, we use the Coulomb gauge-fixing $\partial_\mu A^\mu=0$, with $\mu=1,2$.

By integrating out $A^0$ we get
\begin{equation}
\label{thtaA}
2 \pi j_0 (x)= k \epsilon_{0 \mu \nu} \partial^\mu A^\nu (x),   
\end{equation}
 
and combining this with the Coulomb gauge we obtain with $\mu, \nu = 1,2$
\begin{equation}
\label{Aa}
A_\mu(x) = - \frac{1}{k}\sum_i \epsilon_{\mu \nu} \frac{(x^\nu-x_i^\nu(t))}{|\vec x-\vec x_i(t)|^2}.     
\end{equation}

Inserting (\ref{Aa}) in the interaction term (\ref{inter}) and in the remaining Chern-Simons term (which halves the contribution of the interaction) we recover the previous result (regularizing e.g. on the lattice, see below, to omit the self-interactions) if we set $k = \frac{1}{2 \theta}$. 

\textit{Remark} For a second-quantized Hamiltonian version of the above model, see e.g. \cite{jackiw1990classical}.

By integrating in a space-region $\Sigma$ the equation (\ref{thtaA}) we find a relation between the "electric" charge contained in $\Sigma$, $Q_\Sigma$, and the "magnetic" flux through $\Sigma$, $\Phi_\Sigma$: 
\begin{equation}
\label{QPhi}
 \Phi_\Sigma =4 \pi \theta Q_\Sigma.
\end{equation}

This implies that a point-like anyon in this theory carries both an electric charge and a magnetic flux, so one can think of the anyon as a composite of an electric charge and a magnetic flux.
If, after a Wick rotation (where in Euclidean space-time the Chern-Simons term in the action needs a multiplicative factor $i$),  we put the theory on a lattice,  the lattice spacing $a$ acts as an ultraviolet (UV) regulator. By this regulator, which appears in the Chern-Simons action, the electric worldline of the anyon runs in the lattice, but the magnetic flux line runs in the dual lattice, so that the two lines are shifted by a vector proportional to $a$. This shift between electric and magnetic lines provides a "framing" of the worldlines of anyons which has enabled Wilczek and Zee \cite{wilczek1983linking} to give a very intuitive proof of the spin-statistics theorem in this framework.

\textit{Interlude: Abelian Chern-Simons on the lattice}
\cite{frohlich1989quantum},\cite{muller1991connetion}
To better understand the above statement on the support of electric and magnetic lines let us make a small digression in the lattice. Consider a $d$-dimensional cubic lattice $\Lambda$ of lattice spacing $a$. An oriented $p$-cell, $c_p$, generated by the unit lattice vectors $e_{\mu_i}, i=1, \dots, p$  in the positive $\mu_i=1,2...d$ direction, with origin at the site $x$ of the lattice, is denoted with $(x; \mu_1, \dots, \mu_p)$. In the language of lattice theories for $p=1$ the cell is a link, for $p=2$ is a plaquette. A (real) lattice $p$-form (or $p$-
cochain) $A^p$ is a real valued function on $p$-cells with 
\begin{equation}
\label{lform}
A^p(c_p=(x; \mu_1, \dots, \mu_p)) \equiv  A^p_{\mu_1, \dots, \mu_p}(x)
\end{equation}
totally antisymmetric in $(\mu_1, \dots, \mu_p)$ and satisfying $A^p(-c_p) = -A^p(c_p)$, where $-c_p$ denotes the cell obtained from
$c_p$ by reversing the orientation.
On any function $A$ on the lattice the difference operators $\partial_\mu$ (analogous to the derivative in the continuum  in the $\mu$ direction)  are defined by:
$$
\partial_\mu A(x) = \frac{1}{a}(A(x+ a e_\mu)-A(x)).
$$
On the lattice $p$-forms acts the lattice exterior differential $d$ which maps $p$-forms into $p+1$-forms and if we adopt the definition (\ref{lform}) its expression is formally the same as that of the components of the $p+1$-form $d A$ in the continuum. For example for $A^1 \equiv A$ a 1-form, we have 
$$
dA(x; \mu, \nu)= \partial_\mu A(x; \nu)-\partial_\nu A(x; \mu) \equiv \partial_\mu A_\nu(x)-\partial_\nu A_\mu(x).
$$
With the above definitions for a lattice 1-form $A$ approximating the continuum gauge field $A^\mu$, the lattice approximation of the continuum term (\ref{AdA}) on $\R^3$ is given (with the needed additional $i$) by
\begin{equation}
\label{lAdA}
 \frac{i}{4 \pi} \frac{a^3}{2} \sum_{x \in a \Z^3} \epsilon _{\mu \nu \lambda} (A_\lambda(x-a e_\lambda)+A_\lambda(x - a e_\lambda + a e_0) dA(x; \mu, \nu) ,
\end{equation}
where $\mu,\nu,\lambda=0,1,2$ with 0 corresponding to the Euclidean time direction and $\epsilon$ is the Levi-Civita tensor or symbol. The two terms in (\ref{lAdA}) are necessary for reflection positivity which allows a reconstruction of the Hilbert space of states in real time, see section 5 of these notes for a sketchy discussion. Similarly the lattice approximation of (\ref{inter}) for a lattice current $j$ is given by
\begin{equation}
 -i a^3 \sum_{x \in a \Z^3} A_\mu(x)j_\mu(x) .  
\end{equation}
It is thus seen that while the electric current flows along the links of the lattice, the magnetic flux described by $dA$ flows through the plaquettes  in which the line passing orthogonally through their centres belongs to the links of the dual lattice, see Fig.5 (a). ( The sites of the dual 3-dimensional cubic lattice are at the centers of each cube of the original cubic lattice.)

\begin{figure}[h]
\centering
\includegraphics[width=0.9\textwidth]{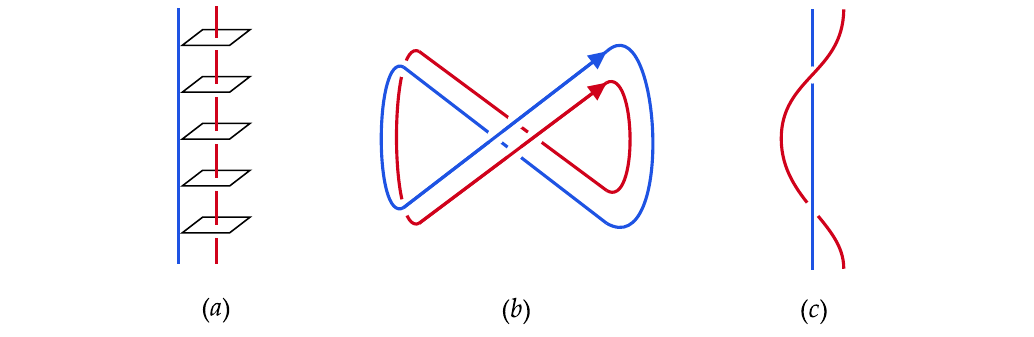}
\caption{(a) Electric (blue) and magnetic (red) flux lines in the lattice; in drawings concerning flux lines for anyons this association of colors will be mantained.  Braids of electric and magnetic flux lines arising from a positively oriented exchange of two anyons (b) and from a $2 \pi$ rotation of one anyon (c).  }
\end{figure}

\subsection{Spin-statistics connection}
First, we consider the worldlines that do not intersect of two  exchanged anyons and formally identify the points at positive and negative time infinity, obtaining closed worldlines. It can be easily verified [see Fig.5 (b)] that the electric worldlines are linked to the magnetic flux lines trough a crossing. The crossing describes an exchange, with orientation
because under- and over-crossing 
are distinct in 2+1 dimensions, and is clearly related to statistics.
Similarly, if we rotate along the time the world line of one anyon and identify as done before the points at positive and negative time infinity, again the electric worldlines are linked to the magnetic flux lines [see Fig.5 (c)], this time with a $2 \pi$ twist. The 2$\pi$ twist  describes a rotation of $2 \pi$, related to spin.
The phase factors produced  by  the crossing and the $2 \pi$ twist can be computed with the Aharonov-Bohm effect \cite{aharonov1959significance} as they formally involve
 the motion of a 
charged particle in the presence of a magnetic flux, whose support is never intersected. The phase factors are
given by the product of the magnetic flux, $\Phi$, enclosed by the closed worldline of the charge, multiplied by the value of the charge itself, q, divided by 2, due to the additional contribution of Chern-Simons, i.e. is given by $q \Phi/2$. It follows that $q \Phi/(4 \pi)$[with  $q=1$ in our units]$\mod \Z$ gives both spin and statistics parameter for such anyons.

Alternatively, the contribution of the closed worldlines of anyons, whose support we denote by $\cal L$, can be considered as Wilson loops \cite{wilson1974confinement},  $\exp{i \int_{\cal L}}A_\mu dx^\mu$, and computed directly by integrating the Chern-Simons gauge field and obtaining their expectation value.

Let us do in detail this: a far-reaching generalization of this calculation  proves that through Chern-Simons theory we can compute topological invariants of knots and links \cite{witten1989quantum}.

A\textit{ knot} $K$ is a smooth embedding of $S^1$ into $\R^3$ (see e.g. \cite{rolfsen2003knots}). Two knots $K_1$ and
$K_2$ are considered the same knot if they are related to each other by an ambient
isotopy.

More generally a \textit{link} is a smooth embedding of a disjoint union of circles into
$\R^3$, i.e. a link is a collection of disjoint knots, possibly linked to each
other. 

We can get a knot or link by connecting  with paths in $\R^3$ the corresponding upper and
lower points of the strand of a braid. This is called a closed braid [see Fig.6 (a)].

\begin{figure}[h]
\centering
\includegraphics[width=0.9\textwidth]{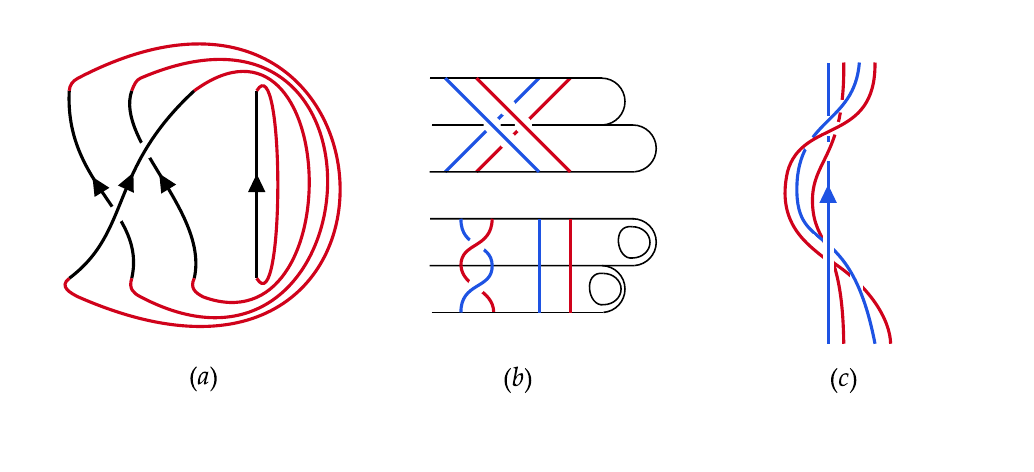}
\caption{(a) A closed braid with the connecting paths in red. (b) A pictorial representation of Finkelstein's rubber band lemma. (c) The braiding of electric and magnetic flux lines arising by a $2 \pi$ rotation of two anyons.}
\end{figure}

A theorem of Alexander \cite{alexander1923lemma} ensures that in fact every link is isotopic to a closed braid.

The \textit{linking number} is a link invariant that counts the
number of times that a closed  curve encircles a second closed curve in a link in $\R^3$. In terms of the crossings of the projection of the link it counts the number of overcrossing minus the number of undercrossings.

Let $\Sigma$ be a surface (possibly not connected) in $\R^3$ whose Poincaré dual in the sense of (de Rham) currents is the 1-current $J_{\Sigma} \equiv (J_{\Sigma })_{\mu} d x^\mu$ [see the Appendix A for basic ideas of the theory of de Rham currents]. The boundary of $\Sigma$, $\partial \Sigma $ is a link $\mathcal{L}$. According to the theory of currents then its Poincaré dual is a 2-current given by $J_{\mathcal{L}} = d J_{\Sigma}$ or in components $(J_{\mathcal{L}})_{\mu \nu} = \partial_\mu (J_{\Sigma})_{ \nu} d x^\mu \wedge d x^\nu$.

Consider the Wilson loop that arises in the path-integral from the trajectories (closed at infinity) of $N$ anyons  from the term (\ref{inter}) in the (Euclidean) action. If $\mathcal{L}$ denotes the link formed by the trajectories and $\Sigma$ a surface whose boundary is $\mathcal{L}$ the  corresponding Wilson loop is given by
\begin{equation}
\label{W}
   W((\mathcal{L}) \equiv e^{i \int_{\mathcal{L}}A_\mu d x^\mu} =  e^{i \int A_\mu \epsilon^{\mu \nu \rho} \partial_\nu (J_\Sigma)_\rho d^3x}.
\end{equation}
Now we calculate its average with respect to the Chern-Simons action (\ref{CS1}) in Euclidean space-time at first formally in the continuum:
\begin{equation}
\label{avW}
  \langle   W(\mathcal{L}) \rangle \equiv \frac{\int {\cal D} A_\mu \exp[\frac{1 }{ 8 \pi i \theta} \int d^3 x \epsilon_{\mu \nu\rho}A^\mu \partial^\nu A^\rho(x)] W(\mathcal{L})}{\int {\cal D} A_\mu \exp[\frac{1 }{8 \pi i \theta}\int d^3 x \epsilon_{\mu \nu\rho}A^\mu \partial^\nu A^\rho(x)]}.
\end{equation}
Inserting (\ref{W}) in (\ref{avW}) we can perform the calculation by making in the numerator the shift $A_\mu \rightarrow A_\mu - 4 \pi \theta \Sigma_\mu$
and completing the square. One obtains
\begin{equation}
\label{Wc}
\langle   W(\mathcal{L}) \rangle =\exp[i 2 \pi\theta \int d^3 x \epsilon_{\mu \nu\rho}J_\Sigma^\mu \partial^\nu J_\Sigma^\rho(x)] .   
\end{equation}
We regularize the expression (\ref{Wc}) by putting it on the lattice as discussed above. The result is the exponential of $i 2 \pi \theta$ multiplied by the number of times with sign that the curve $\mathcal{L}$ crosses the interior of the surface $\Sigma$ , i.e. the linking number $\ell$ of $\mathcal{L}$ with the same curve shifted according to the lattice prescription discussed above.
Since $\ell =1$ both for the link describing the positive exchange of two identical anyons and the $2 \pi$ rotation of one of them, it follows that $\theta$ mod $\Z$ gives both spin and statistics parameter for such anyons. 

\textit{Remark} Note that if we use the projection of the link on a plane behind it, the positively oriented exchange corresponds to an overcrossing, the negatively oriented exchange to an undercrossing. Then reflection with respect to such a plane or to a plane orthogonal to it exchanges over with under crossings. The Chern-Simons  term involves the Levi-Civita symbol $\epsilon$ which in 3 dimensions is not invariant under reflection with respect to a horizontal and vertical plane in Euclidean space-time, assuming as vertical the direction of the Euclidean time. It follows that a priori parity and time-reversal can be broken by Chern-Simons. Actually the  calculation above shows that this is definitely the case if $\theta \notin \Z/2$, because in this case the acquired phase factors for over and undercrossing are different. If more than one Chern-Simons term appears in the lagrangian, although each with $\theta \notin \Z/2$ can still happen that the parity and time reversal may nevertheless be unbroken due to a compensation between the effect of different Chern-Simons. A physical example of this phenomenon is given by the  fractional topological insulators time-reversal symmetric, for a review see e.g. \cite{stern2016fractional}, \cite{neupert2015fractional}. A more abstract example appears in some version of the bosonization of fermionic planar systems made by coupling to Chern-Simons gauge fields, see e.g. \cite{frohlich1992non1}, \cite{lerda1993slave}.

The equality of the linking number associated with spin and with (positively oriented) exchange found above is a simple consequence of the \textit{"rubber-band lemma" } stated earlier by
Rubinstein-Finkelstein : \textit{if a rubber band is wrapped twice about a rod, it exhibits
at least two deformities: a self-crossing and a $ 2 \pi$ twist.} [See Fig.6 (b)] This suggests that
the two are homotopically equivalent and in \cite{finkelstein1968connection} one can also find a proof.

To apply this lemma to the above system of anyons we imagine that the two boundaries
of the rubber band describe the worldlines of electric charge $e$ and magnetic flux, $\Phi$.

In a sense the closure of the worldlines described above corresponds to the hypothesis of the existence of an antiparticle; if this is not assumed the relation between spin and statistics may be violated. For example in non-relativistic quantum systems defined not by a lattice regularization one can have spin 0 particles  that still obey non trivial $\theta$ statistics, as discussed in \cite{lerda2008anyons}. As we prove in the next section this is impossible in a local quantum field theory with anti-particles having the same dispersion of the particle.

With arguments similar to those used above we can demonstrate the 

\textit{Spin addition rule}:

If we denote with $S_N$ the total $(2 d)$ spin of $N$ identical anyons with individual spin $S$, then
\begin{equation}
\label{SN}
    S_N = N^2 S \mod \Z.
\end{equation}

In fact, let's assume that we label anyons with an index that increases with the distance from the fixed point of rotation.  Then for a rotation of $2 \pi$ in the first anyon its electric line wind around the magnetic one once, giving a contribution $S$ to $S_N$. In the second anyon, not only does its electric line move around its magnetic line, but also its electric line wind around the magnetic line of the first anyon and its magnetic line goes around the electric one of the first, by giving an additional contribution to $S_N$ given by 2 $S$ [see Fig.6 (c)]. Iterating, from the windings we get the phase
$$
\exp[ 2 \pi i S_N] = \exp[2 \pi i (N S + 2 \frac{N(N-1)}{2} S)].
$$

\textit{Remark} Previously we discussed the quantum mechanics of anyons on $\R^2$, that on the 2-torus is discussed e.g. in \cite{iengo1992anyon} and on the general Riemann surfaces e.g. in \cite{bergeron1993canonical}.

\section{Quantum Field Theory of Anyons}
 
Let's move on from quantum mechanics to the quantum field theory (QFT) of anyons, as we  will see it is not entirely straightforward in $d=2$...

In this section we discuss some QFT models of anyons (for more details see \cite{frohlich1989quantum} ), we show how a Dirac ansatz solves the localization problem that arises in the construction of anyonic quantum fields, we outline their relation with vortices (details on vortex/particle duality are available in Appendix B) and how anyonic wave functions can be reconstructed from quantum fields.

For later purposes we begin the discussion with a brief sketch of the relation between the Euclidean path-integral and operator approaches to quantum field theory.  It appears that the Euclidean path-integral approach is, in most cases, the most efficient way of dealing with non-perturbative aspects in models. In fact from the Haag theorem \cite{haag1955quantum} (see also e.g. \cite{haag2012local} , \cite{strocchi2013introduction} ) we know that the Hilbert space of states of a translationally invariant interacting theory, $\mathcal{H}$, is disjoint from the Hilbert space, $\mathcal{H}_0$, of a free theory (except for non-relativistic models without vacuum polarization) and starting from the Euclidean correlators of the interacting theory it is possible  to recover $\mathcal{H}$.

\subsection{A look at the OS reconstruction theorem}

The key to the operator/path-integral relation is the so-called Osterwalder-Schrader (OS) reconstruction theorem \cite{osterwalder1973axioms} (and see e.g. \cite{glimm2012quantum}, \cite{Seiler:1982pw}) which we will sketch in a particular and mathematically somewhat imprecise form.

Let $O(S, \alpha)$ be an Euclidean field constructed with the Lagrangian fields of the model with connected support $S$ and with $ \alpha$ that indexes generically all other characteristics identifying it. 
$S$ can be a point in Euclidean space-time, but also a line, e.g. for Wilson loops, or other supports and $\alpha$ can be a set of quantum numbers, but more general situations are allowed, as will be apparent in the following.

Denoting by $< \cdot >$ the expectation value in an Euclidean quantum field theory we consider a full sequence of correlators
\begin{equation}
\{\langle O(S_1, \alpha_1)...O(S_n, \alpha_n) \rangle \}  \label{OS}  
\end{equation}

with $n = 0, 1, 2...$, where the supports $S$ have arbitrary positions in Euclidean space-time.
 Assume that such correlators satisfy:

0) \textit{technical continuity and boundedness properties}

1) \textit{translational invariance }(in lattice models restricted to lattice translations)

2) \textit{OS (or reflection) positivity }

Let $\cal V$ be the vector-subspace of the polynomial algebra of fields  $O(S, \alpha)$ spanned by monomials containing 
only disjoint  supports of the fields.
Let ${\cal V}_\pm$ be the subspaces of $\cal V$ with supports of all the fields contained in the positive (+), negative (-) Euclidean time, $x^0$, subspace. We define an antilinear involution $\Theta_{OS} : {\cal V}_+ \rightarrow  {\cal V}_-$ which reflects all the supports of fields in the Euclidean time w.r.t. $x^0 = 0$ and takes the complex conjugate (for the Grassmann fields there is some modification). 
OS positivity is the property
$$
\langle F \Theta_{OS} F \rangle \geq 0, \quad \forall F\in  {\cal V}_+
$$

3)\textit{ Clustering}

We denote by $F^a, a \in \R^d$ the translation of $a$ of $F \in {\cal V}$.
Then clustering is the statement: $ \forall F, G  \in {\cal V}$, 
$$
\lim_{|a| \rightarrow \infty} \langle G F^a \rangle = \langle G \rangle \langle F \rangle.
$$

\textit{Remark } Exponential clustering occurs if the theory has mass gap, as in the anyon models discussed in this section, but it is also verified for charged correlators even in the presence of gapless photons as in QED (see \cite{frohlich1986magnetic}, \cite{frohlichsuperselection} for a proof in the lattice approximation). 

\textit{Reconstruction Theorem
}
If the correlators (\ref{OS}) satisfy 0)-3) then there are:
1) a separable Hilbert space $\cal H$, 
2) a distinguished vector $\ket{ \Omega} \in \cal H$, called vacuum, 
3) a continuous unitary representation in $\cal H$ of space-time translations with $\ket{ \Omega} $ as unique invariant vector.
4) The generator of time-translations, the Hamiltonian $H$, is a positive semi-definite operator.
5) A dense set of vectors in $\cal H$ can be obtained by applying the polynomials of the field operators $\hat O(S,\alpha)$ to the vacuum.

\textit{Comments on the reconstruction}
To each $ F\in  {\cal V}_+$ we associate a state $\ket{F}$ in $\cal H$, $\ket{\Omega}$ is given by $\ket{1}$ and for $F, G \in {\cal V}_+$ the inner product is defined by
$$
\braket{F | G} = \langle (\Theta_{OS} F) G \rangle
$$
Hamiltonian $H$ and momentum $P$ are defined through
\begin{eqnarray}
    \bra{F}e^{-t H}\ket{G} \equiv \braket{F|G^t} = \langle (\Theta_{OS} F) G^t \rangle, \nonumber \\
     \bra{F}e^{i \vec a \cdot \vec P}\ket{G} \equiv \braket{F|G^{\vec a}} = \langle (\Theta_{OS} F)G^{\vec a} \rangle.
 \end{eqnarray}
 The time evolution is given by $e^{-i t H}$, which exists because $H \geq 0$. 
One can also define field operators $\hat O(S,\alpha)$ for $S$ having support of length $l$ in the Euclidean time direction by 
$$
 e^{-\epsilon H}\hat O(S,\alpha) e^{-(l+\epsilon) H}\ket{G}=\ket{O(S,\alpha)^t G^{l + 2 \epsilon}}, \quad \forall \epsilon >0
$$
More formally, for supports $S_1 ....S_n$, that are time-ordered with positive Euclidean time increasing
$$
\hat O(S_1,\alpha_1)...\hat O(S_n,\alpha_n) \ket{\Omega} = \ket{O(S_1, \alpha_1)...O(S_n, \alpha_n)}.
$$

\textit{Remark} If the correlators are not only translationally invariant but also Euclidean covariant, then in $\cal H$ a unitary representation of the covering of the restricted Poincarè group is recovered, with the spectrum of the energy-momentum operator in the forward light cone.

After this long digression we return to our main issue: quantum fields for anyons.

\subsection{Some troubles}

Let's try to construct anyonic quantum field operators.

Consider first the operator approach.
Suppose that there is a local quantum field  $\hat\phi(x)$ which couples the vacuum of the anyon field theory to a one-anyon state with statistics parameter $\theta$.
I.e. we assume that for equal-time space points $\vec x$ and $\vec y$ the commutation rule 
\begin{equation}
\label{CCRS}
 \hat\phi(\vec x) \hat\phi(\vec y) = e^{i 2 \pi \theta} \hat\phi(\vec y) \hat\phi(\vec x) \end{equation}
is verified.
Since the points in $\R^2$ are not ordered applying the commutation rule again we obtain
$$
\hat\phi(\vec x) \hat\phi (\vec y)= e^{i 4 \pi \theta} \hat\phi(\vec x) \hat\phi (\vec y),
$$
which is consistent only with the usual values $\theta = 0, 1/2$.
In $d=1$ this argument does not apply because the points on the real line are ordered. Therefore the fields localized in points (or more correctly in a bounded region of space-time: $\hat\phi(f) =\int f(x) \hat\phi(x)$, with $f$ a test function of compact support) can have non-trivial braid statistics because one can consistently have
$$
\hat\phi(x^1) \hat\phi (y^1)= e^{\pm i 2\pi \theta} \hat\phi(y^1) \hat\phi (x^1)
$$
with the two signs occurring if $x^1 \gtrless y^1$ and furthermore this inequality is stable under boosts in relativistic systems.
The first example of non usual $\theta$, what will be called a braid statistics
of fields, in $d = 1$ appears for the free fields in \cite{streater1970fermion}. In interacting
theories, with the name of dual algebra, can be found for the first time in nuce in \cite{frohlich1976new}, and is discussed in details in statistical mechanics in \cite{fradkin1980disorder} and in QFT in \cite{frohlich1992non}.

The problem seen in operator formalism in $d=2$ appears also in path-integral formalism.
In QM (and in the semiclassical treatment of QFT) only worldlines of  particles without boundaries appear, because particles cannot 
be created or annihilated. This fact is 
crucial for the derivation of the topological results for spin and statistics discussed in the previous section in terms of paths ("rubber band lemma").

In fully-quantized QFT, the Feynman-Schwinger-representation (see \cite{feynman1950mathematical},\cite{schwinger1951gauge}) of correlators of fields in terms of the worldlines of  particles shows, even in the Euclidean framework, 
open paths with ends corresponding to the insertion of the fields, where particles are created or annihilated

For example, consider the quantum field theory analogous to the relativistic version of the anyon model with Chern-Simons presented in the previous section : a massive charged scalar field $\phi$ coupled to a $U(1)$ gauge field $A_\mu$.

The Euclidean action is given by
\begin{eqnarray}
    \label{CS} 
   & S(A, \phi)=\int d^3x \big[ \frac{1}{2}|(\partial_\mu-A_\mu ) \phi|^2 + \frac{1}{2}m^2|\phi|^2\big] (x) + S_{CS}(A) \nonumber\\
   & S_{CS}(A) =\frac{i k}{4\pi} \int d^3x \big[ 
\epsilon^{\mu\nu\rho}A_\mu \partial_\nu A_\rho \big] (x) .
\end{eqnarray}  

\textit{Remark} The Euclidean approach, discussed here, can be rendered mathematically rigorous e.g. with a UV lattice cutoff plus counterterms.

Based on the argument given in the previous section it is expected that $\hat\phi$ obtained via OS reconstruction creates and annihilates anyon particles. 

Consider the Euclidean two-point correlator of $\phi$, $\langle \bar \phi (y) \phi (x) \rangle$. To analyze it we  introduce
 a Feynman-Schwinger representation, considering $A_\mu$ as an external field; it is then integrated with the Chern-Simons action. 
 
Since the action (\ref{CS}) is quadratic in $\phi$ its correlation function with $A_\mu$ as external field can be computed in terms of the 2-point correlators of $\phi$ which can be written,
 following Schwinger, as:
\begin{equation}
\label{2point}
\langle \bar \phi (y) \phi (x) \rangle (A)= (\Delta_A + m^2)^{-1}(x,y)=\int_0^\infty d s e^{-s m^2} e^{-s \Delta_A}(x,y).
\end{equation}
The covariant 3D Laplacian $\Delta_A$ can be regarded as the “Hamiltonian” of a particle of mass $1/2$
 and charge 1, minimally coupled to the gauge field $A_\mu$. The last term in (\ref{2point})
 can then be viewed as the evolution kernel of this particle with initial position $x$ at
“time” $\tau=0 $ and final position $y $ at $\tau= s$. It admits, therefore, an Euclidean Feynman path integral representation
\begin{equation}
\label{fey}
e^{-s \Delta_A}(x,y)=\int_{q(O)=x, q(s)= y} {\cal D}q(\tau) \exp[-\int_0^s d\tau \frac{1}{4} {\dot q}^2  (\tau) + i \dot q^\mu A_\mu (\tau)].
\end{equation}

Associating to the trajectory $q(\tau)$, which starts from $x$ and ends in $y$, a "worldline" current  $j^\mu(x)=\int d \tau \dot q^\mu \delta(x-q(\tau))$, plugging (\ref{fey}) in (\ref{2point}) one can write for a suitable measure $D \nu(j)$:
 \begin{equation}
 \langle \bar \phi (y) \phi (x) \rangle (A) = \int D \nu(j) \exp[i \int d^3y j^\mu(y) A_\mu(y)].
 \end{equation}
By integrating out (using a gauge-fixing) the gauge field, which appears quadratically in the Chern-Simons action in (\ref{CS}),
 we obtain an interaction between the "worldline"  electric currents appearing in the correlation function with boundary $x$ and $y$ and other closed "worldline"  electric currents introduced by the partition function  of $\phi$ in presence of the external field $A$, denoted $Z(A)$, appearing as a normalization factor:
 \begin{equation}
 \langle \bar \phi (y) \phi (x) \rangle = \frac{1}{Z} \int D A e^{-S_{CS}(A)} Z(A) \langle \bar \phi (y) \phi (x) \rangle (A).    
 \end{equation}

 The question now is: how to extend the topological arguments of the previous section for statistics and spin to this new setting with open paths, where the topological stability associated
 to the paths without boundaries of the previous section disappears?

\subsection{Dirac ansatz}

One way out of this conundrum both in the operator and in the path-integral formalism is provided by the Dirac ansatz \cite{dirac1955gauge} for charged non-local fields and its generalizations. Let us recall
 the physical origin of the
Dirac ansatz.

In gauge theories one cannot have both a local charged field operator and the space of states $\cal V$ (e.g. obtained via reconstruction) of the quantum field theory with positive metric (see \cite{strocchi1977spontaneous}).

In the standard BRS approach this can be understood as follows.

Let $Q_{B}$ be the BRST \cite{becchi1974abelian} \cite{becchi1976renormalization} \cite{tyutin2008gauge} charge selecting out of $\cal V$ the space ${\cal V}_{\mathit{phys}}$ of physical states, $| \mathit{phys} \rangle$, by $Q_{B}| \mathit{phys} \rangle = 0$.

From the equation of motion we get \cite{kugo1979local}:
\begin{equation}
\label{eomB}
 \partial^\mu \hat F_{\mu\nu}-\hat J_\nu =\{Q_B, D_\nu \hat \bar c\}.   
\end{equation}
In (\ref{eomB}) $\hat F_{\mu\nu}$ is in general the operator corresponding to the antisymmetrization of $\frac{\delta {\cal  L}}{\delta \partial_\mu A_\nu}$, where ${\cal  L}$ is the lagrangian; hence in QED is simply the gauge field-strength (or curvature) operator. ${\hat J_\mu}$ is the operator for the Noether current of the global gauge transformations and ${\hat \bar c }$ the antighost field operator.

It immediately follows that
$$
\langle phys|\partial^\mu \hat F_{\mu\nu}- \hat J_\nu |phys \rangle=0.
$$
If $\cal V$ has (semi-definite) positive inner product using Schwartz inequality one can raise the above mean value to an operator identity :
\begin{equation}
\label{strocchi}
(\partial^\mu \hat F_{\mu\nu}- \hat J_\nu )|phys \rangle=0.
\end{equation}
By smearing  the above density $\hat J_0$  with a test function $f(x)$ equal to 1 in a sphere in space of radius $R$ and smoothing interpolating to 0 outside of a sphere of radius $R+1$ one obtains a charge operator $Q_R$ relative to the sphere of radius $R$. If $\hat\phi(x)$ is a charged field operator
\begin{equation}
\label{charge}
\lim_{R \rightarrow \infty} [Q_R, \hat\phi(x)] = \hat\delta \phi(x) \neq 0.
\end{equation} 

But by (\ref{strocchi}) on ${\cal V}_{phys}$ we can replace $\hat J_0$ by $\partial^i \hat F_{i 0}$, and (formally by integration by parts) this show that the above $Q_R$ has support only in the spherical shell with inner radius $R$. Then, by locality, as $R \rightarrow \infty$ the above commutator vanishes in ${\cal V}_{phys}$, against the assumption that $\hat\phi$ is charged, eq.(\ref{charge}). 

Note that the previous argument excludes fields localized not only in points, but in any bounded region of space-time.

Therefore we have two options for the charged field: One can give up the positivity of the space of states obtained by the reconstruction theorem from the correlators  and maintain locality of the fields, and this is  standard approach in perturbative treatments.

Otherwise the positivity of the space of the states reconstructed from correlators is maintained, but one gives up the locality of charged fields, and this happens in the non-perturbative approach of Osterwalder-Schrader reconstruction for charged fields in gauge theories. A similar idea is also at the heart of the algebraic approach to quantum field theory, which we will discuss briefly in the last section.

A physical example of this situation is provided by Quantum Electro Dynamics (QED)
  in the operator approach. 
  Though it is trivial for the topics of statistics and spin this example is easy and familiar and also is the framework in which historically the Dirac ansatz was introduced, so we start from it in the BRS version. 

  Let  $\ket{\Omega}$ denote the physical vacuum and $\hat \psi_\alpha ({\vec x})$ the local interacting (time-0) electron Dirac field operator in QED, with
 ${\vec x} \in \R^3$ and $\alpha$ the spinor index. 
Then one naively  expects that $\hat \psi_\alpha ({\vec x})\ket{\Omega}$ is a charged state. However, it is not a state in the physical Hilbert space, even with a UV cutoff or after smearing with a test function. In fact,
 \begin{equation}          
[Q_{B} ,\hat \psi_\alpha ({\vec x}) ] \neq 0 \label{BRS}
\end{equation} 
which implies, using that $|\Omega \rangle$ is physical,
\begin{equation}  Q_{B} \hat \psi_\alpha ({\vec x})\ket{\Omega} \neq 0  
\end{equation}                
(Note, however, that perturbatively for the charged asymptotic field, $\hat \psi^{as}_\alpha$, assumed to exist,
  $[Q_{B} , \hat \psi^{as}_\alpha({\vec x}) ] = 0$,
 see \cite{kugo1979local}.)   
The basic motivation of (\ref{BRS}) is that $\hat\psi_\alpha({\vec x})$ is not gauge invariant, and 
BRST transformations are naturally related to gauge transformations.
To convert $\hat \psi_\alpha ({\vec x})$ into a gauge-invariant field operator Dirac \cite{dirac1955gauge} proposed the following ansatz: Let $\hat {\vec A}$ denote the quantum photon  gauge field
 and ${\vec E}_{\vec x}$ a classical Coulomb-like electric field satisfying
 \begin{equation}
\label{div}            {\vec  \nabla} \cdot {\vec E}_{\vec x} ({\vec y})= \delta^{(3)}({\vec x}- {\vec y}).
\end{equation}                                                        Then a "physical electron operator" is formally given by 
\begin{equation}
\label{ginv}
 \hat \psi_\alpha ({\vec x}) \exp [i \int d^3{ y} \hat {\vec A} ({\vec y}) \cdot {\vec E}_{\vec x} ({\vec y})].
\end{equation}
 It is gauge-invariant: under a gauge transformation with parameter $\Lambda({\vec y})$ we have
\begin{equation}
\hat \psi_\alpha({\vec x}) \rightarrow \hat \psi_\alpha({\vec x}) e^{i \Lambda ({\vec x})} , \quad
\hat {\vec A}({\vec y}) \rightarrow \hat {\vec A}({\vec y})+{\vec \nabla} \Lambda ({\vec y}), 
\end{equation}
and, by (\ref{div})
\begin{equation}
\int d^3{ y} {\vec \nabla} \Lambda({\vec y})\cdot {\vec E}_{\vec x} ({\vec y})= - \int d^3{y}  \Lambda ({\vec y}) \nabla \cdot {\vec E}_{\vec x} ({\vec y}) 
  = - \Lambda ({\vec x}),
\end{equation}
The field (\ref{ginv}) is clearly non-local.      
 The ${\vec E}$-dependent Dirac dressing phase factor in (\ref{ginv}) describes the Coulomb photon cloud tied to the electron even asymptotically and its support reaches
 space-like infinity.
 
Euclidean correlators of the field  operators $\hat \psi_\alpha ({\vec x}) \exp [i \int d^3{ y} 
\hat {\vec A}({\vec y})\cdot {\vec E}_{\vec x} ({\vec y})]$ according to the previous section have the following form
\begin{equation}
    \langle...\psi_\alpha(x) \exp[i \int d^4y A_\mu (y) E_x^\mu (y)]...\rangle.
\label{psie}
\end{equation}
In (\ref{psie}) $x = (x^0 , {\vec x}) \in {\R}^4 , \mu=0,1,2,3, \psi_\alpha(x)$ is a Grassmann field, $ A_\mu(x)$ the gauge field, $E_x^\mu $ 
is an electric current distribution
 related to ${\vec E}_{\vec x}$ by   
 $E_x^\mu(y) = (0, {\vec E}_{\vec x} ({\vec y}) \delta(y^0 - x^0))$, so that $\partial_\mu E_x^\mu (y) = \delta^{(4)}(x -y) $ and
 $\langle \cdot \rangle$ denotes the vacuum expectation value in the Euclidean path-integral
 measure for QED .
 
 The above result can be made rigorous with a lattice regulator. Furthermore
 an OS reconstruction theorem allows to reconstruct from these correlators (with UV cutoff)
 the corresponding non-local field operators \cite{frohlichsuperselection} (see also \cite{frohlich1999gauge}).
 
By integrating out $\psi_\alpha$ in Euclidean QED one can obtain a Feynman-Schwinger representation of correlators in terms of
 worldlines of two
 kinds: closed, corresponding to virtual particle-antiparticle pairs, and open with boundary on the points of the field insertions, 
corresponding to creation and  annihilation of  electrons/positrons. 
The crucial point is that because of the Dirac dressing at these boundary points the  electric flux flowing through the open worldlines is spread out through the electric current distributions 
$E$ , thus preserving current conservation, consequence of gauge invariance, in spite of particle creations and annihilations [see Fig.7 (a)]. 

\begin{figure}[h]
\centering
\includegraphics[width=0.9\textwidth]{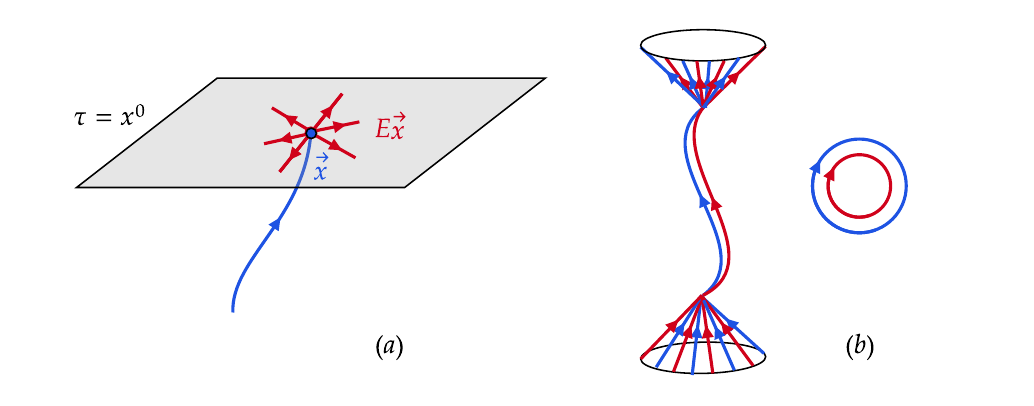}
\caption{(a) Worldline of an electron (in blue) and Coulomb field associated with its annihilation (in red) in the Euclidean version of Dirac ansatz drawn in d=2+1. (b) Worldlines that appear in the Euclidean two-point correlator of an anyonic field: one worldline associated with creation and annihilation (left) and the other to a virtual anyon-antianyon pair.}
\end{figure}

Note that the only constraint on the classical electric field ${\vec E}_{\vec x}$ is (\ref{div}), so its support can by restricted to a space-like cone of arbitrary opening angle $\delta$.  Formally in the limit $\delta \rightarrow 0$, neglecting problems of divergence (except for topological QFTs), it reduces to a "Mandelstam string \cite{mandelstam1962quantum}" (or "Wilson line").

\subsection{Anyonic quantum fields}

We now discuss how the Dirac ansatz solves the puzzle mentioned at the beginning of this section, first in the path-integral formalism and then in the operator formalism.

For simplicity we start considering  the model with Euclidean lagrangian (\ref{CS}) but many considerations made in this section apply to general quantum field theories of anyons and a different interesting model is briefly discussed later.

The anyons in the considered model are its charged particle excitations.
The non-local physical anyonic field is constructed as follows: let
$E_x^\mu , x \in {\bf R}^3$  denote a 3-dimensional Coulomb-like electric field distribution satisfying  $\partial_\mu E_x^\mu (y) = \delta^{(3)}(x-y)$.
It is assumed that the support of $E_x^\mu$
is given by a three-dimensional cone $C_x$ with apex $x$, taken in the 
positive (negative) time half-space if 
$x^0>0$ $ (x^0 <0)$, in order to have positive metric in the OS-reconstructed Hilbert space of states.  
Gauge-invariant  "physical" Euclidean anyon fields are then given by
\begin{equation}
\label{anyon}
\phi(E_x) =   \phi(x)  \exp[i \int d^3y A _\mu(y) E_x^\mu (y)].
\end{equation}
Anti-anyons fields are defined similarly by taking the complex conjugation of  (\ref{anyon}). We use the word "anyon" generically to mean both anyons and anti-anyons.

Via a variant of OS reconstruction theorem  one can (formally) obtain \cite{frohlich1989quantum} non-local anyon field operators
 $\hat\phi(E_{x}), {x} \in {\R}^3, x^0 > 0$. 
Exponential clustering applies because the Chern-Simons gauge field in $\R^2$ has no physical degrees of freedom (and on Riemann surfaces its states belong to a finite-dimensional Hilbert space), its only effect is to transmute the statistics of the  massive scalar field. Furthermore in the lattice approximation one can prove that  $\hat\phi(E_{x})$ couples the vacuum to a one-particle state, i.e. it creates and annihilates anyon particles.
 
 The local field $\phi(x)$ carries electric charge 1, for the minimal coupling to $A_\mu$, and magnetic flux $2 \pi \theta$, for  the Chern-Simons interaction,
 as can be easily verified using the equations of motion for $A_\mu$.
So along the support of
$E_x^\mu$ flow both
 an electric and a magnetic flux; with a UV lattice regulator the support of the two fluxes is splitted, as discussed in the previous section [see Fig.7 (b)]. Below we show that these two kind of flux lines
 play the role of the two boundaries of the band in a QFT version of the "rubber band lemma", thus exhibiting spin-statistics connection.
 
We first discuss the spin and then the statistics with similar methods \cite{frohlich1989quantum} \cite{frohlich1988quantum} .

Since the “physical” charged  fields are non-local with a tail reaching infinity, a rotation by an angle $\varphi$ should be defined as 
 the $R \rightarrow \infty $ limit of rotation with InfraRed (IR) cutoff $R$, $U(\varphi)^R$ (at positive Euclidean time), acting as a  rotation by $\varphi$ within a ball
 of radius $R$ centered at the origin,
  smoothly interpolating to the identity for distances between $R$ and $R + 1$ and acting  trivially outside a ball of radius $R + 1$.
 This limit procedure
ensures that the rotation is generated by a local current. 
A $2 \pi$-rotation with IR cutoff acts, as discussed in the previous section, on the dressing factor in the anyon field $\phi(E_x)$ producing a 
phase factor proportional to the 
linking number $\ell$ of electric and magnetic flux lines as $R \rightarrow \infty$. The difference $U(2 \pi )^R (E_x^\mu )- E_x^\mu$ contains only closed flux lines, so that it has zero divergence [see Fig.8 (a)]. 

\begin{figure}[h]
\centering
\includegraphics[width=0.9\textwidth]{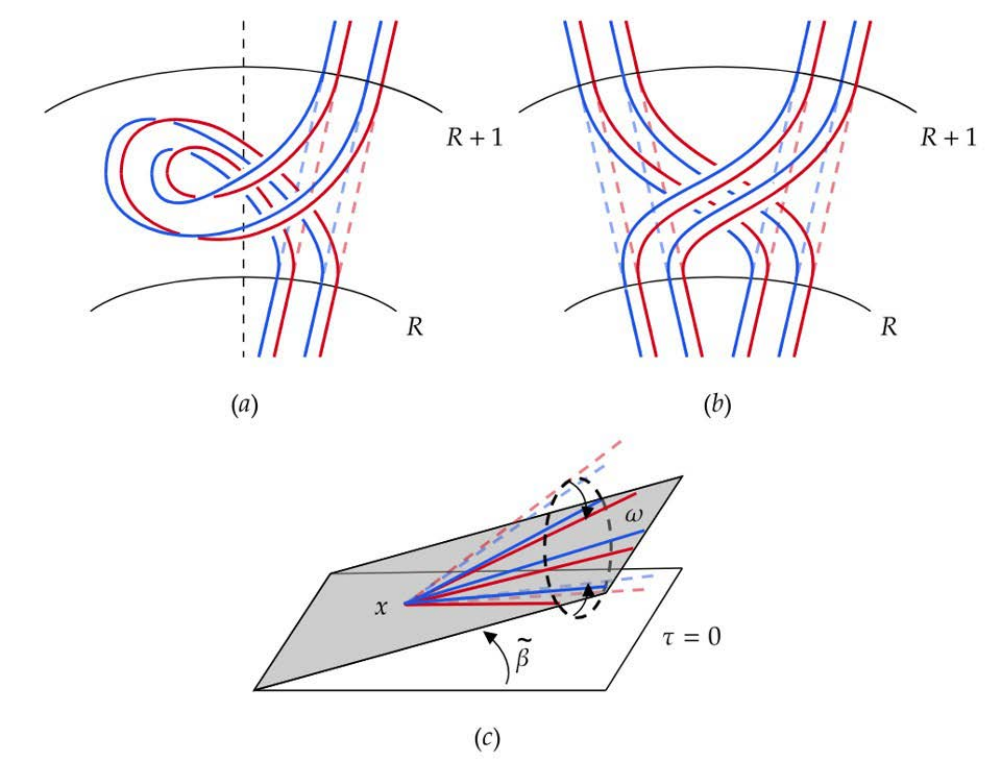}
\caption{ (a) Electric and magnetic flux lines appearing in the Euclidean cone for an anyon corresponding to $U(2 \pi )^R (E_x^\mu )$. (b) The same for $U(\sigma )^R (E_x^\mu +  E_y^{\prime \mu})$ in an exchange of two anyons. (c) Shrinking from an Euclidean cone to an Euclidean wedge rotated by an angle $\tilde{\beta}$ in the Euclidean time direction.}
\end{figure}

Then, since $\R^3$ is contractible, by the Poincaré lemma there exists a 1-form whose components we denote with  $a_\mu^R$ so that
\begin{equation}
\label{spin}
U(2 \pi )^R (E_x^\mu )- E_x^\mu = \epsilon^{\mu\nu\rho} \partial_\nu a_\rho^R .
\end{equation}
 Then
\begin{eqnarray}
\label{ast}
&\lim_{R\rightarrow \infty} \langle U(2 \pi )^R (\phi(x)  \exp[i \int d^3y A_\mu (y) E_x^\mu(y)])...\rangle= \nonumber\\
&\lim_{R \rightarrow \infty} \langle.... \phi(x)  \exp[i \int d^3y A_\mu (y) (E_x^\mu(y)+  \epsilon^{\mu\nu\rho} \partial_\nu a_\rho^R (y)]...\rangle,
\end{eqnarray}
where $\langle \cdot  \rangle$ denotes the vacuum expectation value in the Euclidean path-integral measure for the model. As in section 5.2 we first introduce
 the Feynman-Schwinger representation for $\phi$ in terms of currents $j^\mu$, considering $A_\mu$ as an external field and then we integrate
 on it with Chern-Simons action. 
 We then perform the change of variable $A_\mu \rightarrow A_\mu + a_\mu^R$.
 By extracting the contribution coming from the deformation induced by the twist of $2 \pi$
 one can reconstruct the original path-integral measure obtaining:
\begin{eqnarray}
\label{fin}
 &\lim_{R \to \infty}  \exp[i \frac{\pi}{k} \int d^3y \epsilon^{\mu\nu\rho} a_\mu^R \partial_\nu a_\rho^R (y)+O(1/R)]
\langle.... \phi(x)  \\&\exp[i \int d^3y A_\mu (y) E_x^\mu (y)])...\rangle
= e^{i 2\pi\theta} \langle... \phi(x)  \exp[i \int d^3y A_\mu (y)E_x^\mu (y)])...\rangle.\nonumber
\end{eqnarray}
The term denoted with $O(1/R)$ in (\ref{fin}) comes from the interaction of  $a_\mu^R$ with $E_x$ and 
$\int d^3y \epsilon^{\mu\nu\rho} a_\mu^R \partial_\nu a_\rho^R (y)$ gives the linking number of the electric and magnetic flux 
lines of the field $a_\mu^R $, yielding $\ell=1$.

\textit{Remark} To be convinced that $\ell =1$ one can break down the flux into $n$ fluxes of intensities $1/n$, then use the reasoning made for the spin addition rule eq. (\ref{SN}): the factor $1/n \times 1/n $ is compensated by the $n^2$ which comes from an argument similar to that used to derive the quantum-mechanical spin addition rule.

The phase obtained clearly measures the $2d$ spin or more precisely the spin-type, i.e. the spin modulo integers.
Hence, although the local field $\phi$ has spin 0, the physical anyon field has spin-type $S = \theta = 1/(2k)$.

The statistics calculation then proceeds in a similar way to the spin.
Similarly a $\sigma$ exchange on non-local fields should be defined as the $R \rightarrow \infty$ limit of an exchange with IR cutoff $R$,
$U(\sigma )^R $. This acts by exchanging the local fields and distorting the electric current distributions
of the two Dirac dressing of the physical fields that are exchanged in such a way that within a ball of radius $R$ the two distributions
 are exchanged,
 the deformation smoothly interpolates to the identity for distances between $R$ and $R+1$ and leaves the distributions unchanged
 for distances greater than $R+1$ [see Fig.8 (b)]. 

Let us consider $E_x$ and $ E^\prime_y$
 two electric current distributions with disjoint supports; as $R\rightarrow \infty$ the above procedure for the definition 
of oriented exchanges gives rise to a phase factor proportional 
to the linking number $\int d^3y \epsilon^{\mu\nu\rho} a_\mu^R \partial_\nu a_\rho^R (y)=\pm 1$ of electric and magnetic flux lines  of
$U(\pm\sigma )^R (E_x^\mu +  E_y^{\prime \mu})- (E_x^\mu +  E_y^{\prime \mu}) )\equiv \epsilon^{\mu\nu\rho} \partial_\nu a^R_\rho $ that produces
 a statistics transmutation from  the bosonic statistics of $\phi$ to a statistics parameter $\theta = 1/(2k)$ for the physical anyonic field. 
 
 The spin-statistics connection, $S= \theta$, 
follows simply from the
 "rubber band lemma" applied to the bands of electric and magnetic flux in the  "Dirac dressing"
 which describes the difference between the deformed and undeformed Coulomb fields $E$, pushed to infinity by the $R\rightarrow \infty$ limit.
   A priori both the $2 \pi$ rotation and the exchange  have two contributions: one local describing the “bare” spin and statistics parameter (of $\phi$) and the other 
topological, located at
infinity, from the Dirac dressing factor. The second contribution as we have seen for anyons can transmute spin(-type) and statistics of the charged local (non-gauge invariant) field. 

\textit{Remark} This basic idea \cite{marchetti2010spin} works not only for anyons, as discussed in this section,  but also for dyons \cite{lechner2001spin}, where a modification is needed to take into account Dirac quantization
condition, and skyrmions \cite{froehlich1990quantum}, 
when the above procedure has to be adapted to their soliton nature. 

  With the strategy used for spin and statistics one can also  derive for the anyon fields the spin addition rule seen in quantum mechanics.
 Let us assume that the supports of $E_i$ and $E_j$ for Euclidean anyon fields do not intersect for $i \neq j, i , j =1, \dots, N$. We can then define $a_i^R$ by
$$
E^\mu_i-U(2 \pi)^R E^\mu_i= \epsilon^{\mu \nu \rho} \partial_\nu a_{i \rho}^R.
$$

Summing the contributions of all the fields we need to compute
$$
\lim_{R \rightarrow \infty } \sum_{i,j} \int \epsilon^{\mu \nu \rho}  a_{i \mu}^R\partial_\nu  a_{j \rho}^R
$$

It is easy to understand that if $\Sigma_{i}$ is a surface whose boundary is given by
a flux line of $E_i$ , then for $R$  sufficiently large, adopting the same strategy for the flux lines used above for the one-anyon spin,  the resulting linking number is $\ell = N^2 $. Therefore as in quantum mechanics the 2d spin of $N$ anyons
is not the sum of the spins $S$ of the individual anyons but $N^2 S$.

\textit{Remark: Maxwell-Chern-Simons}
The addition to the lagrangian (\ref{CS}) of the Maxwell term for the gauge field, which gives rise  to the so-called Maxell-Chern-Simons theory, is discussed in the Hamiltonian formalism e.g. in \cite{semenoff1989exotic}, neglecting the non-perturbative problems of cutoff.

\subsection{Anyons as vortices}

The anyons in the specific model considered above have unit electric charge and fractional $2 \theta$ magnetic flux (in our units $\hbar = 1 = e$ the magnetic flux quantum is $2 \pi$ , so that $\Phi/(2 \pi)$ gives the number of magnetic flux quanta). One can also consider a model where anyons instead have unit magnetic flux and fractional electric charge $2 \theta$. Particle excitations with unit magnetic flux are vortices, so a model in which they appear is of Ginzburg-Landau (see paper 73 in \cite{ter2013collected}) or Abelian Higgs \cite{higgs1964broken} type with broken $U(1)$ electromagnetic global symmetry.
 In a model with electrically charged vortices, the Abelian Higgs model with Chern-Simons term with gauge potential $A_\mu$ and charged scalar field $\phi$, the Euclidean action is given by
\begin{equation} 
\label{HiggsCS}
    S(A, \phi)=\int d^3x \big[ \frac{1}{4}(F_{\mu\nu} )^2 + \frac{1}{2}|(\partial_\mu- i A_\mu ) \phi|^2 +{g}(|\phi|^2 -\phi_0^2)^2 + \frac{i \theta }{2 \pi } 
\epsilon^{\mu\nu\rho}A_\mu \partial_\nu A_\rho \big] (x) .
\end{equation}

To understand how anyons appear in this model we need a small detour on the quantum field theory of vortices.
The vortices are defects of the model, but let's try to specify more precisely in what sense.

The word defect in QFT (or statistical field theory) has recently acquired a wider meaning than originally used. In a somewhat vague definition a defect  correspond in correlators to an "inhomogeneity"
 localized on a connected submanifold of space-time (or space). The dimension of such submanifold (or of the smallest submanifold to which it is contractible within its support) is called the dimension of the defect, so one speaks of point, line, surface ... defect. 

In terms of the Euclidean path-integral formalism for a field theory defects were originally defined for field configurations. For a fixed field configuration let $\Gamma$ be a region in the Euclidean space-time where the configuration differs strongly
from the global minimum of the action (i.e. from the vacuum) but is close to a local minimum (usually corresponding to a classical solution of the equations of motion). If $\Gamma$ is connected it is said
to be the support of a defect (see e.g. \cite{Seiler:1982pw},\cite{frohlich1981higgs}). 

In a partition function in $\R^d$ for a translationally invariant (pure) vacuum, only closed defects appear, since at "infinity" the field configurations must approach the vacuum configuration.

The vortices in the Abelian Higgs model in the three-dimensional Euclidean space-time in the Higgs phase are line defects in the sense indicated above. Indeed, it is well known that for a vortex configuration the first three terms in (\ref{HiggsCS}) are large only in a compact
region of space outside of which they converge to 0 in exponentially fast way (see e.g. \cite{jaffe1980vortices}). The Chern-Simons term has the effect of providing these vortices with an electric charge, as can be seen already classically from the equation of motion 
\begin{equation}
\label{Higgseq}
 \partial^\mu F_{\mu 0} + \frac{\theta}{\pi} \epsilon_{0 \mu \nu} F^{\mu \nu}  = j_0.
\end{equation}
 By integrating in space (\ref{Higgseq}) we get the relation between electric charge $Q$ and magnetic flux $\Phi$ of the vortices:
 \begin{equation}
 \label{Qphiv}
     Q  = \frac{2 \theta}{2 \pi} \Phi.
 \end{equation}

 As a general rule Euclidean correlation function of quantum fields associated to line defects can be obtained by "opening"
 closed line defects appearing in the partition function, as we have done for the "worldlines" of electric currents in (\ref{2point}).
 
 A natural source of a magnetic flux is a monopole. In order to construct the Euclidean two-point function of a quantum vortex field with a vortex created at $x \in \R^3$ and annihilated at $y \in \R^3$, obtained by introducing a open 
vortex line with boundary $x$ and $y$,  we then proceed as follows.

Let us first consider (\ref{HiggsCS}) without Chern-Simons term. We construct a modified version of the action obtained by shifting $F_{\mu \nu}$ by a field strength $\mathcal{F}_{\mu\nu}$ corresponding to a curvature in the cohomology class of a monopole in $x$ and an antimonopole in $y$, and replacing the derivative in the second term of (\ref{HiggsCS}) with a covariant derivative with the corresponding connection, locally given by $\mathcal{A}_\mu$. In particular we can choose the difference $\mathcal{F}_{\mu\nu}-(\partial_\mu \mathcal{A}_\nu -\partial_\nu \mathcal{A}_\mu)$ as a de Rham current carrying magnetic flux with support on a line from $x$ to $y$.
Then the two-point function is the ratio between the partition function with the modified action and the original one (suitably UV renormalized).
Note that the magnetic current from $x$ to $y$ creates a line-defect in the two-point function.

When we add the Chern-Simons term the vortex lines become electrically charged and due to the Gauss law we have to implement a Dirac ansatz, so that the electric lines have no source or sink. To implement the Dirac ansatz in this context we consider the  $E_x^\mu$ and $E_y^\mu$ electric distributions introduced before.
Since
\[
\epsilon^{\mu \nu \rho} \partial_\mu(\mathcal{F}_{\nu\rho}-\epsilon_{\nu \rho \tau}(E_x^\tau-E_y^\tau))=0
\]
 by the Poincaré lemma it exists a current $\alpha_\mu(E_x,E_y)$ such that
\[
(\mathcal{F}_{\nu\rho}-\epsilon_{\nu \rho \tau}(E_x^\tau-E_y^\tau))= \partial_\nu \alpha_\rho(E_x,E_y) -\partial_\rho \alpha_\nu(E_x,E_y) .
\]
After having also modified the Chern-Simons term by shifting in it $A_\mu$ by $\alpha_\mu(E_x,E_y)$, taking again the ratio between the partition function with the modified action and the original one (appropriately UV renormalized) we obtain the Euclidean two-point function of the quantum vortex field. More precisely, generalizing the above construction to an arbitrary number of anyon field insertions and applying the OS reconstruction theorem (formally in the continuum and rigorously with a lattice regularization) we reconstruct the quantum field operator of the model. Exponential clustering applies, because the Maxwell gauge field acquires a gap by the Anderson-Higgs mechanism \cite{anderson1963plasmons},\cite{higgs1964broken},\cite{englert1964broken},\cite{guralnik1964global}.

In the lattice one can also prove that the anyonic field operator also in this model couples the vacuum to a one particle state, i.e. it creates and destroys anyon particles \cite{frohlich1989quantum}.

\textit{Remark} Since the role of electric charge and magnetic flux in this model of anyonic vortices is exchanged compared to the previously considered model, one might guess that the two models are approximately related by a duality transformation. In Appendix B we provide an heuristic "proof" that this is indeed the case at an appropriate limit.

\subsection{Anyonic quantum field operators}

Let us discuss briefly the operator formalism.

Quantum field operators can be obtained via OS reconstruction from the correlators of the Euclidean fields discussed above, based on the electric distributions $E$ supported in cones $C$ in the Euclidean space-time. Unfortunately the reconstructed fields are completely non-local in Minkowski space-time, therefore we cannot discuss their statistics that need space-like separated supports. To construct operators localized in Minkowski space-like cones that can have space-like separated supports one should proceed as follows. 
 
First we construct Euclidean anyon correlation functions in which the support $C$ of each electric flux distribution $E$ is shrunk to a wedge $\mathcal{W}$ contained in a two-dimensional plane [see Fig.8 (c)] rotated by an angle $\tilde{\beta}$ in the Euclidean time direction.
(With the exception of topological QFTs, a renormalization of the self-energy of $E$ is necessary, as it  diverges logarithmically, while the cone shrinks into a two-dimensional wedge.)

From these correlation functions via OS reconstruction and analytic continuation in $\tilde{\beta}$ (analogous to the inverse Wick rotation in time) we obtain (distribution-valued) non-local quantum fields $\hat\phi(\mathcal{W}_x,E(\beta))$ with the electric distribution $E(\beta)$ localized in a wedge $\mathcal{W}_x$ with apex at $x$ and support in a space-like plane  boosted by $\beta$ with respect to the time zero plane in the  Minkowski three-dimensional space.
With $f$  a test function in general for ultradistributions in $\R^3$\cite{constantinescu1979ultradistributions} (for example, for anyon vortices the UV singularity is stronger than that allowed by the tempered distributions), and with $g \in \mathcal{S}(\R)$, 
\begin{equation}
\hat\phi(\mathcal{C}(f,g;E))= \int d^3x f(x) \int d \beta g(\beta) \hat\phi(\mathcal{W}_x,E(\beta))
\end{equation}
 is a well defined operator, with support in a space-like cone $\mathcal{C}(f,g;E)$ which can be obtained from the straight string  (half-line) at the center of the support of $E(\beta)$ by acting on it with a neighbourhood of the identity of the Poincaré group determined by  $f,g, E$. 
 For simplicity, the explicit reference to $(f,g;E)$ is from now on omitted in the notation of the cone.
 
 Ideally shrinking the cone to the above quoted string (ignoring the derived UV problems) it can be considered as an analogue of the Mandelstam string (or Wilson line) introduced in gauge theories.  Actually, a deep theorem of Buchholz and Fredenhagen \cite{buchholz1982locality} essentially proves that in relativistic QFT with all the states generated by one-particle states, without massless particles (and with zero particle density) the support of fields coupling the vacuum to a one-particle state can be localized in space-like cones of arbitrarily small opening angle.
 Of course this choice is not optimal for the energy, which is expected to decrease by widening the angle. 
 
 How does this localization avoid the problems encountered for local fields in (\ref{CCRS})? The point is that in two spatial dimensions the manifold of directions at infinity $S^1_\infty$ is not simply connected being isomorphic to $S^1$, and  similarly the manifold of space-like directions in Minkowski 2+1 dimensional is contractible to $S^1_\infty$. If we define the action of rotations of an angle $\alpha$ on $\hat\phi(\mathcal{C})$ with a limit procedure as done in the Euclidean framework:
 \[
 \hat\phi(\mathcal{C(\alpha)})= \lim_{R \to \infty} U(\alpha)^R  \hat\phi(\mathcal{C})
 \]
 (the limit $R \to \infty$ being taken in the weak sense), because of the non-simply connected nature of the direction at infinity a priori 
 \begin{equation}
 \label{Calpha}
\hat\phi(\mathcal{C(\alpha)}) \neq \hat\phi(\mathcal{C}(\alpha + 2 \pi)).   
 \end{equation}
 
 The difference between the two members of the equation must 
be an operator that commutes with all observables since a rotation of $2 \pi$ leaves all observables unchanged. In fact, using the representation of the rotation on the states obtained applying to the vacuum $n$ anyonic fields, it is deduced from the spin addition rule that for an anyon of spin $S$ that the second member of (\ref{Calpha}) is equal to the first times $e^{i 2 \pi ((n+1)^2-n^2) S}$.

 As it happens when multivaluedness appears, we can work with field operators whose asymptotic direction is in the universal covering space of $S^1_\infty$, allowing all possible real $\alpha$ without defining them modulo $2 \pi$. However, if we want to make the field single-valued we must use a simply-connected chart.
 
 Here  we discuss the statistics for non-local quantum fields $\hat\phi(\mathcal{W}_x,E(\beta))$ which are the analogues of the distribution-valued field operators, but with obvious changes the following discussion applies to the field operators localized in space-like cones $\mathcal{C}$. 
 
 Given a wedge $\mathcal{W}$  we define as($\mathcal{W}$) the projection in the time-zero plane of the asymptotic direction of the axis of the wedge.
 We fix a forbidden direction at infinity, $\xi$, defining an associated chart and consider only wedges $\mathcal{W}$ whose projection in the time-zero plane of their directions at infinity does not intersect $\xi$ and with as($\mathcal{W}$) between $\xi -2 \pi$ and $\xi$. In this range all the fields are single-valued. Therefore we put an upper index $\xi$. The introduction of the direction $\xi$ determines a natural order among space-like separated space-like wedges.
 Then for space-like separated wedges $\mathcal{W}^1_x$ and $\mathcal{W}^2_y$ one has similarly to what happens for local quantum fields in one space dimension (see (\ref{CCRS}))
 \begin{equation}
 \label{statqft}
 \hat\phi^\xi(\mathcal{W}^1_x,E(\beta_1))   \hat\phi^\xi(\mathcal{W}^2_y,E(\beta_2)) =  e^{\pm i 2\pi \theta} \hat\phi^\xi(\mathcal{W}^2_y,E(\beta_2)) \hat\phi^\xi(\mathcal{W}^1_x,E(\beta_1))
 \end{equation}
 with the two signs that occur if as$\mathcal{W}^1_x \gtrless $ as$\mathcal{W}^2_y$.
 
 In the non-relativistic QFT with gap and zero-anyon density, a similar situation occurs is expected to occur without the action of boosts i.e. for non-local fields $\hat\phi(\mathcal{W}_x,E)$ localized in wedges in the time-zero plane.

\subsection{Anyon wave functions from anyon field operators}

A question which naturally arises is: what is the relation between the anyonic field operators discussed above and the wave functions we have seen in section 2.

The link is provided in momentum space (as discussed in section 3, a spatial description is not allowed in RQFT) by the Haag-Ruelle scattering theory \cite{haag1958quantum},\cite{ruelle1962asymptotic} (see also e.g. \cite{glimm2012quantum}).
We just sketch here its basic ideas adapted to the anyons \cite{frohlich1989quantum}\cite{frohlich1991spin}(see also \cite{fredenhagen1996scattering}).

As done in section 3 we denote by $V_m$ the forward hyperboloid of mass $m > 0$ of the one-particle state to which the anyon field operator couple the vacuum.

 Let $\mathcal{C}$ be a space-like cone with apex in the origin in a Lorentz reference system where the axis of $\mathcal{C}$ is at constant time and we denote by  $\mathcal{C}(\alpha)$ the cone rotated by an angle $\alpha$.  With an appropriate average of a field operator $\hat\phi^\xi(\mathcal{C(\alpha)})$ we construct field operators with definite 3-momentum $p$:
\begin{equation}
 \hat\phi^\xi(\mathcal{C}(\alpha;p)) \equiv \int d^3 x U(x) \hat\phi^\xi(\mathcal{C}(\alpha)) U(x)^\dagger e^{i p x},
\end{equation}
where $U(x)$ denotes the representation of the translation of $x \in \R^3$. 
Let us now consider the
subspace $\mathcal{S}(\mathcal{C}) \subset \mathcal{S}(\R^3)$ defined as the set of test functions $h$ whose Fourier transform $\tilde{h}(p)$ satisfies the following conditions:
1) the intersection of the support of $\tilde{h}(p)$ with the spectrum of the 3-momentum of the theory is contained in $V_m$, 
2) the velocities $\vec p/(m^2+ \vec p^2)$ in the support of $h$ are contained in the interior of $\mathcal{C}$, so that their directions are contained in the asymptotic space directions of $\mathcal{C}$.

If we now average  with functions in $\mathcal{S}(\mathcal{C}_i), i=1, \dots, n$ a set of fields  whose cones   $\hat\phi^\xi(\mathcal{C}_i(\alpha_i;p_i))$ are space-like separated and we let them evolve with the hamiltonian of the system for a time $t$, for $t \to \pm \infty$ the distance between their supports (up to exponentially small tails) tends to infinity and thank to the mass gap the interaction among them is  damped exponentially, so we can "send them back" with the free time evolution. In formulas this  means that one can prove the existence of the strong limits
\begin{equation}
    \lim_{t \to \pm \infty} \int \prod_i d^3p_i\tilde h_i(p_i) e^{i(p^0_i-\sqrt{\vec{p_i}^2 + m^2})t} \hat\phi^\xi(\mathcal{C}_1(\alpha_1;p_1)) \dots \hat\phi^\xi(\mathcal{C}_n(\alpha_n;p_n)\ket{\Omega}.
\end{equation}

The corresponding wave functions will be denoted $\psi^{as} (p_1, \dots. p_n), n \in \Z$, where $as=in$/$out$ for $t \to -$/$+ \infty$.
The space of wave functions  for anyons, $\mathcal{H}^{as}$, is the closure of the linear span of these wave functions. The fact that the directions of $p_i$ are linked to the asymptotic direction of $\mathcal{C}_i(\alpha_i)$ from the averaging with $\tilde h_i$
forces the one-particle wave functions to have spin $S=\theta$ and the $n$-particle wave functions to satisfy the anyon spin-addition rule. In addition, the oriented exchanges of fields produce factors $e^{\pm i 2 \pi \theta}$. Therefore these asymptotic momentum wave-functions have all the properties of the wave-functions for anyons discussed in section 2.

 \section{Where anyons appear in nature}

 Anyons appear in nature in the fractional quantum Hall effect (FQHE), as demonstrated experimentally completely satisfactorily only in 2020 \cite{nakamura2020direct}.
 In this section we first outline the main features of the quantum Hall effect (QHE)\cite{tsui1982two},\cite{laughlin1983anomalous}, see also e.g. \cite{prange1989quantum}]. Then we give some more details on the theory of the fractional quantum Hall effect and the associated edge currents with an additional brief comment on the fractional spin QHE and finally we turn to a short exposition of the experiments proving the fractional charge and statistics of the charge carriers in the FQHE.

\subsection{A look at the quantum Hall effect}

The effect appears in two-dimensional electron systems in the presence of a
strong transverse magnetic field.

Experimentally, the relevant two-dimensional electron systems
are realized in so-called inversion layers, formed at the interface between a semiconductor
and an insulator 
(e.g. Ga As- Al${}_{1-x}$ Ga${}_x$ As). 
In the
above systems an electric field is produced, perpendicular to the interface, which attracts
electrons from the semiconductor to the interface [see Fig.9 (a)]. These electrons then move
in a quantum well created by this field and the interface. The motion perpendicular
to the interface is quantized and so degrees of freedom in this direction are frozen. The result is an effective two-dimensional electron system (2DES), see Fig.9.

\begin{figure}[h]
\centering
\includegraphics[width=0.9\textwidth]{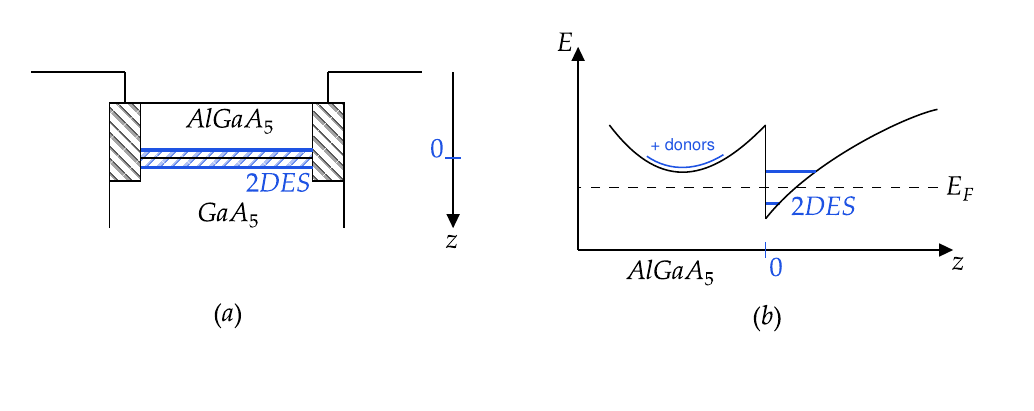}
\caption{Schematic representation of an inversion layer (a) and of the associated potential (b).}
\end{figure}

In this two-dimensional electron system we introduce a magnetic field $B$ perpendicular to the plane and an electric field $E$ parallel to the plane. This induces a Hall current perpendicular to $E$ in the plane, due to the Lorentz
force. Suppose also that  the longitudinal conductivity $\sigma_L$ vanishes, then the relation between the current in the sample and the electric field is given by the Hall law
\begin{equation}
\label{j}
j_\mu = \sigma_H \epsilon _{\mu \nu}E^\nu
\end{equation}
where $\sigma_H$ is the Hall conductance (or conductivity) and the indices label the two directions of the plane.
Note that if $\sigma_H \neq 0$, the vanishing of longitudinal conductivity $\sigma_L$ implies the vanishing of longitudinal resistivity or resistance $R_L$.

\textit{Remark} In fact, the standard geometry for QHE experiments is the Hall bar schematically shown in Fig.10 (a), where the two-dimensional electron gas is located.
The way to use the Hall bar is to drive a current $I$
along the direction let's $x$, corresponding to a current density $j_x=(I/W)$ where $W$
is the width of the bar. There is no current density in the perpendicular direction, so $j_y = 0$.
The electric field $E_x, E_y$ is measured from the voltage drops $V, V_H$ in Fig.10 (a). We can calculate the conductivities solving the set of equations $j_x=\sigma_LE_x+\sigma_HE_y$ and $j_y=\sigma_LE_y-\sigma_HE_x=0$, in particular obtaining 
$\sigma_H=j_xE_y/(E_x^2+E_y^2)$.

In the 2-dimensional electron systems  of the above-mentioned inversion layers at very low temperature $(T \sim  1$ K)
 and in a very strong magnetic field
$(B \sim 10$ T), the Hall conductivity shows experimentally the following behaviour. If it is plotted as a function of the filling $\nu = n h/( eB)$, with $n$ the electron density, one gets
a series of plateaux at the values
\begin{align}
\label{sigmaH}
    \sigma_H = \frac{e^2}{h}
 \nu , \quad \nu=\begin{cases}&1,2,3, \dots\\
 & \frac{1}{3},\frac{2}{3},\frac{1}{5},\frac{2}{5}, \dots
  \end{cases}  
\end{align}
and more generally for $\nu = m/p$,  $m$ and $p$ integers without common divisor and $p$
odd; see Fig.10 (b) for the corresponding behaviour of the Hall resistivity $R_H$.
(Some plateaux with even denominators have also been observed in some material, but we do not discuss them here.)

At the plateaux it was found that the longitudinal resistivity (and therefore also conductivity) vanishes so that we recover the Hall law (\ref{j}).
(In the following, unless explicitly written, we set $e =1 = \hbar)$.

The appearance of plateaux at integer values in (\ref{sigmaH}) is called Integer
QHE (IQHE), the effect was discovered by von Klitzing \cite{klitzing1980new} and for it he was awarded with the Nobel prize in 1985. The appearance of plateaux at fractions with an odd denominator is 
called Fractional QHE (FQHE).

Let us try to describe in a heuristic way why the longitudinal conductivity vanishes
at the plateaux. 

In the presence of the magnetic field the
single particle states of the electron system group together into Landau levels (see paper 4 in \cite{ter2013collected}). Then the presence of impurities  broadens the levels, but also has a much stronger effect. As suggested by Anderson \cite{anderson1958absence}, beyond a critical amount of impurity scattering, the diffusive  motion of the electrons is not only reduced, but stops completely: the electron is trapped near an impurity and the conductivity vanishes.

\begin{figure}[!htb]
\begin{center}
  {\includegraphics[width= 6.0cm]{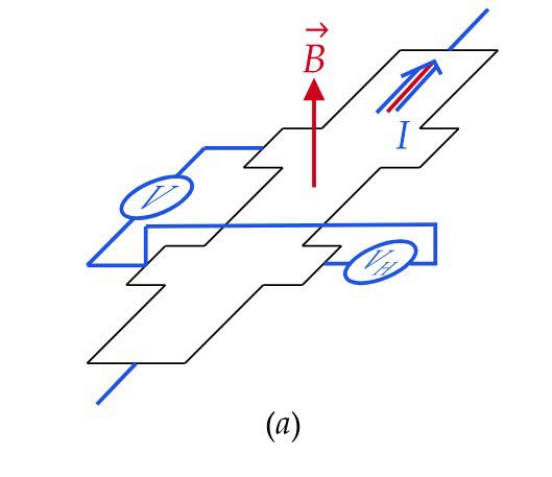}}
  
  \hspace*{4pt}
  {\includegraphics[width= 7.0cm]{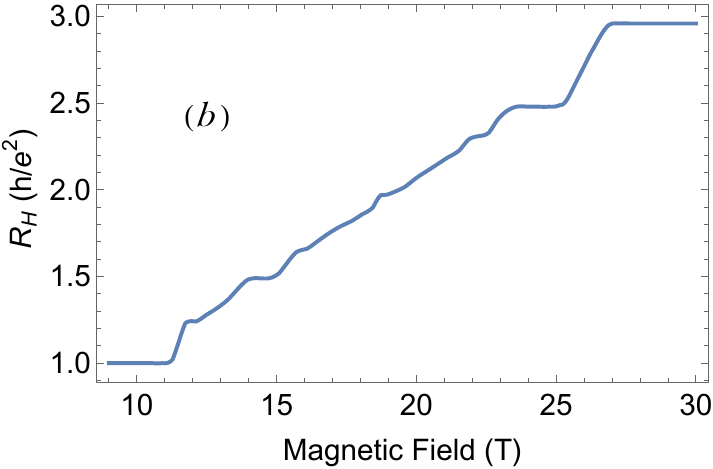}}
  
  \caption{(a) Schematic representation of a Hall bar used in QHE experiments. (b) Hall resistivity $R_H$ in the FQHE between the plateaux at $\nu=1$ and $\nu=1/3$, from \cite{stormer1999nobel}.}
\end{center}
\end{figure}

In fact, even if in the random potential generated by the impurities a priori the electron can hop from one potential well created by one impurity to another, propagating therefore as in a lattice, because of the disorder the energy levels of nearby wells have a random distribution that causes an interference and levels with similar energy in value are statistically very far apart so that the motion is exponentially suppressed. The corresponding one-electron states are called localized.

So if the Fermi energy of our electron systems is in the above mobility gap, then there is no contribution to the longitudinal current even in the presence of an electric field. The current does not change when the Fermi energy moves in the region of localized states, and this
explains the plateaux. 

In the corresponding parameter region then one can use safely the equation (\ref{j}) to analyze the physics. Note that the origin of the vanishing of longitudinal conductivity is irrelevant to this analysis. The equation (\ref{j}) can also be used in cases where the simple framework outlined above involving the one-particle electron states, which apply to the IQHE, cannot be applied.  In particular equation (\ref{j}) can be applied to the FQHE \cite{frohlich2024fractional} where the Coulomb interaction is crucial.
Taking the divergence of (\ref{j}) and using current conservation and Faraday's law we get
\begin{equation}
\label{CSe}
\frac{\partial \rho}{\partial t}= -\partial^\mu j_\mu = -\sigma_H \partial^\mu \epsilon _{\mu \nu}E^\nu = \sigma_H \frac{\partial B}{\partial t}.
\end{equation}
Defining the charge $Q = \int \rho(\vec x) d^2x$ and integrating  on the sample and over time both sides of (\ref{CSe}) we get
\begin{equation}
\label{Delta}
    \Delta Q = \sigma_H \Delta \Phi,
\end{equation}
where $\Phi$ is the magnetic flux and $\Delta$ refers to change over time.

The equation (\ref{Delta}) is similar to that found in the model discussed above with anyon vortices, (\ref{Qphiv}), identifying $\sigma_H$ with $\frac{2 \theta}{2 \pi}$.

Using (\ref{Delta}) a simple argument due to
Laughlin \cite{laughlin1983anomalous},\cite{laughlin1981quantized}, outlined below, suggests that at the filling $\nu = 1/p, p$ odd, the charge carriers have a charge
$e/p$, where $e$ is the charge of the electron.

Consider an annulus (Corbino disk) and insert at its center a magnetic flux adiabatically increased by one quantum flux $h/e$ [see Fig.11 (a)].
The microscopic quantum mechanical hamiltonians for the $2d$ system on the surface with the initial and final flux are gauge equivalent, so they have the same spectrum. If
the Fermi energy of the system is in a mobility gap, then the adiabatic process described above does
not alter the total energy of the system and its only effect should be a charge transfer between the two edges. 
\begin{figure}[h]
\centering
\includegraphics[width=0.7\textwidth]{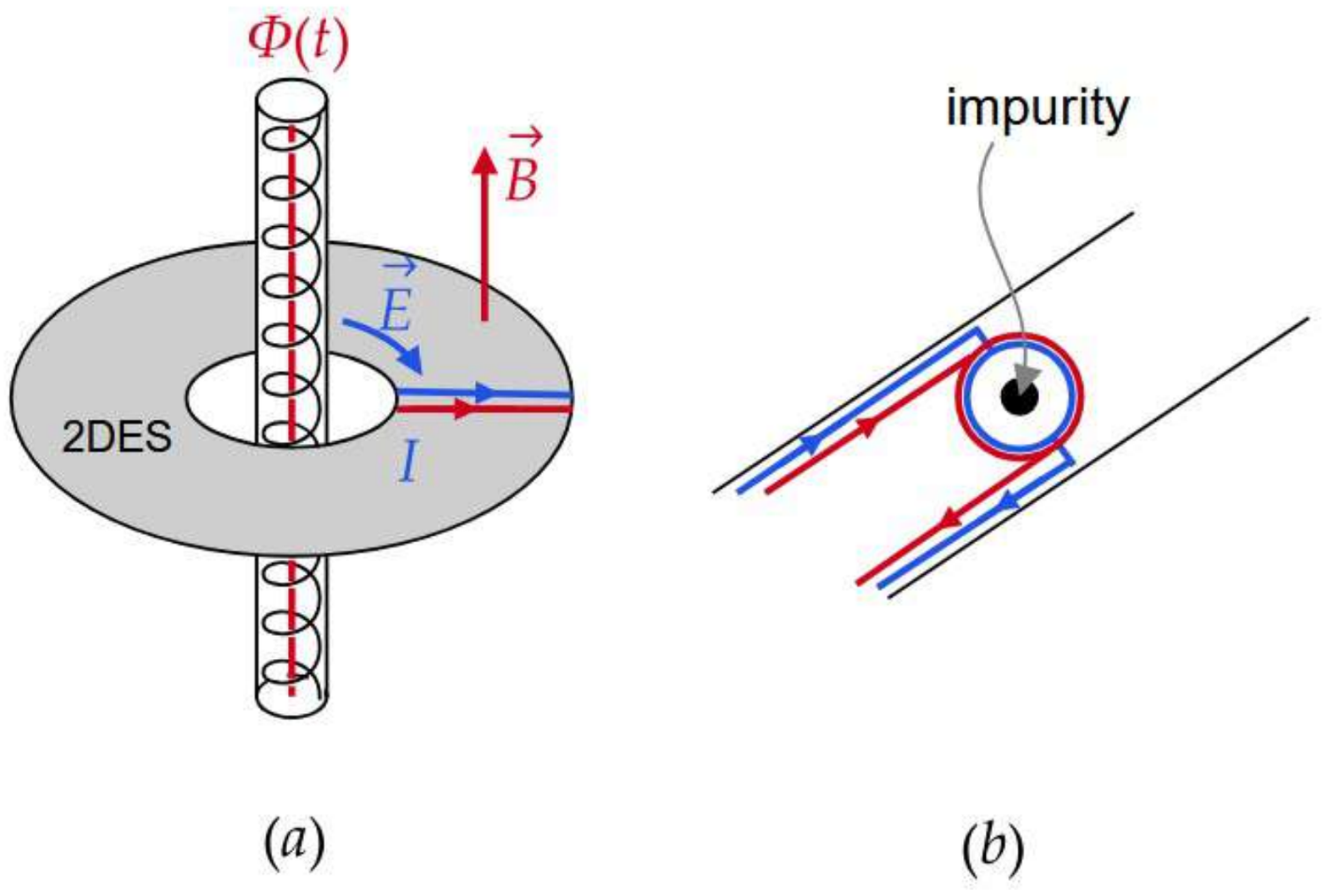}
\caption{(a) The Corbino disk for the Laughlin argument in \cite{laughlin1983anomalous},\cite{laughlin1981quantized}.(b) Schematic representation of the geometry of the experiment in ref. \cite{tsui19922}}
\end{figure}

Suppose that, as a result of the electric field of Faraday's law combined with Hall's law (\ref{j}), during such process $m$
charge carriers of possibly fractional charge charge $f e$,  move from one edge to the other of the surface of the annulus. Then from (\ref{Delta}) one immediately gets $\sigma_H = m f e^2 /h$.

For example, in the plateau at $\nu = 1/3$ where $\sigma_H = 1/3 e^2 /h$ this suggests as the simplest possibility  that the charge carriers have charge $e/3$.
Laughlin's theory \cite{laughlin1983anomalous} in fact predicts that at $\nu = 1/p$, with odd $p$, the ground state of the $2d$ electron system considered is
an \textit{incompressible quantum liquid} now called \textit{Hall fluid} and slight deviations from a stable state at $\nu = 1/p$ state
create elementary charged excitations with fractional charge $e/p$ and elementary magnetic flux $h/e$, called Laughlin vortices, in agreement with (\ref{Delta}). 

In the presence of impurities and at low concentrations, these
quasi-particles are expected to be trapped in localized states such as those described for electron gas. As discussed above the simultaneous presence of charge and magnetic flux suggests that they are anyons, as first observed in \cite{halperin1984statistics},\cite{arovas1984fractional}, which are somewhat analogous to the vortices of the model (\ref{HiggsCS}) with $\theta = \frac{\nu}{2}$.

\subsection{FQHE and edge currents}

Assuming incompressibility, as predicted in Laughlin's theory, there is another reason for the anyon nature of Laughlin vortices, which is also key to understanding the experimental verification firstly of the fractional electric charge and later of the braid statistics of such quasi-particle excitations.

We denote by $S_{\mathit{eff}}(a)$ the Euclidean effective  action  for a gauge field $a_\mu$, perturbation of the classical electromagnetic potential considered above, obtained from the full action of the Hall system, $S(a, \psi)$ integrating the other varying fields, generically indicated by $\psi$, i.e.
\begin{equation}
\label{seff}
e^{-S_{\mathit{eff}}(a)}=\int {\mathcal D} \psi e^{-S(a, \psi)}    
\end{equation}

Since the Hall fluid is incompressible the connected n-point functions of the current $j_\mu$ coupled to $a_\mu$ must have a fast decay in space so that in the scaling limit they must converge to local distributions \cite{frohlich1993gauge}. 

From the general properties of distributions such distributions can only be derivatives of the Dirac delta function of some order. On the other hand the connected correlation functions of the current evaluated for $a_\mu =0$ at non-coinciding points are also the coefficients of a Taylor expansion of the effective action of $a_\mu$ in powers of $a$. 

In fact from (\ref{seff}) it follows that the connected correlation functions of the current $j_\mu (x)= \frac{\delta S(a, \psi)}{\delta a_\mu(x)}$ in the presence of $a_\mu$ at non-coinciding points are given by 
\begin{equation}
 \langle j_{\mu_1}(x_1)...j_{\mu_n}(x_n) \rangle_c = (-1)^{n+1} \prod_{i=1}^n \frac{\delta }{\delta a_{\mu_i}(x_i)}S_{\mathit{eff}}(a).  
\end{equation}

 From the decay of the connected correlators of the current, it is deduced that the effective action of an incompressible fluid is local in $a$.
Because the gauge
field has the same scaling dimension of the momentum, i.e. of the spatial derivative,  at large spatial scales the leading contribution to effective action comes from the lowest non-zero term of the Taylor expansion.

One can use the gauge invariance
\[
S_{\mathit{eff}}(a_\mu + \partial_\mu \lambda) = S_{\mathit{eff}}(a_\mu),
\]
with $\lambda(x)$ an arbitrary function, to determine the leading term of $S_{\mathit{eff}}(a)$.

Usually the leading term is Maxwell's action, of scaling dimension 4, but in 2 space dimensions if parity is broken, as in the case of the Hall effect by the magnetic field, the leading term is the Chern-Simons term of scaling dimension 3(!) i.e., for the Euclidean lagrangian, the term $\frac{i k}{4 \pi} \epsilon_{\mu\nu\rho}a^\mu \partial^\nu a^\rho$ for some real $k$.

Adding the interaction term with the current $j_\mu$, i.e the lagrangian term $-i j_\mu a^\mu$, and deriving the equation of motion, we see immediately that to satisfy the Hall law (\ref{j}) the coefficient of the Chern-Simons action must be $\sigma_H/2$ i.e. $k= 2 \pi \sigma_H$.
We can say more with a gimmick: since this is a gauge theory, by Gauss law the current is conserved without using the equation of motion for the matter fields (see e.g. \cite{strocchi1981gauss}). Thus, from $\partial^\mu j_\mu = 0$ for the Poincarè lemma there is a gauge field $b^\mu$ such that $j_\mu = \frac{1}{2 \pi} \epsilon_{\mu \nu \rho} \partial^\nu b^\rho$. In terms of $b^\mu$ it is easy to write the leading term of the action in the scaling limit which produces by integration over $b^\mu$ in the Euclidean path-integral the leading term of the effective action $S_{eff}(a_\mu)$: 
\begin{equation}
\label{ba}
-\frac{i}{4 \pi \nu} \int \epsilon^{\mu\alpha\beta}b_\mu \partial_\alpha b_\beta +\frac{i}{2 \pi} \int\epsilon^{\mu\alpha\beta}b_\mu \partial_\alpha a_\beta +(...)
\end{equation}
where $\nu$ is the filling and (...) refers to irrelevant terms and boundary terms. This action can be used to derive the possible values of $\sigma_H$ if the electron system is completely polarized so that the 3-dimensional spin coincides with the 2-dimensional one. The  gauge theory described by (\ref{ba}) must have
excitations (static and point-like in the scaling limit) of charge $\pm 1$  (in units where $e = 1$), corresponding to a hole or electron. These excitations must be fermions and this implies that
\[
\nu= \frac{1}{2m +1}, \quad m \in \N.
\]
In fact, from the equation of motion for $b$ derived from (\ref{ba}) we conclude that their charge is given by $q= 1 = \nu n$, where $n$ is the number of elementary magnetic fluxes they carry. An exchange of two electrons gives a factor $-1$, but computed with the Aharonov-Bohm effect this is given by $\exp (i q \Phi/2) = \exp (i \pi  \nu n^2) = \exp (i \pi n)$, hence $n$ must be odd.

But...the Chern-Simons action for $a$ is gauge invariant when considered in $\R^3$ with fields vanishing at infinity; however, if we consider it in a region $\R \times \Lambda \subset \R^3$, where $\R$ is in the direction of time and $\Lambda$ is a finite region of space, as in a real sample, it under a gauge transformation changes by an anomalous boundary term
\begin{equation}
\label{anombd}
 \frac{i \sigma_H }{2}  \int_{\R \times \partial \Lambda} d \Sigma_\mu \epsilon^{\mu \nu\rho} \lambda(x) \partial_\nu a_\rho. 
\end{equation}

As we said, gauge invariance is crucial for physical behavior, therefore there must be a boundary term that changes under gauge transformation in the opposite way. We know that gauge-invariance implies conservation of the current corresponding to global transformations, so to find the structure of the boundary term consider once again the Hall law (\ref{j}) combined with a local version of (\ref{Delta}), with charge density and magnetic field corresponding to the perturbation of the ground state. We obtain  the covariant equation for the bulk current $j_\mu^b$:
\begin{equation}
\label{sigmaHx}
    j_\mu^b(x)= \sigma_H(x) \epsilon_{\mu\nu\rho} \partial^\nu a^\rho,
\end{equation}
with $\sigma_H(x)$ vanishing outside of the sample. 

We choose a local coordinate system in space with index 1 labelling the direction locally along the edge of the sample and 2 that orthogonal. Then taking the divergence of (\ref{sigmaHx}) we get 
\begin{equation}
 \partial^\mu j_\mu^b(x) = (\partial_2 \sigma_H(x))   \epsilon_{2 \nu \rho}  \partial^\nu a^\rho. 
\end{equation}
It follows that there is a current at the edge of the sample, $j_\mu^e$ which satisfies on the edge:
\begin{equation}
\label{anom}
\partial^\mu j_\mu^e(x) = -\sigma_H (\partial^0 a^1-\partial^1 a^0)=-\sigma_H E^1,
\end{equation}
where $E^1$ is the electric field along the edge. We recognize (\ref{anom}) as the equation of the chiral anomaly (see e.g. \cite{strocchi2013introduction}) with a modified coefficient $\sigma_H$.

How can we physically interpret this boundary contribution?

The fact that the behavior at the edge of a quantum Hall system is different from that in the bulk can already be suggested by a
 classical reasoning. We know that charged particles in a plane in the presence of a perpendicular magnetic field travel through cyclotron orbits with a definite orientation.
At the edge of the sample, the orbits are reflected,  the particles bounce back with
 a skipping motion along the boundary with the orientation induced by the motion in the bulk.
 
 Suppose now that the quantum analogue of (at least some of) cyclotron orbits are pinned by disorder, then under a potential difference the corresponding charged particles cannot produce current, but the motion on the boundary always can.
 This is, in a semiclassical version, the origin of the edge currents. 
 
 As firstly argued in \cite{halperin1982quantized} a more quantum-mechanical picture of the phenomenon is as follows: we assume  first that these charged particles are fermions described by a Fermi liquid. The boundary of the sample can be represented by a confining potential which raises outwards near the edge  so that states corresponding to the Fermi energy are then located at the boundary. Therefore there are empty states at the energy immediately above the Fermi energy, so that the states close to the Fermi energy are gapless and can move under the influence of magnetic field and potential difference in a fixed direction, that is they are chiral. 
 
 Since the Pauli principle also applies to anyons in magnetic field, although in a modified form leading to what is called Haldane liquid in ref.\cite{iguchi1998generalization}, there is an analogue of Fermi energy also for them and the same reasoning can be applied \cite{wen1990chiral},\cite{frohlich1993gauge}.

 \textit{Remark: Haldane's exclusion statistics}
 The Haldane exclusion statistics generalizes the Pauli exclusion principle for fermions \cite{haldane1991fractional}. 
 A priori it is not related to the exchange statistics, permutation or braid, that is to say the statistics discussed so far.
 For quasi-particles in finite-density quantum systems of identical particles, it is identified by a parameter $g \in [0, 1]$ which measures the effective interaction between quasi-particles occupying the same state in the one-particle Hilbert space. A way of characterizing this statistics at $T = 0$ is the following \cite{wu1994statistical}: Let us consider the Fermi surface of the Fermi gas with density $n_0$. A quasi-particle obeys the exclusion statistics with parameter $g$ if the density $n_g$ of the quasi-particles having the volume enclosed by that Fermi surface, satisfies $n_0/n_g = (1 - g)$. It follows that $g = 0$ for standard fermions, and other intermediate values of $g \in (0,1)$ define a fractional exclusion statistics. In 2D in general there is no relation between braid and exclusion statistics, for example free nonrelativistic anyons do not show evidence of fractional exclusion statistics. However, in \cite{ye2015hall} it is proved that for an incompressible anyonic fluid with braid statistics parameter $\theta$ obtained from a fermionic system with Hall conductivity $\sigma_H$ by adding the appropriate Chern-Simons interaction, the relation $g = 2 \pi \sigma_H (2 \theta -1)$mod 1 is satisfied. In the case of anyons in the first Landau level, $ 2 \pi \sigma_H=1$ and therefore $g = 2 \theta$, as already demonstrated for these systems in \cite{ma1991quantum}, \cite{de1994equation}.
 
 The Chern-Simons term has been derived by assuming a gap for charged excitations, so it does not apply to the boundary states, because they are gapless and chiral, according to the above picture, as also shown by the chiral anomaly (\ref{anom}).
 We denote by $V_\nu$ the propagation speed of chiral modes of the plateaux labelled by $\nu$, which give rise to the edge current and we label with the indices $+/-$ the left-/right-handed modes. For $\nu = 1$ we know that they are chiral fermions and therefore their effective action is given by  the so-called JR  action \cite{jackiw1985vector}. To write it one can use the coordinates "light-cone" $x_{\pm} = (x^0 \pm x^1)/\sqrt{2} $ on the (1+1)-dimensional
boundary space-time, $\Sigma \simeq \R \times \partial \Lambda$, where $x^1$ is the spatial coordinate along $ \partial \Lambda$ and $x^0= V_\nu t$. The corresponding components of the gauge field are $a_{\pm}= (a_0 \pm a_1)//\sqrt{2}$. In terms of these quantities the effective action for the chiral fermion is given by 
\begin{equation}
\Gamma[a]= \frac{1}{4 \pi} \int_\Sigma\Big[ a_+a_- - a_\pm \frac{\partial_\mp^2}{\partial_+\partial_-} a_\pm\Big] dx_+ dx_-.
\end{equation}
Rotating to the Euclidean space-time only yields a multiplicative $i$. A gauge transformation with parameter $\lambda(x)$ generates the change of the Euclidean effective action given by
\begin{equation}
 \pm  \frac{i}{4 \pi} \int_\Sigma\Big[ a_+ \partial_-\lambda-a_- \partial_+\lambda \Big] dx_+ dx_-= \mp \frac{i}{4 \pi} \int_\Sigma\Big[ \epsilon^{\mu\nu} \lambda \partial_\mu a_\nu\Big] d^2x.
\end{equation}
Thus, the contribution of the right-handed edge current cancels out the anomalous contribution of the bulk (\ref{anombd}) for $\nu=1$, i.e $\sigma_H=1/(2 \pi)$.
For the case of fractional charges it is clear that by multiplying the boundary effective action $\Gamma$ by the appropriate $\sigma_H$ the corresponding bulk contribution (\ref{anombd}) is cancelled.
So in the FQHE necessarily at the boundary of the sample run chiral modes and the above derivation suggests that they obey the same statistics of Laughlin vortices in the bulk. 

\subsection{Fractional quantum spin Hall effect} 

There is also a time-reversal symmetric version of the FQHE, called the Fractional Quantum Spin Hall Effect  (FQSHE). It appears if electrons with spin along one direction are in a FQH state of fractional filling $\nu$ and electrons with spin in the opposite direction are in a FQH state of filling $-\nu$, as first suggested by a general analysis of Hall fluids in \cite{frohlich1992u}.
For a system of this type in a sample two sets of modes which carry spin along opposite directions run on the edge of the sample propagating in opposite directions, so that the charge  current vanishes and only remains a spin current; in addition these two sets are transformed into each other by time reversal. It is the fractional version of the (integer) Quantum Spin Hall Effect. This was first suggested in a specific condensed matter model in \cite{kane2005quantum}, then predicted to occur in 
HgTe quantum wells \cite{bernevig2006quantum} and subsequently observed experimentally there\cite{konig2007quantum}.
Assume that the system is in the plane $x y$, denoting as before with $a_\mu$ the gauge field coupled to the 3-current of charge and with $w_\mu$ the analogous field coupled to the 3-current of spin. The time component of  $w_\mu$ is derived from the Zeeman term and thus proportional to the magnetic field perpendicular to the plane, the space components are derived from the spin-orbit coupling and are therefore proportional to the electric field assumed in the plane, so only the $z$ component of the spin enters the Hamiltonian. Choosing appropriately the spatial dependence of the electric field the spin current can be made proportional to the charge current times the Pauli matrix $\sigma_z$. Thus imitating the argument presented for the FQHE in the previous section the leading term in the scaling limit of the bulk Euclidean effective action produced by the electrons with the two spin directions is given by (see e.g. \cite{bernevig2006quantum1}, \cite{frohlich2024fractional}):
\begin{align}
\label{BF}
 &\frac{i \nu}{4 \pi} \int \epsilon_{\mu\nu\rho}(a^\mu + w^\mu) \partial^\nu (a^\rho + w^\rho)-\frac{i \nu}{4 \pi} \int \epsilon_{\mu\nu\rho}(a^\mu - w^\mu) \partial^\nu (a^\rho -w^\rho) = \nonumber\\  
& \frac{i 2 \nu}{4 \pi} \int \epsilon_{\mu\nu\rho}(a^\mu ) \partial^\nu (w^\rho) + \frac{i 2 \nu}{4 \pi} \int \epsilon_{\mu\nu\rho}(w^\mu ) \partial^\nu (a^\rho),
\end{align}
 
where the second expression is in the so-called BF form (see e.g.\cite{blau1991topological} and for applications in condensed matter \cite{hansson2004superconductors}).
Because from their definition it follows that under time reversal the fields are transformed as $a^0 \rightarrow a^0, a^i \rightarrow -a^i, w^0 \rightarrow -w^0, w^i \rightarrow w^i, i=1,2$,
one can easily see from the BF form that the action (\ref{BF}) is time-reversal invariant.  One can also easily derive the form of the bulk action in the scaling limit: Introducing two gauge  fields for contribution of the electrons with up and down spin , $b^\mu_\uparrow, b^\mu_\downarrow$, it is given by
\begin{align}
\label{BF1}
 &-\frac{i}{4 \pi \nu} \int \epsilon_{\mu\nu\rho} b^\mu_\uparrow\partial^\nu b^\rho_\uparrow+\frac{i}{2 \pi} \int \epsilon_{\mu\nu\rho} b^\mu_\uparrow  \partial^\nu (a^\rho + w^\rho) + \frac{i}{4 \pi \nu} \int \epsilon_{\mu\nu\rho}b^\mu_\downarrow \partial^\nu b^\rho_\downarrow  \nonumber\\ 
& + \frac{i }{2 \pi} \int \epsilon_{\mu\nu\rho}b^\mu_\downarrow \partial^\nu (a^\rho-w^\rho)= 
-\frac{i}{2 \pi \nu} \int (\epsilon_{\mu\nu\rho} b^\mu_c\partial^\nu b^\rho_s+\epsilon_{\mu\nu\rho} b^\mu_s  \partial^\nu b^\rho_c)  + \nonumber\\ 
&\frac{i}{\pi} \int \epsilon_{\mu\nu\rho}b^\mu_c\partial^\nu a^\rho + \frac{i }{\pi} \int \epsilon_{\mu\nu\rho}b^\mu_s \partial^\nu w^\rho,
\end{align}
where $b^\mu_\uparrow = b^\mu_c  + b^\mu_s,b^\mu_\downarrow = b^\mu_c - b^\mu_s $.
 For a more complete discussion and for generalizations see e.g. the previously cited reviews on fractional topological insulators.

\subsection{Experiments}

The experimental verification of the fractional charge of the charge carriers of the FQHE at filling $\nu = 1/3$ is based on measuring the resistivity fluctuations of a narrow sample where the dissipationless current flow at the edges is broken by a resonant tunneling from one edge to the other caused by an impurity in the bulk (backscattering), see Fig.11 (b).

The quantum mechanical trajectories  of the charge carrier of the edge current can go around the impurity an arbitrary number $n$ of times, thus producing a longitudinal resistance whose modulus is given by some $r^n$. Since the charge carriers in their path  enclose a region with magnetic flux, $\Phi = B A$, where $A$ is the enclosed area, and carry an unknown charge $e^*$ they also acquire through the Aharonov-Bohm effect a phase $e^{i \Phi  e^* n}$.

Adding all the contributions the longitudinal resistivity is proportional to
\begin{equation}
\abs{ \frac{1}{1-r e^{i \Phi e^*}}}^2= \frac{1}{1+ r^2 -2 r \cos( \Phi e^*)},
\end{equation}
having therefore its maxima at a distance corresponding to $ \Delta \Phi e^* = 2 \pi $.
\begin{figure}[!htb]
\begin{center}
  {\includegraphics[width= 10.0cm]{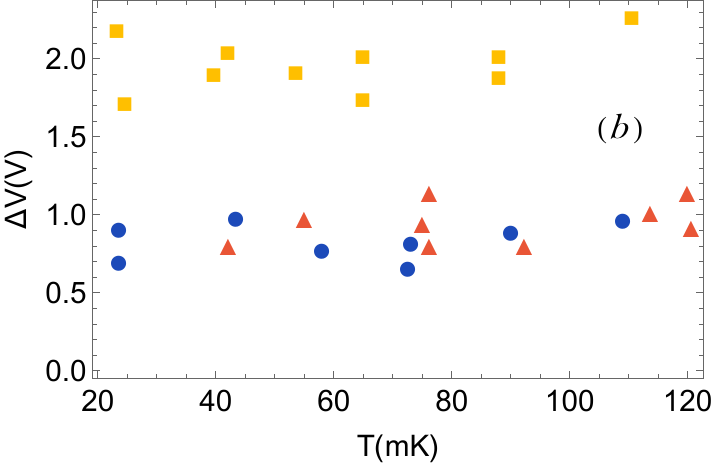}}
  
  \hspace*{4pt}
  {\includegraphics[width= 10.0cm]{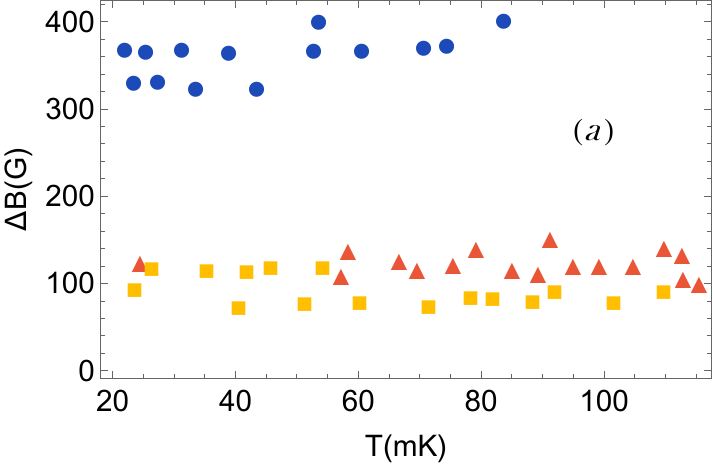}}
  
  \caption{Experimental data from \cite{tsui19922}, with blue circles for $\nu=1/3$, red triangles for $\nu=1$, yellow squares for $\nu=2$. }
\end{center}
\end{figure}
In the experiment it was also possible to modify the applied potential difference applied $V$, with a consequent modification of the charge carrier density, itself proportional to $\nu B$.
Therefore the resistivity maxima are for $\Delta B = \frac{2 \pi}{e^* A}$ and $\Delta V = \frac{2 \pi \nu}{e^* A}$.

The result of the experiment \cite{tsui19922} were:
$\Delta B (\nu =1/3) \simeq 3 \Delta B (\nu =1) \simeq 3 \Delta B (\nu =2)$ and $\Delta V (\nu =1/3) \simeq  \Delta V (\nu =1) \simeq \frac{1}{2} \Delta V(\nu =2)$ [see Fig.12] thus
verifying that at $\nu =1/3$ the charge carriers have electric charge $e^* = e/3$.

The experimental verification of their braid statistics theoretically predicted with $\theta = 1/6$ required a more sophisticated apparatus:
a Fabry-Pérot interferometer \cite{halperin2011theory}.

It consists of a two-dimensional electron system in the FQHE regime with 
two quantum point contacts (QPC, narrow constrictions between two wide electrically conducting regions, of a width comparable to the electronic wavelength) at its edge, used for current scattering. The current is carried by the quasi-particles described above which travel in chiral states. In particular, when the edge current reaches
a QPC there are two possibilities: that the charge carriers cross by tunnel effect the QPC (backscattering) or that they are transmitted. The deflected currents from the
two QPCs may interfere with each other: this effect is observed by measuring the total backscattering current.

As a result of the disorder, some number $N_l$ of quasi-particles are
localized in the bulk of the interferometer, 
then the quasi-particles travelling in chiral states, if deflected by QPCs, can go around 
those localized [see Fig.13 (a)].

\begin{figure}[!htb]
\begin{center}
  {\includegraphics[width=9.0cm]{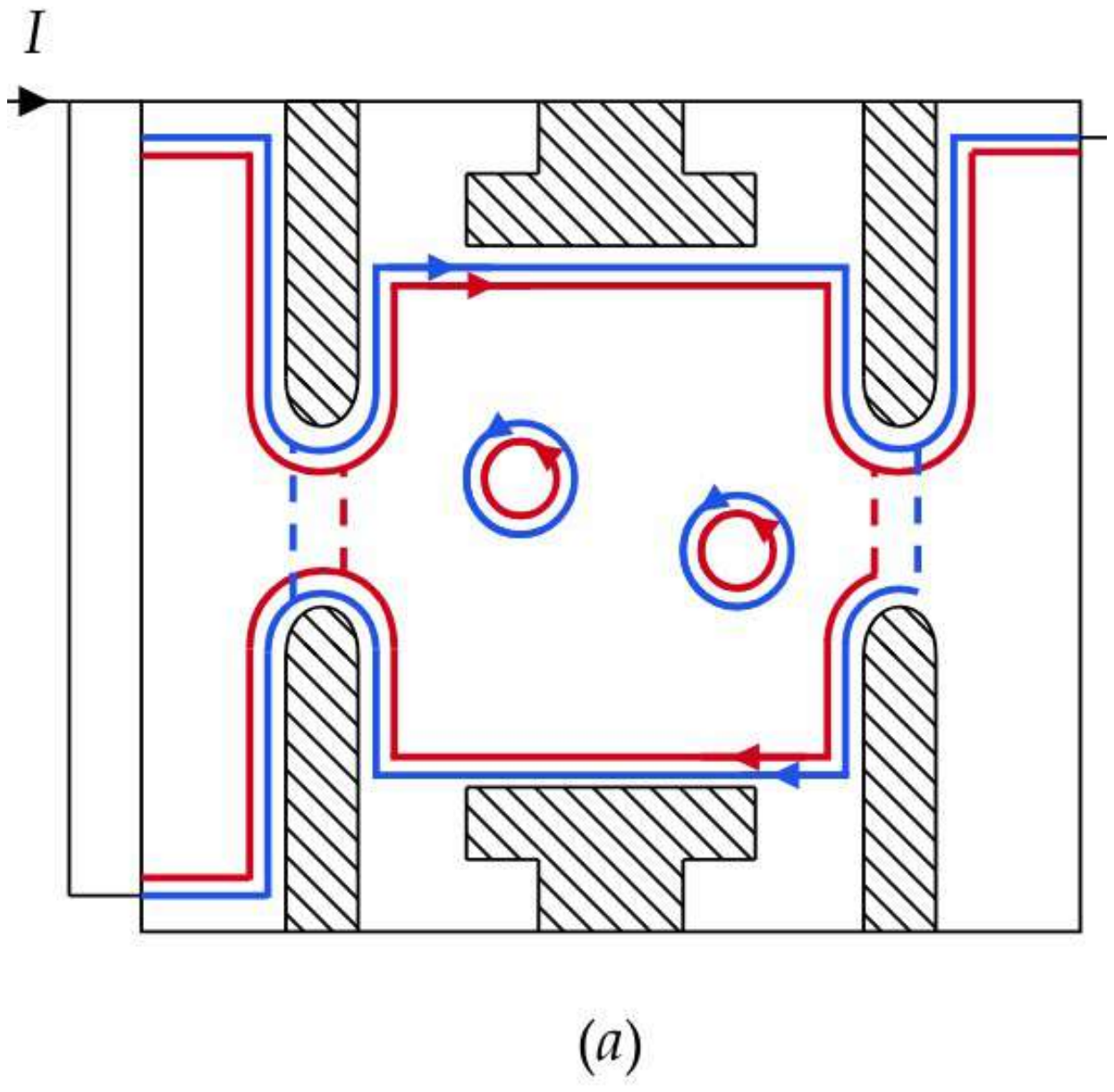}}
  
  \hspace*{4pt}
  {\includegraphics[width= 8.0cm]{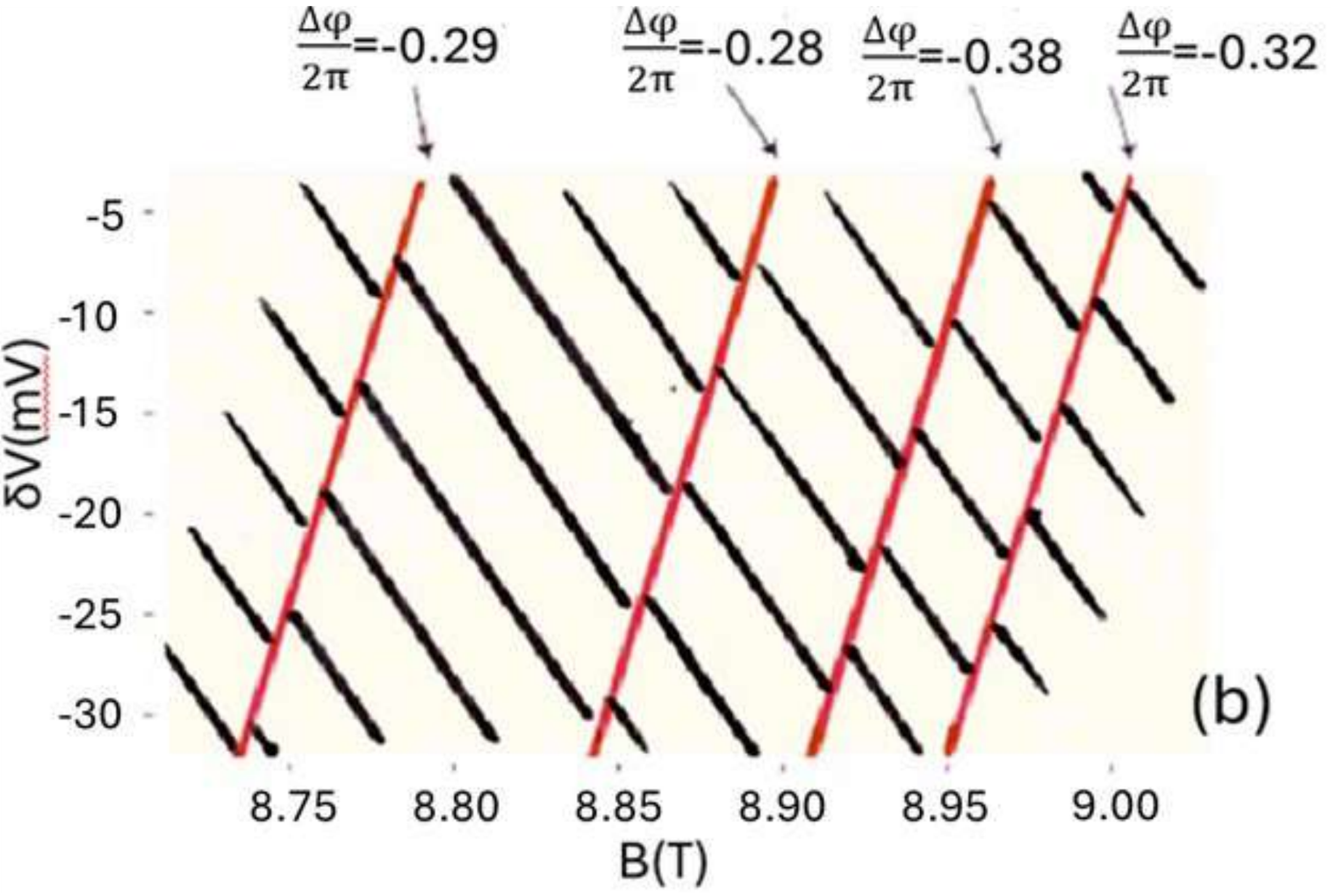}}
  
  \caption{(a) Schematic representation of  the Fabry-Pérot interferometer used in \cite{nakamura2020direct}; the lines at the edge of the darkened regions describe edge currents, the dashed lines describe the quantum tunneling, the lines in the bulk describe the localized anyons. (b) Schematic representation of the experimental results of
  \cite{nakamura2020direct}. The black lines describes least-squares fits of conductivity;  the average
value of the phase jump obtained from the difference between 
the fitted values of $\alpha$
 in adjacent regions along the red lines is shown in the top line of the figure. }
\end{center}
\end{figure}

In this case, the phase difference that occurs between the two paths, $\varphi$, is given by two terms:
\begin{equation}
    \varphi = 2 \pi (e^* B A + N_l 2 \theta),
\end{equation}
the first term is due to the Aharonov-Bohm effect due to impurities as in the previous experiment and the second is due to  braid statistics, where the factor 2 corresponds to the fact that a $2 \pi$ rotation is equivalent to two exchanges with the same orientation.

Variations of  conductivity  are proportional as before to variations of $\cos \varphi$ but the area of interference $A$ depends on the voltage and increases with it, because the distance between the support of the edge modes increases, then $d A/ d V \equiv c>0$.

Suppose first that $N_l$ is constant, then if small variations of $V$ and $B$ give a constant conductivity, that is a $\varphi$ constant, then 
$\frac{dA}{dV} d V + A d B = 0$ i.e $\frac{dV}{dB} = - A/c <0 $, so the lines with constant conductivity are straight with constant negative slope. 

However if changing $V$ or $B$ changes the number of localized charge carriers, we have a jump, shown by the experiment,  with a predicted $\frac{\Delta \varphi}{2 \pi} = 2 \theta$.
Such jumps [see Fig.13 (b)], occur through positive slope lines in the $B-V$ plane, according 
to the theoretical expectations. In fact, by increasing the magnetic field intensity, localized quasi-particles are removed (or quasi-hole created) because the filling $\nu \sim n/B$ is lowered. Increasing $V$, on the contrary, increases the quasi-particle
number, then the filling and for higher $V$, the intensity of the magnetic field at which it is possible to remove a localized quasi-particle must increase. 

To determine the value of phase change associated with each discrete jump, a
fit of experimental conductivity data was made using the formula:
\[
\delta \sigma = \delta \sigma_0 \cos (2 \pi (e^* B A + \alpha))
\]
with $\alpha$ fit parameter corresponding to $\frac{\Delta \varphi}{2 \pi}$. The value of the discrete jumps was determined by calculating the difference of the
parameters $\alpha$ found through the fit in adjacent regions. Since an increase of the magnetic field is expected to reduce the number of 
localized quasiparticles, the change in phase across each jump is 
predicted to be $\-2 \theta$. From an average of these values and
assuming that each discrete phase jump corresponds to the removal of a localized quasi-particle, it was found \cite{nakamura2020direct} the value $\alpha = -0.31 \pm 0.04$ in perfect agreement with the theoretical prediction for the statistics parameter $\theta=1/6$ for the plateau at $\nu =1/3$.

\section{General properties: A look at the Algebraic QFT of anyons}

In section 5 we discussed the QFT of anyons in specific models, a natural question is: how general are the results obtained such as the spin-statistics connection, the spin addition rule, the breaking of parity, the localizability in space-like cones...and how the treatment generalizes to the case of matrices instead of phases $e^{\pm i 2 \pi \theta}$ associated with oriented exchanges?

 To answer these questions, it is clear that we need a framework based on well- founded physical principles, rigorously implemented mathematically. Probably the best framework with such characteristics is the Algebraic QFT approach (see e.g. \cite{haag2012local}). The degree of mathematical sophistication of this approach is generally greater than that used so far, therefore we give only some basic ideas and some results of it applied to 2+1 dimensional relativistic quantum field theory (RQFT), just enough to answer the above questions in this formalism describing systems with infinite degrees of freedom satisfying Einstein's causality and with mass gap in 2+1 dimensions.  For further details of the approach we refer the interested reader to \cite{frohlich1992non}, \cite{frohlich1990braid},\cite{frohlich1990braid1}. 
 We proceed in  the spirit described in the Introduction regarding relativistic considerations. Indeed, although condensed matter systems in which anyons may appear in Nature do not fit into this framework, some of their qualitative characteristics at zero temperature 
and at large scales are expected to be reproduced by a local quantum theory, although not a Poincarè covariant, described by a variant of the above framework.
For some ideas along this line of thought see for the formalism e.g. \cite{dubin1970time},\cite{strocchi1988long} and references therein, and for a more specific application to anyons, based on the replacement of space-like cones with wedges contained in a time slice, see \cite{frohlich1991spin}.

\subsection{Basic structures of Algebraic RQFT}
  
Consider a local relativistic quantum field theory in the 3-dimensional Minkowsky space-time $\R^{1,2}$ (at 0 temperature and density and without zero-mass particles). Local observables of the theory are obtained by smearing with suitable test functions of compact support its non-charged ( i.e. invariant under gauge transformations producing superselection sectors) distributional field operators, e.g. in the models considered above: currents, (suitable combinations of) Wilson loops...to obtain, in general unbounded, self-adjoint operators, let's say $O$.

Given a bounded open region of space-time $\mathcal{O}$ (typically chosen as a double cone obtained intersecting a forward and a 
backward light cone) we say that such observables are localized in $\mathcal{O}$ if the support  (e.g. points, loops,...) of the distributional field  operators on which the test functions average them  are all
contained in $\mathcal{O}$. Since observables are self-adjoint one can construct bounded functions of them, let's say $A$. We denote by $\mathcal{A(O)}$ the von Neumann  (weakly-closed *) algebra generated by all the bounded functions $A$ of all the observables localized in $\mathcal{O}$, and with an abuse of language we call its elements still local observables. Of course it is assumed that if $\mathcal{O}_1 \subseteq \mathcal{O}_2$ then  $\mathcal{A(O}_1) \subseteq \mathcal{A(O}_2)$.

The algebra of all (quasi)-local observables, $\mathcal{A}$ is defined as the closure in norm (denoted by upper index $n$)
\[
\mathcal{A} \equiv \overline{\bigvee_{\mathcal{O} \subset \R^{1,2}} \mathcal{A(O)}}^n.
\]

The key idea of algebraic RQFT is that these algebras of observables in principle contain all the physical informations about the RQFT.

In this approach the basic postulates in a theory without massless particles are:

1) \textit{Einstein causality}: if $\mathcal{O}_1$ is spacelike separated from $\mathcal{O}_2$, and we write $\mathcal{O}_1 \bigtimes \mathcal{O}_2$, then 
\[
[A,B] = 0, \forall A \in \mathcal{A(O}_1), B \in \mathcal{A(O}_2),
\]
expressing the fact that measurements of two observables in causally disjoint domains cannot affect each other.

2) \textit{Poincaré covariance}: $\mathcal{A}$ is the representation space of a representation of the restricted
Poincaré group $\mathcal{P}^\uparrow_+(1,2)$ as a group of ${}^*$ automorphisms $\alpha_{a,\Lambda}, (a,\Lambda) \in \mathcal{P}^\uparrow_+(1,2)$ preserving the local structure of observables, i.e.
 $\alpha_{a,\Lambda}(\mathcal{A(O)})=\mathcal{A(O}_{a,\Lambda})$, where $\mathcal{O}_{a,\Lambda} = \{ x \in \R^{1,2} | \Lambda^{-1}(x-a) \in \mathcal{O}\}$ 

 3) \textit{Vacuum}: recalling that a state of an operator algebra is a positive linear functional on it of norm 1, it is assumed that there is a state called the vacuum state, $\omega$, on $\mathcal{A}$ invariant under the previously mentioned representation of the Poincarè group. It is also assumed that this state is the unique invariant state.
 
 Given $\mathcal{A}$ and $\omega$ the GNS construction (see e.g. \cite{haag2012local}) provides a Hilbert space $\mathcal{H}_1$
 where $\mathcal{A}$ can be represented as an algebra of bounded operators with a cyclic vector, the vacuum vector $\ket{\Omega}$. We identify $\mathcal{A}$ with its representation in $\mathcal{H}_1$. For each region $\mathcal{O} \subset \R^{1,2}$ we denote by $\mathcal{O}^\prime$ its space-like complement.   Motivated by the Buchholz-Fredenhagen theorem we consider space-like cones, $\mathcal{C}$, causal completion of space-like wedges and the associated von Neumann algebra
 \[
 \mathcal{A(C)} \equiv \overline{\bigvee_{\mathcal{O} \subset \mathcal{C} } \mathcal{A(O)}}^w ,
 \]
where the upper index means that the closure is in the weak topology.

 4) \textit{Duality} : $\mathcal{A(C)}^\prime = \mathcal{A(C}^\prime)$,
 i.e. all bounded operators on $\mathcal{H}_1$ commuting with  $\mathcal{A(C)}$ can
be obtained as weak limits of quasi-local observables in $\mathcal{C}^\prime$.

The physics of a system characterized by a pair $\{ \mathcal{A}, \alpha\}$ can be derived from the representation theory of such pair. For zero temperature and zero density systems, the physical representations are the \textit{covariant positive-energy representations}. 
A representation $j$ of $\mathcal{A}$ on a separable Hilbert space, $\mathcal{H}_j$, is a covariant
positive-energy representation, iff there exists a unitary representation, $U_j$ of the
covering group of the Poincaré group $\Tilde{ {\cal P}^\uparrow_+(1 ,2) }$ on $\mathcal{H}_j$ such that
\begin{equation}
\label{Uj}
j(\alpha_{a,\Lambda}(A)) = U_j(a, \tilde \Lambda) j(A)  U_j^\dagger(a, \tilde \Lambda),
\end{equation}
where $(a, \tilde \Lambda)$ is an element of $\Tilde{ {\cal P}^\uparrow_+(1 ,2) }$ projecting onto $(a,\Lambda) \in  {\cal P}^\uparrow_+(1 ,2) $. Furthermore
\[
U_j(a, \id) = e^{i a \cdot P_j}.
\]
where $P_j$ is the 3-momentum on $\mathcal{H}_j$ and it is assumed that it satisfies the relativistic spectrum condition, that is, that its spectrum is contained in the closure of the forward light cone.
The space of physical states of the theory
is defined as
\begin{equation}
\label{Hilb}
 \mathcal{H} \equiv \oplus_{j \in L} \mathcal{H}_j   
\end{equation}
where the sum extends over the list $L$ of all inequivalent and irreducible physical representations of $\mathcal{A}$.
Let us recall a basic property (Schur lemma) of the irreducible representations that we will use
later: if $B$ is a bounded operator in $\mathcal{H}_j$ commuting with all the elements of
$j(\mathcal{A}), j$ irreducible, then $B$ is a multiple of the identity. In symbols:
\begin{equation}
\label{B}
B \in j(\mathcal{A})^\prime  \Leftrightarrow B = \lambda \id,
\end{equation}
 for some $\lambda \in \C$.
We can now make more precise the statement of the Buchholz-Fredenhagen theorem \cite{buchholz1982locality}:  for the physical representations considered (also for those non-irreducible) for an
arbitrary space-like cone $\mathcal{C}$ there is an onto isometry
\[
V_j^\mathcal{C} : \mathcal{H}_1 \to \mathcal{H}_j
\]
such that
\begin{equation}
\label{rhoV}
j(A) = V_j^\mathcal{C} A (V_j^\mathcal{C})^\dagger, \forall A\in   \mathcal{A(C}^\prime)\cap \mathcal{A}
\end{equation}
Using $V_j^\mathcal{C}$ we can construct a representation $\rho_j^\mathcal{C}$ of $ \mathcal{A}$ in $\mathcal{H}_1$ equivalent to $j(A)$ by setting
\[
\rho_j^\mathcal{C}(A) \equiv (V_j^\mathcal{C})^\dagger j(A) V_j^\mathcal{C} ,  \forall A \in  \mathcal{A}
\]
and from (\ref{rhoV}) it follows that 
\begin{equation}
\label{rhoA}
 \rho_j^\mathcal{C}(A)=A, \quad \forall  A\in   \mathcal{A(C}^\prime)\cap \mathcal{A}. 
\end{equation}

\textit{Remark} To link this construction with the field operators in the QFT  models of anyons discussed in section 5, for example one $\mathcal{H}_j$ could be the Hilbert space with one-anyon and $j$ characterizes its charge. Then the corresponding isometry  $V_j^\mathcal{C}$ can be constructed heuristically considering a field operator mentioned in section 5, $\hat\phi(\mathcal{C}^0(f,g;E))$, with support in the space-like cone $\mathcal{C}^0$. If, as expected, $\hat\phi(\mathcal{C}^0)$ is densely defined and 
closed it has polar decomposition
\begin{equation}
\label{polard}
\hat\phi(\mathcal{C}^0)= U^{\mathcal{C}^0} |\hat\phi(\mathcal{C}^0)|
\end{equation}
where $|\hat\phi(\mathcal{C}^0)|$ is a positive operator of charge 0, thus leaving all superselection sectors invariant, while $U^{\mathcal{C}^0}$ is an operator of charge corresponding to $j$  which maps the orthogonal complement of the null space of $|\hat\phi(\mathcal{C}^0)|$ isometrically into a subspace of $\mathcal{H}$. Heuristically we expect the partial isometry $U^{\mathcal{C}^0}$ to commute with $\mathcal{A}((\mathcal{C}^0)^\prime)$ and therefore by duality $(U^{\mathcal{C}^0})^\dagger U^{\mathcal{C}^0}$ is a projection in  $\mathcal{A}(\mathcal{C}^0)$. Then, by a theorem of Borchers \cite{borchers1967remark} (see also e.g. \cite{d1989technical}) there is an element $B \in \mathcal{A}(\mathcal{C})$ with ${\mathcal{C}}$ a cone containing properly $\mathcal{C}^0$ such that $V^{\mathcal{C}} \equiv U^{\mathcal{C}^0} B $ is an onto isometry  that commutes with all the observables in the space-like complement of the cone ${\mathcal{C}}$. The restriction of this isometry to the vacuum sector, $V_j^{\mathcal{C}}$, is the desired isometry.
Note that this isometry is only defined between $\mathcal{H}_1$ and $\mathcal{H}_j$, not in the entire Hilbert space of the theory as was the standard charged field $\hat\phi(\mathcal{C}(f,g;E))$.

There is however a problem with $\rho_j^\mathcal{C}(A)$: although it is a bounded operator it is not a quasi-local observable, because, as also appears from the previous Remark, $\rho_j^\mathcal{C}$ affects the observables in $\mathcal{C}$ which is not a bounded region. It is therefore
natural to enlarge the algebra $\mathcal{A}$ to an algebra $\mathcal{B}$ containing the quasi-local
observables localized in space-like cones. These observables however feel the
non-trivial topology of the manifold of space-like asymptotic directions and the
enlargement from $\mathcal{A}$ to an algebra $\mathcal{B}$ cannot be done globally on the circle of space directions at
infinity. We need a variant of the forbidden direction discussed in section 5 for the models of QFT of anyons. It results from the following construction.
We choose a space-like cone "forbidden"  of arbitrarily small but finite opening angle $\mathcal{C}^a$, the analogue of the forbidden direction $\xi$ of section 5.
We define 
\[
\mathcal{B}^{\mathcal{C}^a} \equiv \overline{\bigvee_{x \in \R^{1,2} } \mathcal{A}((\mathcal{C}^a + x)^\prime)}^n.
\]
Clearly $\mathcal{B}^{\mathcal{C}^a}$ contains  $\mathcal{A}$ for any choice of $\mathcal{C}^a$ but also contains all the
 (weak limits of) quasi-local observables whose asymptotic directions are space-like
separated from the asymptotic directions of $\mathcal{C}^a$.
For a fixed $\mathcal{C}^a$ we call a cone $\mathcal{C}$ admissible
 if $\mathcal{C}$ is spacelike w.r.t. $\mathcal{C}^a + x$ for some $x$.
The crucial point is that if $\mathcal{C}$ is  admissible the map $\rho_j^\mathcal{C}$ extends to a morphism of the algebra $\mathcal{B}^{\mathcal{C}^a}$. For a fixed representation $j$ if we consider a different admissible cone $\Tilde{\mathcal{C}}$ and the associated morphism $\rho_j^{\Tilde{\mathcal{C}}}$ of $\mathcal{B}^{\mathcal{C}^a}$, since the representations $\rho_j^{\mathcal{C}}(\mathcal{A})$ and $\rho_j^{\Tilde{\mathcal{C}}}(\mathcal{A})$ are unitarily equivalent, there exists a unitary operator on $\mathcal{H}_1$,  $\Gamma(\rho_j^{\mathcal{C}},\rho_j^{\Tilde{\mathcal{C}}})$, for simplicity denoted  $\Gamma_j(\mathcal{C},\Tilde{\mathcal{C}})$ and called charge transport operator, such that
\begin{equation}
\label{rhotilde}
\rho_j^{\mathcal{C}}(A) = \Gamma_j(\mathcal{C},\Tilde{\mathcal{C}}) \rho_j^{\Tilde{\mathcal{C}}}(A) \Gamma_j(\mathcal{C},\Tilde{\mathcal{C}})^\dagger, \quad \forall A \in \mathcal{A}.   
\end{equation}
Let $\mathcal{S}$ denote a space-like cone with $\mathcal{C}\cup \Tilde{\mathcal{C}} \subset \mathcal{S} $, then for $A \in \mathcal{A}(\mathcal{S}^\prime) $ from (\ref{rhoA}) we have
\begin{equation}
A = \Gamma_j(\mathcal{C},\Tilde{\mathcal{C}})  A \Gamma_j(\mathcal{C},\Tilde{\mathcal{C}})^\dagger    
\end{equation}
and then by duality $ \Gamma_j(\mathcal{C},\Tilde{\mathcal{C}}) \in \mathcal{A}(\mathcal{S}) $.

\textit{Remark}
Note that one can only say that $\Gamma_j(\mathcal{C},\Tilde{\mathcal{C}})$ is in the algebra related to     $\mathcal{S}$ not to $\mathcal{C}\cup \Tilde{\mathcal{C}}$.
This is clear e.g. from the construction of the rotated cone in the model (\ref{CS}) discussed before (\ref{spin}): for a $\varphi < 2 \pi$ angle the difference between the rotated cone and the original    one involves a smeared Wilson loop connecting the two cones. Its support is in a third cone containing both the original and the rotated cone, not in the union of the original and the rotated cone [see Fig.14 (a)]. Another source of a support not in $\mathcal{C}\cup \Tilde{\mathcal{C}}$ in the model (\ref{HiggsCS}) is a defect line connecting the apexes of the two cones.
\begin{figure}[h]
\centering
\includegraphics[width=0.9\textwidth]{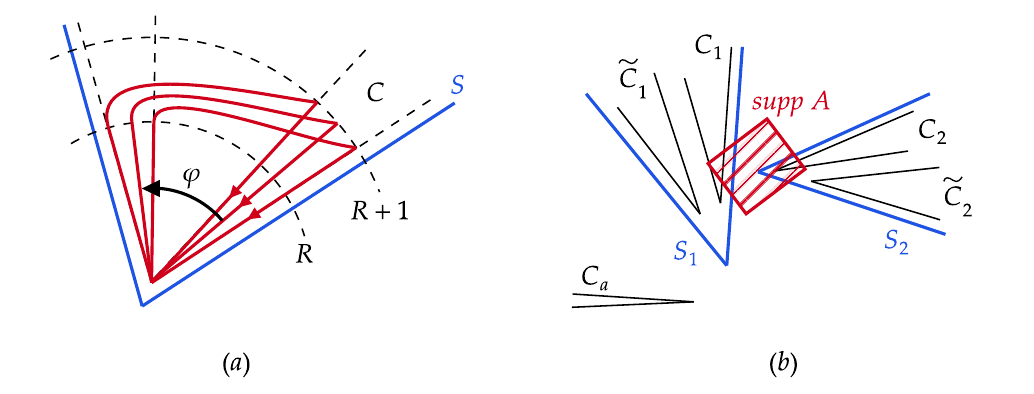}
\caption{(a) Schematic representation of the smeared Wilson loop resulting from the difference between the cone rotated of $\varphi$ and the original one. (b) The position of the cones used in the proof of the equality $ i \times j = j \times i$.}
\end{figure}

\subsection{Charged superselection sectors and intertwiners }  
 
We can now use the composition of morphism $\rho_j^{\mathcal{C}}, j \in L$ of $\mathcal{B}^{\mathcal{C}^a}$ to define a composition of the corresponding representations. For $i,j \in L$ , $\mathcal{C}_1,\mathcal{C}_2$ admissible, $ A \in \mathcal{A}$ we define 
$$
\rho_i^{\mathcal{C}_1} \circ \rho_j^{\mathcal{C}_2}(A) = \rho_i^{\mathcal{C}_1}(  \rho_j^{\mathcal{C}_2}(A)). 
$$
We define $i \times j$ the corresponding representation in $L$. The corresponding representation space,
$\mathcal{H}_{i \times j}$ describes states where a charged particle with superselection quantum
numbers $i$ and a particle with superselection quantum number $j$ are simultaneously
present.
One can then prove the crucial fact:
\begin{equation}
\label{ixj}
  i \times j = j \times i.  
\end{equation}

\textit{Sketch of proof} 
Consider the non-trivial case where both $\rho_i^{\mathcal{C}_1}$ and $\rho_j^{\mathcal{C}_2}$ have non-trivial action on $A$. Let furthermore $\Tilde{\mathcal{C}_1}$ and $\Tilde{\mathcal{C}_2}$ two cones such that
1) $\Tilde{\mathcal{C}_1} \cup \mathcal{C}_1$ is contained in a cone $\mathcal{S}_1$ and $\Tilde{\mathcal{C}_2} \cup \mathcal{C}_2$ is contained in a cone $\mathcal{S}_2$ with $\mathcal{S}_1,\mathcal{S}_2$ admissible, $\mathcal{S}_1 \bigtimes \mathcal{S}_2 $, 
2) $\rho_i^{\Tilde{\mathcal{C}_1}}$ and $\rho_j^{\Tilde{\mathcal{C}_2}}$ act trivially on $A$, see Fig.14 (b). 
Then
\[
\rho_i^{\mathcal{C}_1}(A)  = \Gamma_i(\mathcal{C}_1,\Tilde{\mathcal{C}_1}) \rho_i^{\Tilde{\mathcal{C}_1}}(A) \Gamma_i(\mathcal{C}_1,\Tilde{\mathcal{C}_1})^\dagger
\]
and 
\[
\rho_j^{\mathcal{C}_2}(A)  = \Gamma_j(\mathcal{C}_2,\Tilde{\mathcal{C}_2}) \rho_j^{\Tilde{\mathcal{C}_2}}(A) \Gamma_j(\mathcal{C}_2,\Tilde{\mathcal{C}_2})^\dagger.
\]
Therefore using also a consequence of (\ref{rhoA}) we have:
\begin{align}
\label{rhorho}
& \rho_i^{\mathcal{C}_1} \circ \rho_j^{\mathcal{C}_2}(A) = \Gamma_i(\mathcal{C}_1,\Tilde{\mathcal{C}_1})  \rho_i^{\Tilde{\mathcal{C}_1}} ( \rho_j^{\mathcal{C}_2}(A)) \Gamma_i(\mathcal{C}_1,\Tilde{\mathcal{C}_1})^\dagger = \nonumber\\
 &\Gamma_i(\mathcal{C}_1,\Tilde{\mathcal{C}_1})  \rho_i^{\Tilde{\mathcal{C}_1}} (\Gamma_j(\mathcal{C}_2,\Tilde{\mathcal{C}_2})  \rho_j^{\Tilde{\mathcal{C}_2}}(A) \Gamma_j(\mathcal{C}_2,\Tilde{\mathcal{C}_2})^\dagger) \Gamma_i(\mathcal{C}_1,\Tilde{\mathcal{C}_1})^\dagger  = \nonumber \\
 &\Gamma_i(\mathcal{C}_1,\Tilde{\mathcal{C}_1}) \Gamma_j(\mathcal{C}_2,\Tilde{\mathcal{C}_2}) A  \Gamma_j(\mathcal{C}_2,\Tilde{\mathcal{C}_2})^\dagger \Gamma_i(\mathcal{C}_1,\Tilde{\mathcal{C}_1})^\dagger =  \nonumber\\
 &\Gamma_j(\mathcal{C}_2,\Tilde{\mathcal{C}_2})  \Gamma_i(\mathcal{C}_1,\Tilde{\mathcal{C}_1}) \rho_i^{\Tilde{\mathcal{C}_1}}(A)  \Gamma_i(\mathcal{C}_1,\Tilde{\mathcal{C}_1})^\dagger
 \Gamma_j(\mathcal{C}_2,\Tilde{\mathcal{C}_2})^
 \dagger = \nonumber \\ &  \Gamma_j(\mathcal{C}_2,\Tilde{\mathcal{C}_2}) \rho_j^{\Tilde{\mathcal{C}_2}}(  \rho_i^{\mathcal{C}_1} (A) )\Gamma_j(\mathcal{C}_2,\Tilde{\mathcal{C}_2})^\dagger = \rho_j^{\mathcal{C}_2} \circ  \rho_i^{\mathcal{C}_1} (A).
\end{align}

\textit{Remark} Note that obviously the commutativity of the morphisms does not imply the commutativity of the corresponding "charged fields" or better intertwiners which we will introduce later.

Under reasonable physical assumptions (finite "statistical dimensions" \cite{longo1989index}, \cite{longo1990index}) the following properties apply:

1) $\forall i, j \in L$ the representation $i \times j$ can be decomposed into a finite direct sum of  
representations in $L$: 
\begin{equation}
\label{Nkij}
i \times j \simeq \oplus_{k \in L} N^k_{i j} k
\end{equation}
where $ N^k_{i j}$ is the multiplicity of $k$ in $i \times j$. The families of numbers $\{N^k_{i j} , i,j,k \in L\}$ are called fusion rules.

2) For every irreducible representation $j \in L$ there is a 
unique "charge" conjugate representation $\bar j \in L$, such that $j \times \bar j = \bar j \times j$ contains the vacuum representation $1$ exactly once.

Let us now move on to the definition of the spin of a representation
$j \in L$. By definition in $\mathcal{H}_j$ there is a unitary representation of the covering group of
rotations $\Tilde{SO(2)}$. We denote
$\alpha_\varphi$ the automorphism considered in eq. (\ref{Uj}) corresponding to a rotation
of an angle $\varphi$. From eq. (\ref{Uj}) it follows that
\begin{equation}
\label{U2pi}
U_j(2\pi) j(A) U_j(2\pi)^\dagger = j(\alpha_{2 \pi} (A)) =j(A),
\end{equation}
because, from a physical point of view, quasi-local
observables are invariant under a rotation of $2 \pi$.

Applying (\ref{B}) to the eq. (\ref{U2pi}), since $U_j(2 \pi n) , n \in \Z$ is unitary, one deduces the existence of a real
number $S_j \in [0, 1)$ such that
\[
U_j(2 \pi n) = e^{ i 2 \pi n S_j} \id.
\]
 
$S_j$ is called the spin of the representation $j \in L$ and each sector $\mathcal{H}_j$ is an eigenspace of the operator $U_j(2 \pi n)$ belonging to the
eigenvalue $e^{ i 2 \pi n S_j}$ and a particle with superselection quantum numbers of the
representation $j$ has spin $S_j$ mod 1.

Now we introduce "charged field operators" or more properly "charged intertwiners" performing transitions between different
superselection sectors. 

Given a space-like cone $\mathcal{C}_1$ and an angle variable $\theta$ in a polar coordinate system in the time-zero plane we denote by as($\mathcal{C}_1 ) = \theta_1$ the angle of the projection in the time-zero plane of the axis of $\mathcal{C}_1$ in the polar system described above. We fix a space-like admissible reference cone which in the following we denote by $\mathcal{C}$, with apex at the origin and as($\mathcal{C}$) = 0, see Fig.15 (a).

\begin{figure}[h]
\centering
\includegraphics[width=0.9\textwidth]{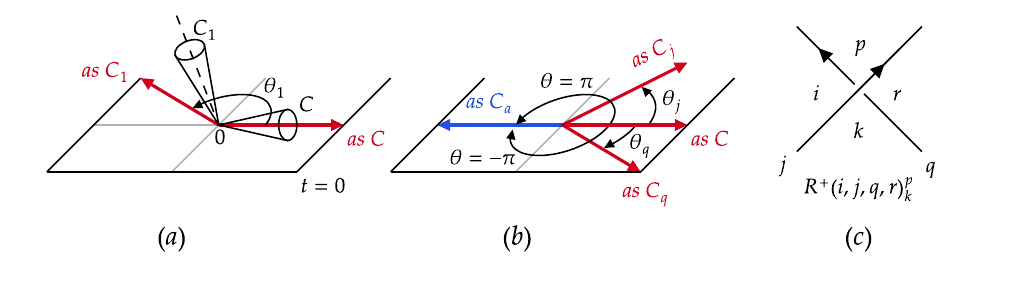}
\caption{(a) Space-like cones  $\mathcal{C}_1$, $\mathcal{C}$ and the angle $\theta_1$. (b) Geometry of the relative positions of the asymptotic directions of the cones used in the proof of braid statistics. (c) The labels associated with an $R^+$ matrix.}
\end{figure}

Given $i,j,k \in L$ with $N^k_{ij} \neq 0$ we define
charged intertwiners  $V^{ik}(\rho_j^\mathcal{C}) : \mathcal{H}_k \rightarrow \mathcal{H}_i$ as bounded operators
satisfying for $A \in \mathcal{A}$ the intertwining relations
\begin{equation}
\label{Vik}
i(\rho_j^\mathcal{C}(A))V^{ik}(\rho_j^\mathcal{C})= V^{ik}(\rho_j^\mathcal{C}) k(A).
\end{equation}
For fixed $\rho_j^\mathcal{C}$ these operators form a complex vector space, $\mathcal{V}^{ik}(\rho_j^\mathcal{C})$
of dimension $N^k_{ij}$ which can be equipped with a scalar product $( \cdot, \cdot)$ as follows. For $V_1^{ik}(\rho_j^\mathcal{C}), V_2^{ik}(\rho_j^\mathcal{C}) \in \mathcal{V}^{ik}(\rho_j^\mathcal{C})$, $V_2^{ik}(\rho_j^\mathcal{C})^\dagger V_1^{ik}(\rho_j^\mathcal{C})$ commute with $k(\mathcal{A})$, therefore it is a multiple $\lambda \in \C$ of the identity, that we identify as their scalar product. We denote by $\{V_\alpha^{ik}(\rho_j^\mathcal{C}), \alpha =1, \dots, N^k_{ij}\}$ an orthonormal basis of  $\mathcal{V}^{ik}(\rho_j^\mathcal{C})$.
More generally for $(a, \tilde \Lambda) \in \Tilde{ {\cal P}^\uparrow_+(1 ,2) }$ we define 
\begin{equation}
\label{Vika}
 V^{ik}(\rho_j^\mathcal{C},(a, \tilde \Lambda)) = U_i(a, \tilde \Lambda) V^{ik}(\rho_j^\mathcal{C} ) U_k^\dagger(a, \tilde \Lambda).
\end{equation}
Using eqs. (\ref{Uj}),(\ref{Vik}) we see that
\begin{equation}
\label{Vikal}
i(\alpha_{a,\Lambda} \circ \rho_j^\mathcal{C} \circ \alpha^{-1}_{a,\Lambda}(A)) V^{ik}(\rho_j^\mathcal{C},(a, \tilde \Lambda))= V^{ik}(\rho_j^\mathcal{C},(a, \tilde \Lambda)) k(A)    
\end{equation}
and setting $\alpha_{a,\Lambda} \circ \rho_j^\mathcal{C} \circ \alpha^{-1}_{a,\Lambda} \equiv \rho_j^{\mathcal{C}_{a,\Lambda}}$ eq. (\ref{Vikal}) proves: $ V^{ik}(\rho_j^\mathcal{C},(a, \tilde \Lambda)) \in \mathcal{V}^{ik}(\rho_j^{\mathcal{C}_{a,\Lambda}})$. In addition, if $\mathcal{C}_{a,\Lambda}$ is admissible eq.(\ref{rhotilde}) proves that there is a charge transport operator $\Gamma_j(\mathcal{C}_{a,\Lambda},\mathcal{C})$ such that
\[
V^{ik}(\rho_j^\mathcal{C},(a, \tilde \Lambda))=i(\Gamma_j(\mathcal{C}_{a,\Lambda},\mathcal{C})) V^{ik}(\rho_j^\mathcal{C}).
\]
For the special case where $(a, \tilde \Lambda) = (0, \varphi)$, which corresponds to a rotation of $\varphi$, from eqs. (\ref{Vika}),(\ref{U2pi}) it follows
\begin{equation}
\label{vikss}
V^{ik}(\rho_j^\mathcal{C}, (0,  2 \pi n) \circ (a, \tilde\Lambda)) = e^{ i 2 \pi n (S_i - S_k)} V^{ik}(\rho_j^{\mathcal{C}}, (a, \tilde\Lambda)) ,   
\end{equation}
thus identifying $V^{ik}(\rho_j^{\mathcal{C}},(a, \tilde \Lambda))$ as a section of a $\pi_1(S^1) \simeq \Z$-bundle over the space
of space-like asymptotic directions homeomorphic to $S^1$, with transition functions
$n \in \Z \rightarrow e^{ i 2 \pi n (S_i - S_k)}$. This result generalizes to the charged intertwiners the multivaluedness related to spin found for quantum field of anyons in specific models discussed in previous sections.

A deep result in relativistic theories is the equality of  the spin of a sector and of its conjugate, more precisely for $j \in L$
\begin{equation}
\label{ssbar}
  e^{ i 2 \pi S_j }= e^{ i 2 \pi  S_{\bar j}}, 
\end{equation}
see e.g.\cite{frohlich1990braid} for a proof (and the remark after (\ref{thetajj}) for a heuristic comment for abelian anyons).

\subsection{Braid statistics of charged intertwiners}

Let $\rho_j^{\mathcal{C}_j},\rho_q^{\mathcal{C}_q}, j, q \in L$ be two morphisms localized in spacelike separated cones $\mathcal{C}_j,\mathcal{C}_q$. According to (\ref{ixj}),(\ref{rhorho}) due to the commutativity of the morphisms
$\mathcal{V}^{ir}(\rho_j^{\mathcal{C}_j} \circ \rho_q^{\mathcal{C}_q}  )=\mathcal{V}^{ir}(\rho_q^{\mathcal{C}_q} \circ \rho_j^{\mathcal{C}_j})$.
From equation (\ref{Vik})
applied twice it follows also that if  $\{V_\alpha^{ik}(\rho_j^{\mathcal{C}_j}, \alpha =1, \dots, N^k_{ij}\}$ is a basis of $\mathcal{V}^{ik}(\rho_j^{\mathcal{C}_j})$ and $\{V_\beta^{pr}(\rho_q^{\mathcal{C}_q}), \beta =1, \dots, N^r_{kq}\}$ is  a basis of $\mathcal{V}^{pr}(\rho_q^{\mathcal{C}_q})$, both $\{V_\alpha^{ik}(\rho_j^{\mathcal{C}_j}) V_\beta^{kr}(\rho_q^{\mathcal{C}_q}),  \alpha =1,$
$\dots, N^k_{ij},
\beta =1, \dots, N^r_{pq}\}$ and $\{V_\gamma^{ip}(\rho_q^{\mathcal{C}_q}) V_\delta^{pr}(\rho_q^{\mathcal{C}_q}),  \gamma =1, \dots, N^p_{iq} ,\delta =1, \dots, N^r_{pq}\}$ are in $\mathcal{V}^{ir}(\rho_j^{\mathcal{C}_j} \circ \rho_q^{\mathcal{C}_q})$ and, actually, since $N^r_{i j \times q}=N^r_{i q \times j} = \sum_{p \in L} N^p_{i j} N^r_{p q}$,
they are orthonormal bases of $\mathcal{V}^{ir}(\rho_j^{\mathcal{C}_j} \circ \rho_q^{\mathcal{C}_q})$.
It follows that it exists a unitary $N^r_{i j \times q} \times N^r_{i j \times q} $ matrix
$R(i,\rho_j^{\mathcal{C}_j},\rho_q^{\mathcal{C}_q},r)$ such that 
\begin{equation}
\label{Rpre}
 V_\alpha^{ik}(\rho_j^{\mathcal{C}_j}) V_\beta^{kr}(\rho_q^{\mathcal{C}_q})= \sum_{p, \gamma,\delta,  } R(i,\rho_j^{\mathcal{C}_j},\rho_q^{\mathcal{C}_q},r)_{k \alpha \beta}^{p \gamma \delta}  V_\gamma^{ip}(\rho_q^{\mathcal{C}_q}) V_\delta^{pr}(\rho_q^{\mathcal{C}_q}).
\end{equation}
Suppose that [see Fig.15 (b)]
\begin{equation}
\label{asp}
\text{as}(\mathcal{C}^a) - 2 \pi < \text{as}(\mathcal{C}_q)<\text{as}(\mathcal{C}_j)<\text{as}(\mathcal{C}^a),   
\end{equation}
then with a similar argument to the one used in (\ref{rhorho}) one can modify arbitrarily $\rho_j^{\mathcal{C}_j}$,  $\rho_q^{\mathcal{C}_q}$ (and , the implicitly present $\mathcal{C}^a$) in (\ref{Rpre}) without changing $R(i,\rho_j^{\mathcal{C}_j},\rho_q^{\mathcal{C}_q},r)$ provided the condition (\ref{asp}) is always satisfied during the modification. We denote the corresponding $R$-matrix $R^+(i, j, q, r)$. A similar situation occurs if we impose the condition
\begin{equation}
\label{asm}
\text{as}(\mathcal{C}^a) - 2 \pi < \text{as}(\mathcal{C}_j)<\text{as}(\mathcal{C}_q)<\text{as}(\mathcal{C}^a),   
\end{equation}
we denote the corresponding $R$-matrix $R^-(i, j, q, r)$. Therefore the statistics of charged intertwiners is given by
\begin{equation}
\label{Rint}
 V_\alpha^{ik}(\rho_j^{\mathcal{C}_j}) V_\beta^{kr}(\rho_q^{\mathcal{C}_q})= \sum_{p \in L, \gamma = 1, \dots  N^p_{iq},\delta =1, \dots, N^r_{pj}  } R^\pm(i, j, q, r)_{k \alpha \beta}^{p \gamma \delta}  V_\gamma^{ip}(\rho_q^{\mathcal{C}_q}) V_\delta^{pr}(\rho_j^{\mathcal{C}_j}).   
\end{equation}
for $\text{as}(\mathcal{C}^a) - 2 \pi < \text{as}(\mathcal{C}_j),\text{as}(\mathcal{C}_q)<\text{as}(\mathcal{C}^a)$ and $\text{as}(\mathcal{C}_j) \gtrless \text{as}(\mathcal{C}_q)$ and $R^\pm$ are called statistics $R$ matrices.
Equation (\ref{Rint}) generalizes the braid statistics found for anyonic quantum field operators in (\ref{statqft}).
The $R$-matrices satisfy a natural generalization of the identities related to the braid groups encountered before:
\begin{equation}
\label{GB}
\sum_p R^+(i, j, q, r)^p_k R^-(i, q, j, r)^l_p = \delta^l_k,
\end{equation}
and the generalization of Yang-Baxter equations:
\begin{align}
\label{GYBE}
&\sum_p R^+(i, j, q, r)^p_k R^+(p, j, m, n)^s_r R^+(i, q, m, s)^u_p = \nonumber\\ &\sum_p R^+(k, q, m, n)^p_r R^+(i, j, m, p)^u_k R^+(u, j, q, n)^s_p
\end{align}
and a similar equation obtained replacing $R^+$ with $R^-$.
To derive these equations it is useful to use a graphical notation where the equations from left to right for the $R$-symbols correspond to the drawings from bottom to top. We introduce in the drawings labels in $L$ as follows [see Fig.15 (c) ]: At each crossing we associate ordered labels to the two strands that make the crossing, corresponding to the two central labels of $R$, another label to the left of the crossing and another one to the right associated, respectively, to the initial and final labels in $R$. We also introduce a label below and above the crossing corresponding to the lower and upper indices of the $R$-matrix respectively.  
 Applied to $N$ strands in the graphical notation, equations (\ref{GB}),(\ref{GYBE}) identify the $R$-matrices as generators of representations of the groupoid of coloured braids on $N$-strands obtained from the braid group $B_N$ in its geometric definition by labelling i.e. "coloring" (in our case with elements of $L$)  consistently the strands, as e.g. the graphical representation of the Yang-Baxter equation in Fig.16
 suggests.
 \begin{figure}[h]
\centering
\includegraphics[width=0.9 \textwidth]{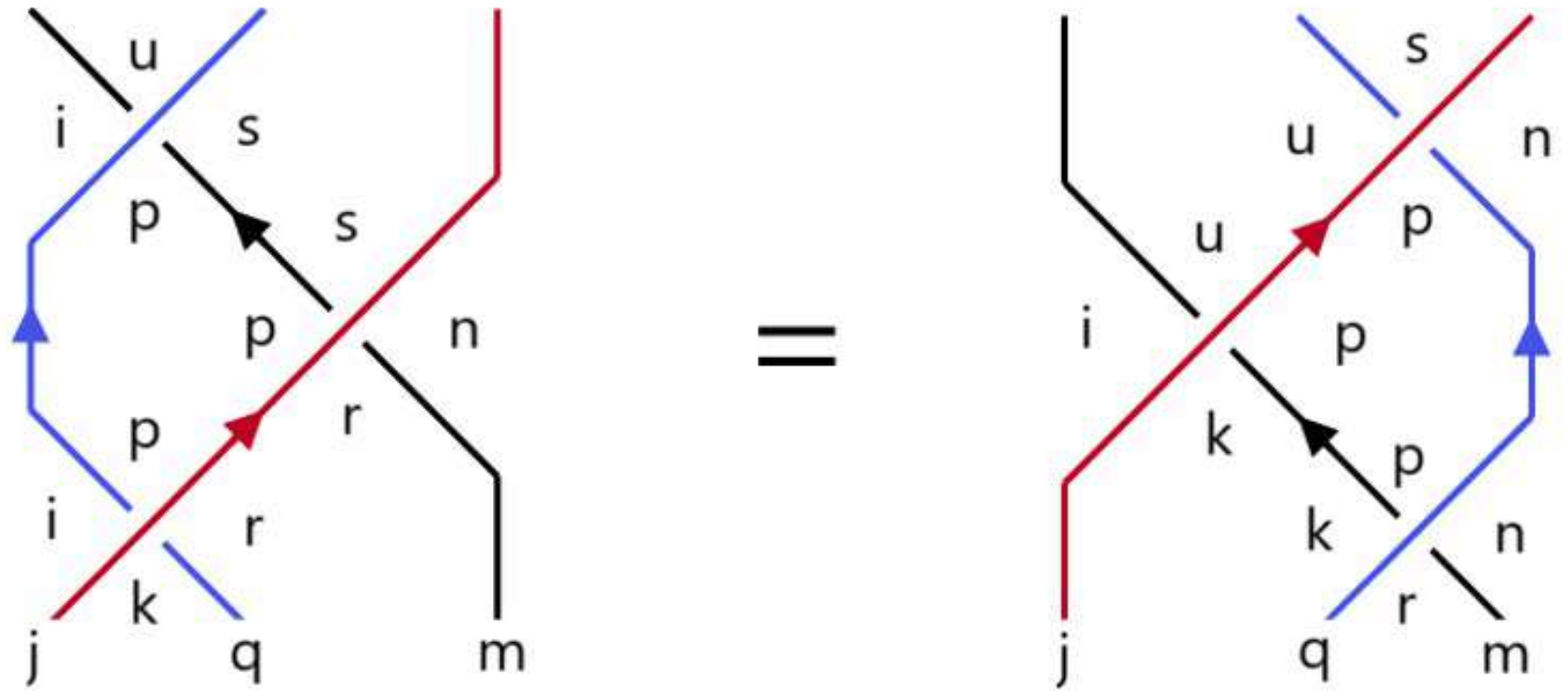}
\caption{ Graphical representation of the Yang-Baxter equation (\ref{GYBE}), where the sum of $p$ on $L$ is implicitly assumed.}
 \end{figure}
 
 If the statistics $R$ matrices satisfy $R^+ = R^- $ they provide representations of the permutation groups and the QFT obeys the permutation statistics.
 
 In particular if a morphism $\rho$ of a QFT can be localized in a bounded region, ${\cal O}$ then, since that region
is contained in  cones of arbitrary asymptotic direction, the corresponding intertwiner satisfies the permutation statistics. Therefore, in this general algebraic framework we recover the result discussed in section 3.5 that to have a non-trivial braid statistics a localization in cones and not in bounded regions is a necessary condition.

 Using the relation (\ref{Rint}) one can also answer the question on the need of parity breaking. 
 We denote $P$ the reflection with respect to a space axis, e. g. $x^1$, and we suppose that $P$ is represented on the algebra of observables ${\cal A}$ by an automorphism $\alpha_P$ and as in (\ref{Vikal}) we set $\alpha_P \circ \rho_j^\mathcal{C} \circ \alpha_P \equiv \rho_j^{\mathcal{C}^P}$. 
 
We have the following statement: if $P$ is implemented unitarily in all superselection sectors $j \in L$, i.e. $j(\alpha_P(A))= U_j(P)j(A)U_j(P)^\dagger$, with $U_j(P)$ 
unitary, then $R^+ = R^-$. 

In fact, consider in (\ref{Rint}) the case with $\text{as}(\mathcal{C}_j) >0>\text{as}(\mathcal{C}_q)$ so that the two cones are on the opposite sides of the $x^1$ axis. Clearly, the $P$ exchanges their order i.e. $\text{as}(\mathcal{C}_j^P) <0<\text{as}(\mathcal{C}_q^P)$ [see Fig.17 (a)].  

\begin{figure}[h]
\centering
\includegraphics[width=0.9\textwidth]{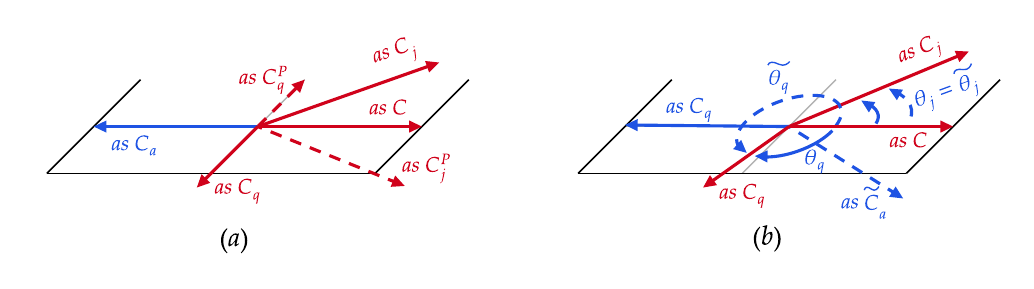}
\caption{ Geometry of the relative positions of the asymptotic directions of the cones for discussing the parity-breaking (a) and for proving the spin-statistics connection (b).}
 \end{figure}

Applying in (\ref{Rint}) $U_i(P)$ from left and $U_r(P)^\dagger$ from  right we get from the left side of the equation
\begin{align}
& U_i(P) V_\alpha^{ik}(\rho_j^{\mathcal{C}_j})U_k(P)^\dagger U_k(P) V_\beta^{kr}(\rho_q^{\mathcal{C}_q}) U_r(P)^\dagger=  V_\alpha^{ik}(\rho_j^{\mathcal{C}_j^P}) V_\beta^{kr}(\rho_q^{\mathcal{C}_q^P}) = \nonumber\\ &\sum_{p , \gamma ,\delta  } R^-(i, j, q, r)_{k \alpha \beta}^{p \gamma \delta}  V_\gamma^{ip}(\rho_q^{\mathcal{C}_q^P}) V_\delta^{pr}(\rho_j^{\mathcal{C}_j^P}).  
\end{align}
But applying them to the right side of the equation(\ref{Rint}) we get
\begin{align}
&  \sum_{p , \gamma ,\delta  } R^+(i, j, q, r)_{k \alpha \beta}^{p \gamma \delta}  U_i(P) V_\gamma^{ip}(\rho_q^{\mathcal{C}_q}) U_p(P)^\dagger U_p(P) V_\delta^{pr}(\rho_j^{\mathcal{C}_j})U_r(P)^\dagger= \nonumber \\
&\sum_{p , \gamma ,\delta  } R^+(i, j, q, r)_{k \alpha \beta}^{p \gamma \delta}  V_\gamma^{ip}(\rho_q^{\mathcal{C}_q^P}) V_\delta^{pr}(\rho_j^{\mathcal{C}_j^P}). 
\end{align}
Therefore if $R^+ \neq R^-$ the parity cannot be implemented unitarily in all superselection sectors $j \in L$.

\textit{Remark} It can still happen that $R^+ \neq R^-$ and the parity is implemented in the total Hilbert space of physical states by a unitary operator $U(P)$ if it exchanges two charged superselection sectors, in a similar way to that which takes place for the time-reversal in the FQSHE. 

 \subsection{Spin-statistics connection for charged intertwiners}

 To establish a relation between spin of charged sectors and statistics of charged intertwiners we consider two "forbidden" cones $\mathcal{C}^a$ and $\tilde{\mathcal{C}^a}$ and two morphisms, $\rho_j^{\mathcal{C}_j}$ and $\rho_q^{\mathcal{C}_q}$ of $\mathcal{B}^{\mathcal{C}^a}$ with ${\mathcal{C}_j} \bigtimes {\mathcal{C}_q}$, with both cones admissible both for $\mathcal{C}^a$ and $\tilde{\mathcal{C}^a}$.
 We also consider [see Fig.17 (b)]
 \begin{equation}
 \label{astilde1}
  \text{as}(\mathcal{C}^a) -2 \pi < \text{as}(\mathcal{C}_q) < \text{as}(\tilde{\mathcal{C}^a}) <  \text{as}(\mathcal{C}_j) < \text{as}(\mathcal{C}^a)\end{equation}
  and
  \begin{equation}
 \label{astilde2}
  \text{as}(\tilde{\mathcal{C}^a}) -2 \pi < \text{as}(\mathcal{C}_j) < \text{as}(\mathcal{C}^a) <  \text{as}(\mathcal{C}_q) < \text{as}(\tilde{\mathcal{C}^a}).\end{equation}
 Since $\mathcal{C}_q$ and $\mathcal{C}_j$ are also admissible w.r.t. $\tilde{\mathcal{C}^a}$ the corresponding morphism can be considered also for $\mathcal{B}^{\tilde{\mathcal{C}^a}}$ and we denote the corresponding charged intertwiners with a tilde.
 As a result of the relative position of the cones in eqs.(\ref{astilde1}), (\ref{astilde2}) we have
 \begin{equation}
 \label{vv}
   V_\alpha^{ik}(\rho_j^{\mathcal{C}_j}) V_\beta^{kr}(\rho_q^{\mathcal{C}_q})= \sum_{p , \gamma ,\delta} R^+(i, j, q, r)_{k \alpha \beta}^{p \gamma \delta}  V_\gamma^{ip}(\rho_q^{\mathcal{C}_q}) V_\delta^{pr}(\rho_j^{\mathcal{C}_j})   
 \end{equation}
 and
 \begin{equation}
 \label{vvtilde}
 \tilde{V}_\alpha^{ik}(\rho_j^{\mathcal{C}_j}) \tilde{V}_\beta^{kr}(\rho_q^{\mathcal{C}_q})= \sum_{p , \gamma ,\delta} R^-(i, j, q, r)_{k \alpha \beta}^{p \gamma \delta}  \tilde{V}_\gamma^{ip}(\rho_q^{\mathcal{C}_q}) \tilde{V}_\delta^{pr}(\rho_j^{\mathcal{C}_j}).
 \end{equation}
 In addition, the position of the cones suggests that $\tilde{V}_\alpha^{ik}(\rho_j^{\mathcal{C}_j})=V_\alpha^{ik}(\rho_j^{\mathcal{C}_j})$ and $ \tilde{V}_\gamma^{ip}(\rho_q^{\mathcal{C}_q})$ is obtained from $ V_\gamma^{ip}(\rho_q^{\mathcal{C}_q})$ by a $2 \pi$ rotation. In fact this is proved in \cite{frohlich1990braid}, so that from eq.(\ref{vikss})
 \begin{equation}
 \label{tildevv}
  \tilde{V}_\gamma^{ip}(\rho_q^{\mathcal{C}_q}) = e^{ i 2 \pi  (S_p - S_i)}V_\gamma^{ip}(\rho_q^{\mathcal{C}_q})   
 \end{equation}.
 Plugging (\ref{tildevv}) in (\ref{vvtilde}) we get
  \begin{equation}
   V_\alpha^{ik}(\rho_j^{\mathcal{C}_j}) V_\beta^{kr}(\rho_q^{\mathcal{C}_q})= \sum_{p , \gamma ,\delta} R^-(i, j, q, r)_{k \alpha \beta}^{p \gamma \delta}e^{ i 2 \pi  (S_p - S_i - S_r + S_k)}  V_\gamma^{ip}(\rho_q^{\mathcal{C}_q}) V_\delta^{pr}(\rho_j^{\mathcal{C}_j})  
 \end{equation}
 and comparing with (\ref{vv}) we get
 \begin{equation}
 \label{RR}
 R^+(i, j, q, r)_{k \alpha \beta}^{p \gamma \delta}= R^-(i, j, q, r)_{k \alpha \beta}^{p \gamma \delta}e^{ i 2 \pi  (S_p - S_i - S_r + S_k)}.
\end{equation}
Consider now the special case of the equation (\ref{RR}) for $i =1 = r, q = \bar j$. From the property of the conjugate sectors $N^{j}_{1 j}= 1 = N^{ 1}_{j \bar j} $ and from equation (\ref{Rint}), setting $i = 1, q=\bar j, r =1$, we get 
\begin{equation}
R^\pm (1, j, {\bar j}, 1)_{k \alpha \beta}^{p \gamma \delta} = R^\pm (1, j, {\bar j}, 1)_{{\bar j}, 1, 1}^{j, 1, 1} \delta^{\bar j}_k \delta^p_j \delta^\gamma_1 \delta^\delta_1 \delta_\alpha^1 \delta_\beta^1    
\end{equation}
Since $R$ is unitary one can set
\[
R^+ (1, j, {\bar j}, 1)_{{\bar j}, 1, 1}^{j, 1, 1} \equiv e^{ i 2 \pi  \theta_{j, {\bar j}}}.
\]
In relativistic theories a result similar to (\ref{ssbar}) can be proved \cite{frohlich1990braid} (see also the remark after (\ref{thetajj}) for a heuristic comment for abelian anyons) :
\begin{equation}
\label{thetajjbar}
 e^{ i 2 \pi  \theta_{j, {\bar j}}} =   e^{ i 2 \pi  \theta_{{\bar j},j}}  
\end{equation}
From the eqs.(\ref{thetajjbar}), (\ref{GB}) we then get
\[
R^- (1, j, {\bar j}, 1)_{{\bar j}, 1, 1}^{j, 1, 1} =  e^{ -i 2 \pi  \theta_{j, {\bar j}}}.
\]
and finally from (\ref{RR}) one obtains
\begin{equation}
\label{spinstat}
 e^{ i 4 \pi  \theta_{j, {\bar j}}} = e^{ i 4 \pi  S_j}.     
\end{equation}
This is the desired spin-statistics connection.
An immediate consequence of (\ref{spinstat}) is that if the QFT we are considering has permutation statistics, i.e. $R^+ = R^-$, then all its spins $s_j \in \Z/2$, because $e^{ i 4 \pi  S_j} = 1$.

\textit{Remark} The spin-statistics connection (\ref{spinstat}) depends only on the local structure discussed in section 7.1  and on  (\ref{ssbar}) and (\ref{thetajjbar}) but without using the full Poincaré covariance.
Using explicitly Poincaré covariance \cite{frohlich1991spin} (see also \cite{guido1995algebraic}) it can  be sharpened to
\[
 e^{ i 2 \pi  \theta_{j, {\bar j}}} = e^{ i 2 \pi  S_j}.
\]
as in the models considered in section 5.

 \subsection{Back to anyonic quantum fields}
 
In this subsection we return to "abelian" anyons considering the theories where all  morphisms  $\rho_j^{\mathcal{C}_j}, j \in L$ are automorphisms, i.e. $\rho_j^{\mathcal{C}_j}$ admits an inverse $(\rho_j^{\mathcal{C}_j})^{-1}$. From property 2) after eq. (\ref{Nkij}) it is  immediately obtained that the above inverse is just the charge conjugate, i.e. $(\rho_j^{\mathcal{C}_j})^{-1} = \rho_{\bar j}^{\mathcal{C}_j}$. Moreover since morphisms are now bijective, the compositions of representations $j \in L$ are still irreducible.
Thus all irreducible representations can be obtained from a
subset $L_0 \subset L$ by taking tensor powers and composing them. Also, since the composition $\times$ now admits the inverse and $\times$ is commutative, the set $L$ becomes isomorphic to an abelian discrete group $G$ (see \cite{doplicher1969fields}). Any discrete abelian group $G$ can be identified as the group (under composition) of irreducible representations of a dual abelian compact group, called the Pontrjagin dual, $\hat G$. For example, for $G \simeq \Z$ we have $\hat G \simeq U(1)$. 
Since the non trivial irreducible representations of an abelian group are one-dimensional, the non trivial spaces of intertwiners $\mathcal{V}^{ik}(\rho_j^{\mathcal{C}_j}) , j \in L$ are one-dimensional in the present case. 
The fusion rules then become simply $N_{i j}^k = 1$ if $k = i \times j$ and 0 otherwise.
The commutations rule for charged intertwiners therefore takes a simple form: since the spaces of intertwiners are one-dimensional the $R$ matrices are simply phase factors. Using the existence of an inverse one can prove that if all the fusion rules involved are satisfied so that  $R^+(i, j, q, r)^p_k$ does not vanish ( so $p = i \times q, k = i \times j, r = i \times q \times j$), then it is independent of $i$ (and $ r , p, k$) and therefore can be written as 
\begin{equation}
\label{Rab}
 R^\pm(i, j, q, r)^p_k =   e^{ \pm i 2 \pi  \theta_{j, q}},
\end{equation}
for some $\theta_{j, q} \in [0, 1)$.
In fact, let $\mathcal{C}_i, \mathcal{C}_j, \mathcal{C}_q$ (admissible) cones mutually space-like separated.  Using (\ref{Vik}) and the commutativity of $\times$  is easily proved that
\begin{equation}
\label{VVV}
V^{i i\times j} (\rho_j^{\mathcal{C}_j}) = V^{i i\times j} ( \rho_{\bar i}^{\mathcal{C}_i} \circ  \rho_j^{\mathcal{C}_j} \circ \rho_{i}^{\mathcal{C}_i} ) = V^{i 1} (\rho_{\bar i}^{\mathcal{C}_i}) V^{1 j} (\rho_j^{\mathcal{C}_j})  V^{j i \times j} (\rho_{i}^{\mathcal{C}_i}).    
\end{equation} 

For $\text{as}(\mathcal{C}^a) - 2 \pi < \text{as}(\mathcal{C}_q)<\text{as}(\mathcal{C}_j) <\text{as}(\mathcal{C}^a)$ we define $\theta_{j, q} $ by
\[
V^{1 j} (\rho_{j}^{\mathcal{C}_j}) V^{j j \times q} (\rho_q^{\mathcal{C}_q}) = e^{ i 2 \pi  \theta_{j, q}} V^{1 q} (\rho_{q}^{\mathcal{C}_q}) V^{q j \times q} (\rho_j^{\mathcal{C}_j}).
\]
Then applying (\ref{VVV}) to $V^{i i\times j} (\rho_j^{\mathcal{C}_j})$ and $V^{i \times j i\times j \times q} (\rho_q^{\mathcal{C}_q})$ one gets $\forall i \in L$
\begin{equation}
\label{VVij}
 V^{i i \times j} (\rho_{j}^{\mathcal{C}_j}) V^{i \times j i \times j \times q} (\rho_q^{\mathcal{C}_q}) = e^{ i 2 \pi  \theta_{j, q}} V^{i i \times q} (\rho_{q}^{\mathcal{C}_q}) V^{i \times q j  i \times j \times q} (\rho_j^{\mathcal{C}_j}).   
\end{equation}
The $R$-matrices in (\ref{Rab}) define abelian representations of the groupoid of coloured
braids and, in the  case where $L_0$ contains only one non-trivial element, abelian representations of the braid group.

Thanks to the simple fusion rules in the case of abelian braid statistics it is possible to combine the charged intertwiners $V^{ik}(\rho_j^{\mathcal{C}_j})$  into a
more standard quantum charged (anyonic for $R^+ \neq R^-)$ field $\phi(\rho_j^{\mathcal{C}_j})$ with superselection quantum numbers $j$,
defined on the entire physical Hilbert space (\ref{Hilb}) [$ \mathcal{H} \equiv \oplus_{i \in L} \mathcal{H}_i$]  by
\begin{equation}
 \phi(\rho_j^{\mathcal{C}_j}) = \oplus_{i\in L}  V^{i i \times j} (\rho_{j}^{\mathcal{C}_j}).
\end{equation}

\textit{Remark} The construction of the charged quantum field in the case of non-abelian braid statistics is much more difficult; this issue is discussed in \cite{frohlich2006quantum}.

Clearly $ \phi(\rho_j^{\mathcal{C}_j}): \mathcal{H}_{i \times j} \to \mathcal{H}_i$ and if $\pi(B)$ is a representation of $B \in \mathcal{B}^{\mathcal{C}^a}$ in $ \mathcal{H}$; from (\ref{Vik}) we have
\begin{equation}
\pi(\rho_j^{\mathcal{C}_j}(B))  \phi(\rho_j^{\mathcal{C}_j}) = \phi(\rho_j^{\mathcal{C}_j}) \pi(B). 
\end{equation}
Therefore $ \phi(\rho_j^{\mathcal{C}_j})$ plays the role of the isometry $(V^{\mathcal{C}})^\dagger$ discussed in the context of the polar decomposition (\ref{polard}).

In addition, as a consequence of eq. (\ref{VVV}) for $\text{as}(\mathcal{C}^a) - 2 \pi < \text{as}(\mathcal{C}_j),\text{as}(\mathcal{C}_q)<\text{as}(\mathcal{C}^a)$ and $\text{as}(\mathcal{C}_j) \gtrless \text{as}(\mathcal{C}_q)$ 
\begin{equation}
\label{abebra}
 \phi (\rho_{j}^{\mathcal{C}_j}) \phi (\rho_q^{\mathcal{C}_q}) = e^{ \pm i 2 \pi  \theta_{j, q}} \phi (\rho_{q}^{\mathcal{C}_q}) \phi(\rho_j^{\mathcal{C}_j}).    
\end{equation}
Equation (\ref{abebra}) is the model-independent general structure  of the abelian braid statistics for anyonic quantum fields, generalizing the results in specific models of sect. 5.

Let us add a final comment on the spin. Consider the case where $j$ is the only non-trivial element of $L_0$. We denote for simplicity by $nj, n \in \Z$ the $j^{\times n}$  sector.
Defining a representation of $(a, \tilde \Lambda) \in \Tilde{ {\cal P}^\uparrow_+(1 ,2) }$ in the physical Hilbert space setting $U(a, \tilde \Lambda) = \oplus_n U_{nj}(a, \tilde \Lambda)$ and defining as in  (\ref{Vika}) a charged field $ \phi(\rho_j^\mathcal{C},(a, \tilde \Lambda)) = U(a, \tilde \Lambda) \phi(\rho_j^\mathcal{C} ) U^\dagger(a, \tilde \Lambda)$, we see from (\ref{vikss}) that such field on the space satisfies
\begin{equation}
\label{phisec}
\phi(\rho_j^\mathcal{C}, (0,  2 \pi m) \circ (a, \tilde\Lambda)) = (\oplus_n e^{ i 2 \pi m (S_{nj} - S_{(n+1)j})}) \phi(\rho_j^{\mathcal{C}}, (a, \tilde\Lambda)).
\end{equation}
Equation (\ref{phisec}) identifies $ \phi(\rho_j^\mathcal{C},(a, \tilde \Lambda))$ as a section of a $\Z$-bundle over the space of space-like asymptotic directions, with transition functions $m \in \Z \to \oplus_n e^{ i 2 \pi m (S_{nj} - S_{(n+1)j})}$.

In this general framework we now prove the spin-statistics connection and  spin addition rule (\ref{SN}) for a single anyon species:
\begin{equation}
\label{sn}
S_{nj}= n^2 S_j \mod \Z.    
\end{equation}

We need a preliminary result:
\begin{equation}
\label{thetajj}
e^{ i 2 \pi  \theta_{j, j}} = e^{ - i 2 \pi  \theta_{j, {\bar j}}}    
\end{equation}

Before proving this result we give an heuristic intuition of it: let's look at Fig. 5 (b); it is associated with the positively oriented exchange of two identical anyons, so to $e^{ i 2 \pi  \theta_{j, j}}$. If we interpret the reversed arrow as describing the anti-anyon, and we rotate the figure of $\pi/2$ we see that it describes the negatively oriented exchange of an anyon and an anti-anyon, then associated with $e^{ - i 2 \pi  \theta_{j, {\bar j}}} $. Similarly, comparing Fig. 5 (b) rotated by $\pi/2$ and $-\pi/2$ heuristically suggests the equality (\ref{thetajjbar}). Furthermore, using the rubber band lemma, the comparison of Fig. 5(b) with its rotation of $\pi$ suggests ({ssbar}).

To prove (\ref{thetajj}) we consider eq. (\ref{VVij}) first with $i = {\bar j}$, $q = j$, replacing $\mathcal{C}_q$ by $\tilde{\mathcal{C}_j}$ and we use $ j \times {\bar j} = 1$ obtaining
\begin{equation}
\label{theta1}
 V^{ {\bar j} 1} (\rho_{j}^{\mathcal{C}_j}) V^{1  j } (\rho_j^{\tilde{\mathcal{C}_j}}) = e^{ i 2 \pi  \theta_{j, j}} V^{{\bar j} 1} (\rho_{j}^{\tilde{\mathcal{C}_j}}) V^{1  j } (\rho_j^{\mathcal{C}_j}).   
\end{equation}
Then we consider eq. (\ref{VVij}) with $i=1$, $q = {\bar j}$ and again replacing $\mathcal{C}_q$ by $\tilde{\mathcal{C}_j}$; we obtain
\begin{equation}
\label{theta2}
 V^{1 j} (\rho_{j}^{\mathcal{C}_j}) V^{j  1 } (\rho_{\bar j}^{\tilde{\mathcal{C}_j}}) = e^{ i 2 \pi  \theta_{j, {\bar j} }} V^{1 {\bar j} } (\rho_{\bar j}^{\tilde{\mathcal{C}_j}}) V^{ {\bar j} 1 } (\rho_j^{\mathcal{C}_j}).   
\end{equation}
Multiplying equation (\ref{theta2}) from the right by 
$(V^{j 1} (\rho_{\bar j}^{\tilde{\mathcal{C}_j}}))^\dagger =  V^{1 j} (\rho_{j}^{\tilde{\mathcal{C}_j}})$ and from the left by $(V^{1 {\bar j} } (\rho_{\bar j}^{\tilde{\mathcal{C}_j}}))^\dagger = V^{ {\bar j} 1 } (\rho_j^{\tilde{\mathcal{C}_j}})$
as a result we find (\ref{theta1}) with the two cones exchanged and $\theta_{j, j}$ replaced by $ \theta_{j, {\bar j}}$, thus proving
(\ref{thetajj}).
Now we apply (\ref{RR}) with $r= (n+1)j, q=j, i= (n-1)j$ and thn $p= nj=k$; we get
\begin{equation}
\label{sadd}
e^{ i 2 \pi  \theta_{j, j}} = e^{- i 2 \pi  \theta_{j, j}}  e^{ i 2 \pi  (2 S_{nj} - S_{(n-1)j} - S_{(n+1)j})}.  
\end{equation}
Setting $n=0$, using $S_0= 0$,$S_j=S_{-j}$ and (\ref{thetajj}) we get the spin-statistics connection (\ref{spinstat}) and inserting it in (\ref{sadd}) we obtain
\begin{equation}
\label{1s}
1 =  e^{ i 2 \pi  (2 (S_{nj}+ S_j) - S_{(n-1)j} - S_{(n+1)j})}
\end{equation}
and by induction (\ref{sn}).

\textit{Remark} To make contact with anyonic quantum fields in sect. 5, one should identify $\theta_{j, j}$ with $\theta$. But since  $ \phi(\rho_j^{\mathcal{C}_j})$ plays the role of the isometry $(V^{\mathcal{C}})^\dagger$ (and not $V^{\mathcal{C}}$) in  the polar decomposition (\ref{polard}), in view of (\ref{phisec}) we should identify $S_j$ with $-S$, so that still $\theta = S$ mod $\Z$, see \cite{marchetti1992anyon}.

\textit{Final Remark} Many of the mathematical structures discussed in this section have an analogue in conformal field theories in 1+1 dimensions, as one can see in the contribution by C. Northe \cite{northe2024young} in this volume, and are discussed with greater similarity  in the algebraic approach in \cite{fredenhagen1989superselection},\cite{fredenhagen1992superselection},\cite{gabbiani1993operator}. From a physical point of view this is not entirely unexpected, as we have seen in section 4 that a Chern-Simons theory in a two-dimensional bulk hosting anyons has boundary chiral modes. 
We close this section by listing some open problems of the algebraic approach presented: 1) To clarify the connection between the formalism in 2+1 dimensional systems with boundary and the theory 1+1 dimensional emerging on that boundary. The analogous theory on the edge in the FQHE, as is well known, is conformal. A further generalization is obtained by interpreting the boundary as  defect, which implies the extension to 2+1 dimensions of the ideas and formalism discussed for the conformal theory in 1+1 dimensions e.g. in \cite{bischoff2016phase} and in references therein. 2) To give a formulation for the non-relativistic theory which covers the description of models of systems where anyons can be found experimentally, perhaps making more precise and extending the ideas presented at the end of ref.\cite{frohlich1991spin}.

\section{Appendix A: de Rham currents}

Since the theory of de Rham currents \cite{de2012differentiable} may not be familiar to all readers, in this appendix are presented the basic ideas useful for the applications made in the text.
In somewhat vague terms a de Rham $p$-current is a distributional generalization of a $p$-form.
We recall in a brief and mathematically somewhat imprecise way  what a distribution is and few of its properties that we need. The historically paradigmatic example
of distributions are the Dirac delta function and its derivatives.
We start by considering a space of test-functions. For more concreteness we discuss the case of the Schwartz space ${\cal S}(\R^m)$ of test functions
of rapid decrease, that is the vector space of ${\cal C}^\infty$ functions $\varphi : \R^m \rightarrow \C$  which decay at infinity together with all their derivatives more rapidly
than the inverse of any power of the coordinates. (Other interesting spaces of test functions are obtained by requiring a compact support for $\varphi$ and these test functions can also be introduced on manifolds.)
The space of (tempered) distributions ${\cal S}'(\R^m)$ is defined as the set of all linear and continuous (in a suitable topology) functionals $F$ on  ${\cal S}(\R^m)$
\begin{align*}
F :&{\cal S}(\R^m) \rightarrow \C \\
& \varphi \mapsto F(\varphi).    
\end{align*}
Two distributions $F_1, F_2$ are equal iff $F_1(\varphi)=F_2(\varphi), \forall \varphi \in {\cal S}(\R^m) .$
A distribution $F$ is called \textit{regular} if there is a function $f$ on $\R^m$ such that
\begin{equation}
F(\varphi) = \int 
f(x) \varphi(x) d^m x, \forall \varphi \in {\cal S}(\R^m) .    
\end{equation}
A typical non-regular distribution is the Dirac delta function $\delta_a$ in $\R^m$, where $a \in \R^m$ and
\begin{equation}
\label{delta}
\delta_a (\varphi) = \varphi(a).
\end{equation}
For a generic distribution, the value in a point $x$ is not a well-defined quantity.
However, it is customary, especially among physicists, to use the so-called symbolic notation, i.e. the introduction of a formal function
$F(x)$ to write
\begin{equation}
F(\varphi) = \int 
F(x) \varphi(x) d^m x
\end{equation}
also for non-regular distribution. For example (\ref{delta}) in symbolic notation is written
\begin{equation}
\int 
\delta(x -a) \varphi(x) d^m x = \varphi(a).
\end{equation}
In the space of distributions one can introduce distributional derivatives. Partial derivatives, $\partial_\mu F$, of $F \in {\cal S}'(\R^m)$, still belong to ${\cal S}'(\R^m)$ and are defined by
\begin{equation}
(\partial_\mu F)(\varphi) = -F(\partial_\mu \varphi ), \forall \varphi \in {\cal S}(\R^m) .
\end{equation}
This definition can be formally obtained in symbolic notation by integration  by parts, but even for regular distributions it does not always coincide with the ordinary derivative on $f$ if $f$ is not sufficiently smooth.
Since $\varphi$ is ${\cal C}^\infty$, then $F$ is  infinitely differentiable in the sense of distributions and distributional derivatives always
commute. In fact,
\begin{equation}
\label{com}
(\partial_\mu \partial_\nu F)(\varphi) = -(\partial_\nu F)(\partial_\mu \varphi) = F(\partial_\nu \partial_\mu \varphi) = F(\partial_\mu \partial_\nu \varphi) = (\partial_\nu \partial_\mu F)(\varphi), \forall \varphi.
\end{equation}

We are ready for the 
\begin{definition}[$p$-current]
A (de Rham) $p$-current (or current of
degree or rank $p$) on $\R^m$ is a functional, $T_p$, defined on the space of all smooth
$(m-p)$-forms, $\alpha$, with compact support, that is linear and continuous in
the sense of distributions. $m-p$ is called the dimension of $T_p$.
\end{definition}
Currents of rank 0 can be identified as the Hodge dual of ordinary distributions and currents of dimension 0 are distributions.
If $T_{\m_1\cdots \m_p}, \m_1< \m_2\cdots <\m_p$ are $ \binom{m}{p}$ distributions, then 
\be\label{pcurrent}
 T_p=\frac{1}{p!}\, T_{\m_1\cdots \m_p}dx^{\m_1}\!\wedge\cdots\wedge dx^{\m_p}\,
 \ee
is a $p$-current and each current in $\R^m$ can be represented by a differential form whose
coefficients are distributions.
 The exterior product between two currents
is not always defined, but the linear algebraic properties of the $p$-forms naturally extend to the $p$-currents.  The exterior derivative $d$ is still well defined and nilpotent of order 2, i.e. $d^2 = 0$, because distributional partial derivatives always commute, see (\ref{com}).

The notion of current generalizes the notion of form, in fact a $p$-form $A$ defines a $p$-current by setting
\be
\label{AA}
A(\alpha)= \int A \wedge \alpha.
\ee
As with distributions, it is customary to use a symbolic  notation for currents which adopt the notation (\ref{AA}) also for currents not induced by forms.
But even more important in the space of currents can be defined a map sometimes called Poincaré dual (extending in fact Poincaré
duality for closed forms), which associates to each $p$–dimensional surface $\Sigma_p$ (or more generally $p$-chain in algebraic topology) with $p \leq m$ a $(m-p)$ current, i.e. a current of dimension $p$, $J_{\Sigma_p}$, defining
\be
J_{\Sigma_p}(\alpha) = \int_{\Sigma_p} \alpha,
\ee
thus justifying the expression "dimension $p$".
In symbolic notation 
\be
J_{\Sigma_p}(\alpha)=\int J_{\Sigma_p} \wedge \alpha
\ee
where, parametrizing $\Sigma_p$ with  $y^\mu = y^\mu(\sigma^1, \dots, \sigma^p)$ 

\begin{align}
 J_{\Sigma_p}(x) =   &  \frac{1}{(m-p)!}  \ve_{\m_1\cdots\m_{m-p}\mu_{m-p+1}\cdots\mu_m} \int_{\Sigma_p} \delta(x - y (\sigma)) \frac{\partial y^{\mu_{m-p+1}}}{\partial \sigma^{\nu_1}} \dots \frac{\partial y^{\mu_m}}{\partial \sigma^{\nu_p}} \nonumber\\
     & d \sigma^{\nu_1} \dots d \sigma^{\nu_p} dx^{\m_1}\wedge \dots dx^{\m_{m-p}}.
\end{align}

For $p=1$ one recognizes the representation of the current of a charged point-like particle, justifying the expression  "current".

If $\partial\Sigma_{p} =\emptyset$, then $d J_{\Sigma_p} =0$, that is, the closed surfaces are sent in closed currents and if the boundary $\partial \Sigma_p = \Sigma_{p-1}$ , then  $ J_{\Sigma_{p-1}} = (-1 )^p d J_{\Sigma_p}$.

\textit{Remark} The Poincaré lemma also applies to currents. This gives rise to an interesting situation for physics: it may happen that a $p$-current $T_p$ in $\R^m$ is equal to a smooth $p$-form $A_p$ except for some set of points $\varsigma$ where it exhibits distributional singularities. $A_p$ is therefore defined only in $\R^m \setminus \varsigma$. Since $\R^m \setminus \varsigma$ is not contractible if $A_p$ is closed there, it may not be exact. However $T_p$ is defined in $\R^m$, which is contractible, so as a current if it is closed it is exact. That is, even if we cannot be sure that there is a $(p-1)$-form $A_{p-1}$ which in $\R^m \setminus \varsigma$ satisfies $A_p = d A_{p-1}$, we are sure that there is a $(p-1)$-current $T_{p-1}$ such that $T_p = d T_{p-1}$.

 The concept of $p$-currents on a manifolds $M$ of dimension $m$ is obtained from that given above by replacing $\R^m$ by $M$. Of course the representation as forms with coefficients distributions applies only in the charts of $M$.
 All information on the cohomology of $M$ is already
contained in the subspace of currents given by smooth forms, as stated in the following

\begin{theorem}
    Each closed current in $M$ is cohomologous to a smooth form and if given a smooth form $\alpha$, there exists a current $J$ such that $\alpha = d J$ , then there exists a smooth form $\beta$ such that $\alpha = d\beta$. 
\end{theorem}

\section{Appendix B: Vortex/particle duality}

In this appendix we provide a heuristic "proof" that, in an appropriate limit, the models with action (\ref{HiggsCS}) and (\ref{CS}) with $\theta = 1/(2k)$ for a constant module of the complex field, 
$|\phi|(x) \equiv \rho(x)= \lambda$
are related by a vortex/particle duality, when they are in the so-called St{\"u}ckelberg \cite{stuckelberg1938interaction}, or, in the lattice, Villain \cite{villain1975theory} gauged formulation.
To justify this formulation note that if we write 
\begin{equation}
\label{rho}
\phi(x)= \rho(x) e^{i \varphi(x)},  
\end{equation}
the angular variable $\varphi(x)$ is well defined only where $\rho(x) \neq 0$, i.e. outside the center of the vortex configurations of the model. Such centers in 2+1 dimensions in the partition function are sets of closed lines, i.e. links $\mathcal{L}(\phi)$. To have well defined configurations of $\varphi(x)$ let us first eliminate these links from the Euclidean space-time. Then if we insert (\ref{rho}) into (\ref{HiggsCS}) with $\rho(x)= \lambda= \phi_0$ in the corresponding lagrangian the third term vanishes and the second term becomes
\begin{equation}
\label{XY}
\frac{\lambda}{2}(\frac{1}{i}e^{-i \varphi(x)}\partial_\mu e^{i \varphi(x)}-A_\mu(x) )^2.  
\end{equation}
If we integrate $\frac{1}{2 \pi i}e^{-i \varphi(x)}\partial_\mu e^{i \varphi(x)}$ along a knot $\omega$ in $\R^3 \setminus \mathcal{L}(\phi) $ the result is the integer homotopy class of the restriction of
$e^{i \varphi(x)}$ to $\omega$. We can reproduce this result in all $\R^3$ by introducing an integer 1-current $n_\mu(x)$ with support on a surface whose boundary is $\mathcal{L}(\phi)$, taking into account the homotopy class. Therefore we can replace (\ref{XY}) with a term 
\begin{equation}
\label{Villain}
\frac{\lambda}{2}(\partial_\mu  \varphi(x)+ 2 \pi n_\mu(x)-A_\mu(x) )^2.  
\end{equation}
where $\varphi(x)$ is globally defined in $\R^3$ and takes value in ]$-\pi,\pi$], and we also sum on integer 1-currents $n_\mu$ the Boltzmann weight corresponding to (\ref{Villain}). If we replace the matter term in (\ref{HiggsCS}) with (\ref{Villain}) the result is the St{\"u}ckelberg model with Chern-Simons term.

In the path integral formalism there is a simple way to derive in this formulation the vortex/particle duality, using  generalizations of the Fourier representations in the context of field theory.
The Fourier representations below are to be understood as excluding an irrelevant multiplicative constant, which disappears in correlation functions due to a cancellation between numerator and denominator.
Here they are formally written in the continuum, but can be made rigorous in the lattice approximation.
For Maxwell, Chern-Simons and matter terms in the lagrangian we use the following Fourier representations.
Let $B_{\mu\nu}$ be a real antisymmetric tensor field of rank 2, then
\begin{equation}
\label{AMax}
e^{-\frac{1}{2}F_{\mu\nu}^2}    = \int d B_{\mu\nu} e^{-\frac{1}{2}B_{\mu\nu}^2+i B_{\mu\nu}\partial^\mu A^\nu }.
\end{equation}
Let $\tilde{A}_\mu$ be a real vector field, then
\begin{equation}
\label{ACS}
e^{-\frac{i \theta}{2 \pi}\epsilon_{\mu\nu\rho} A^\mu \partial^\nu A^\rho}    = \int d \tilde{A}_{\mu} \exp[i \epsilon_{\mu\nu\rho}(\frac{1}{8 \pi \theta}\tilde{A}^\mu \partial^\nu \tilde{A}^\rho +\frac{1}{2 \pi}A^\mu \partial^\nu \tilde{A}^\rho)].
\end{equation}
For the matter term we use the Poisson summation formula:
\begin{equation}
\label{AVi}
\sum_{n_\mu} \exp[-\frac{\lambda}{2}(\partial_\mu  \varphi+ 2 \pi n_\mu-A_\mu )^2]= \sum_{j_\mu} \exp[-\frac{1}{2 \lambda}j_\mu^2+ i j^\mu (\partial_\mu  \varphi-A_\mu)]
\end{equation}
where $j_\mu$ are integer 1-currents.

Using (\ref{AMax}),(\ref{ACS}),(\ref{AVi}) we can write the partition function of the above St{\"u}ckelberg model as:
\begin{eqnarray}
\label{Zl}
Z(\lambda)=& \int {\cal D} A_\mu  \int {\cal D} \varphi  \sum_{n_\mu} e^{-\int\frac{1}{2}F_{\mu\nu}^2}e^{-\int\frac{\lambda}{2}(\partial_\mu  \varphi+ 2 \pi n_\mu-A_\mu )^2}  e^{-\frac{i \theta}{2 \pi}\int\epsilon_{\mu\nu\rho} A^\mu \partial^\nu A^\rho} \delta(\partial^\mu A_\mu)\nonumber\\= &\int {\cal D} A_\mu  \int {\cal D} \varphi  \int {\cal D} B_{\mu\nu}  \int {\cal D} \tilde{A}_\mu  \sum_{j_\mu} e^{-\int\frac{1}{2}B_{\mu\nu}^2+i\int B_{\mu\nu}\partial^\mu A^\nu }  e^{-\int\frac{1}{2 \lambda}j_\mu^2+ i\int j^\mu (\partial_\mu  \varphi-A_\mu)}\nonumber\\
&
e^{\frac{i}{8 \pi \theta}\int\epsilon_{\mu\nu\rho}\tilde{A}^\mu \partial^\nu \tilde{A}^\rho }e^{\frac{i}{2 \pi }\int\epsilon_{\mu\nu\rho}A^\mu \partial^\nu \tilde{A}^\rho } \delta(\partial^\mu A_\mu).
\end{eqnarray}
Integrating over $\varphi$ in (\ref{Zl}) we get the constraint $\partial^\mu j_\mu =0$ which can be solved by the Poincaré lemma for currents by introducing an integer 2-current $m_\mu[j]$ such that $j_\mu= \epsilon_{\mu\nu\rho}\partial^\nu m^\rho[j]$. By integrating then on $A_\mu$ we get the constraint
\begin{equation}
\label{Bm}
\partial^\nu(B_{\mu\nu} +\epsilon_{\mu\nu\rho} (\frac{1}{2 \pi }\tilde{A}^\rho + m^\rho[j]))=0   
\end{equation}
Again we can solve this constraint by introducing a real scalar field $\chi$ via
\begin{equation}
\label{Bphi}
(B_{\mu\nu} +\epsilon_{\mu\nu\rho} (\frac{1}{2 \pi }\tilde{A}^\rho + m^\rho[j]))=\frac{1}{2 \pi }\epsilon_{\mu\nu\rho} \partial^\rho \chi.
\end{equation}
Finally, we can write $\chi$ as the sum of a field $\tilde{\varphi}$ with values in $]-\pi, \pi]$ and an integer scalar field $\ell$ multiplied by $2 \pi$. Then the currents $m^\rho[j]+\partial^\rho \ell$ varying both $j$ and $\ell$ describe all integer 1-currents, which we denote with $m$. So we can replace the sum on $j$ for $m[j]$ and on $\ell$ simply with a sum on $m$. Taking this change into account and
by inserting (\ref{Bm}) and (\ref{Bphi}) into (\ref{Zl}), after integration on $\varphi$ and $A_\mu$, using also the contraction of the Levi-Civita $\epsilon$ symbols, one obtains
\begin{eqnarray}
\label{Zf}
Z(\lambda)=&  \int {\cal D} \tilde{A}_\mu \int {\cal D} \tilde{\varphi} \sum_{m_\mu}  e^{-\int\frac{1}{8 \pi^2}(\partial_\mu\tilde{\varphi}+2 \pi m_\mu +\tilde{A}_\mu)^2} e^{-\int\frac{1}{2 \lambda}(\partial_\nu m_\rho-\partial_\rho m_\nu)^2}\nonumber\\
&
e^{\frac{i}{8 \pi \theta}\int\epsilon_{\mu\nu\rho}\tilde{A}^\mu \partial^\nu \tilde{A}^\rho } \delta(\partial^\mu \tilde{A}_\mu).
\end{eqnarray}
Therefore in the limit $\lambda \rightarrow  +\infty$  we recover a continuum Villain version of the model (\ref{CS}) for $|\phi|= \frac{1}{2 \pi}$. The same strategy can prove the duality between the correlators of the anyonic fields in the two models.

\textit{Acknowledgments} 

I thank gratefully J{\"u}rg Fr{\"o}hlich for the joy of a long collaboration on many of the topics discussed in this review, the Organizers of the Young Researchers School 2024 Maynooth for the nice atmosphere they managed to create in the School and Farbod Rassouli who very kindly made the drawings.

\bibliographystyle{ieeetr}
\bibliography{bibliography_PAM}

\end{document}